\documentclass[journal]{IEEEtran} 									

\usepackage[latin1]{inputenc}
\usepackage{amsmath}
\usepackage{amsfonts}
\usepackage{amssymb}
\usepackage{graphicx}
\usepackage{color}
\usepackage{dsfont}
\usepackage{diagbox}
\usepackage[table]{xcolor}
\usepackage{scalerel}
\usepackage{booktabs}
\usepackage{tcolorbox}
\usepackage{mathrsfs,manfnt,graphicx,bbding,stackrel}
\usepackage{hyperref,stackrel}

\usepackage{amssymb,tikz,fancyhdr,cancel,enumerate,array,booktabs,setspace,pgf,tikz,fancyhdr,graphicx,color}

\usepackage[T1]{fontenc}
\usepackage{subfigure}
\usetikzlibrary{patterns,arrows,decorations.pathreplacing}
\usetikzlibrary{shapes,arrows,positioning}
\usepackage{pgfplots}
\usepackage{arrayjobx}
	
\newsavebox{\smlmat}
\savebox{\smlmat}{$\left[\begin{smallmatrix}1&1&0&0\\0&1&1&1\\0&0&0&1\end{smallmatrix}\right]$}

\usepackage{mathcomSTEv4}

\newcommand{\dbar}{d\hspace*{-0.08em}\bar{}\hspace*{0.1em}}

\renewcommand{\nuv}{\boldsymbol{\nu}}
\newcommand{\xiv}{{\boldsymbol{\xi}}}
\newcommand{\nuov}{\overline{\boldsymbol{\nu}}}

\newtheorem{thm}{Theorem}

\newtheorem{lem}{Lemma}

\newtheorem{defi}{Definition}
\newtheorem{rem}{Remark}

\newtheorem{example}{Example}
\newtheorem{Cor}[thm]{Corollary}

\newcommand{\rank}{{\rm rank}}
\newcommand{\spark}{{\rm SPARK}}
\newcommand{\spanM}{{\rm span}}
\newcommand{\rhom}{\rho_m(\scaleto{A_m, \boldsymbol{\alpha}}{7pt})}
\newcommand{\rhotm}{\tilde{\rho}_m(\scaleto{A_m, \xiv}{7pt})}

\newcommand{\changed}{}
\newcommand{\changedo}[1]{{#1}}

\graphicspath{{./Figures/}}

\newenvironment{noindlist}
{\begin{list}{\labelitemi}{\leftmargin=0em \itemindent=0em}}
	{\end{list}}

\title{Compressibility Measures for \\  Affinely Singular Random Vectors}

\author{Mohammad-Amin~Charusaie,
~Arash~Amini,~\IEEEmembership{Senior,~IEEE},
~and~Stefano~Rini,~\IEEEmembership{Senior,~IEEE}%
\thanks{M. Charusaie was with Sharif university of technology, Tehran, Iran, at the time of writing this manuscript and is now with Max Planck Institute for Intelligent Systems, T{\"u}ingen, Germany. (email: mcharusaie@tuebingen.mpg.de)}
	\thanks{A. Amini is with the Advanced Communication Research Institute (ACRI), department of electrical engineering, Sharif university of technology, Tehran, Iran. (email: aamini@sharif.edu).}
\thanks{S. Rini is with the electrical and computer engineering department, National Chiao Tung university, Hsinchu, Taiwan. (email: stefano@nctu.edu.tw)}
\thanks{A part of this paper was presented in the International Symposium on Information Theory, 21-26 June, 2020.}
}

\begin{document}

\maketitle

\allowdisplaybreaks

\begin{abstract}
	There are several ways to measure the compressibility of a random  measure; they include general approaches such as using the rate-distortion curve, as well as more specific notions, such as the Renyi information dimension (RID).
	The RID parameter indicates the concentration of the measure around lower-dimensional subsets of the space. 
	While the evaluation of such compressibility parameters is well-studied for continuous and discrete measures 
	, the case of discrete-continuous measures is quite subtle. 
	In this paper, we focus on a class of multi-dimensional random measures that have singularities on affine lower-dimensional subsets. This class of distributions naturally arises when considering linear transformation of component-wise independent discrete-continuous random variables. \changedo{
	To measure the compressibility of such distributions, 
	we introduce the new notion of 
	dimensional-rate bias (DRB) which is closely related to the entropy and differential entropy in discrete and continuous cases, respectively. Similar to entropy and differential entropy, DRB is useful in 
	evaluating the mutual information between  distributions of the aforementioned type.
	Besides the DRB, we also evaluate the  
	the RID of these distributions.
	}
	We further provide an upper-bound for the RID of multi-dimensional random measures that are obtained by Lipschitz functions of component-wise independent discrete-continuous random variables ($\mathbf{X}$). The upper-bound is shown to be achievable when the Lipschitz function is $A \mathbf{X}$, where $A$ satisfies {\changed$\spark({A_{m\times n}}) = m+1$} (e.g., Vandermonde matrices). When considering discrete-domain moving-average processes with non-Gaussian excitation noise, the above results allow us to evaluate the block-average RID and DRB, as well as to determine a relationship between these parameters and other existing compressibility measures.
\end{abstract}


\section{Introduction}
Data compression has  been widely utilized for cutting the  storage and transmission costs for various sources of data. The Shannon entropy  \cite{Shannon1948} is possibly the most fundamental notion of compressibility, which presents the minimum achievable rate to describe the outputs of a discrete source in bits using any lossless method. For continuous sources,  the compressibility notion needs to be cautiously defined, in view of the fact that source realizations require an infinite number of bits to be described. {\changed A similar measure} for this scenario is the differential entropy, in which a diverging term is excluded from the entropy of the quantized source outputs.
While the differential entropy is a natural extension of the entropy, it is not defined for singular or discrete-continuous sources. {\changed Further, its applicability as a compressibility measure \changedo{has been challenged in the literature} \cite{kolmogorov1956shannon}. An early alternative of differential entropy is introduced by R\'enyi in \cite{renyi1959dimension}, and studied for discrete-continuous distributions. A recent work \cite{CompressibilityRectifiable} also studied an extension of differential entropy notion for rectifiable distributions. }

Another notion of compressibility which also takes into account the lossy methods is the  Rate-Distortion Function (RDF) \cite{Shannon1959}; the RDF represents the minimum bit-rate required to describe a source within some fidelity criterion. It is well known that when the distortion tends to zero {\changed ($D\to 0$)}, the limiting value of the RDF is closely related to the differential entropy (if it exists) \cite{Zamir1994}. The compressibility notions are not limited to discrete and continuous sources: there has been some recent efforts to define \changedo{a notion of compressibility} for sequences \cite{Amini2011, Silva2014, Silva2015} and random processes \cite{wu2010renyi, Rezagah2016, Jalali2016uni, ghourchian2018}.

The R\'enyi Information Dimension  (RID) defined in \cite{renyi1959dimension} measures the compressibility of a general random variable/vector (RV), beyond the discrete/continuous cases\footnote{For brevity, we use the same acronym for random variables and vectors.}. {\changed For a mixture of discrete and continuous probability measures with real domain, the RID quantifies the fraction of the continuous part.} 
The RID plays an important role in a number of applications such as signal quantization \cite{cover2012elements}, sparse signal recovery \cite{Javanmard2013}, and chaos theory \cite{farmer1982information}.  
Despite the generality of its definition, R\'enyi evaluated the RID only for discrete-continuous RVs and absolutely continuous random vectors;  a class of singular RVs with self-similar measures was later studied in \cite{Geronimo1989}. 
\changedo{Kawabata and Dembo} in \cite{Dembo1994} showed that the RDF in the high-resolution case ($D\to 0$, where $D$ is distortion) has an asymptotic log-scaling behavior with the rate equal to half the  RID.

In this paper, we study the compressibility of $n$-dimensional random distributions that are composed of finitely or countably many components, each \changedo{having} an absolutely continuous distribution on an affine subset of $\mathbb{R}^n$. A typical example is an $n$-dimensional distribution that has singularities on a number of hyperplanes and points. \changedo{This example} is particularly helpful when studying linear transformations of independently distributed discrete-continuous RVs. More specifically, if $X_i$s are independent and have discrete-continuous measures, the random vector $A \Xv^n$, where $A$ is an arbitrary matrix and $\Xv^n = [X_1,\dots,X_n]^{\intercal}$, has singularities over affine subsets that are determined by  
$A$. 
For this class of measures, we investigate the RDF for $D\to 0$,  \changedo{introduce a new compressibility notion accordingly and draw a link between this notion and mutual information between two RVs.}
We further study Lipschitz transformations --instead of  linear mappings-- \changedo{of such RVs} and derive a bound for the resulting RID. 
As \changedo{an interesting application}, we derive various compressibility measures for discrete-domain moving-average processes with discrete-continuous excitation noise. With this result, we establish a link between the studied compressibility measures (e.g., undersampling rate in the compressed sensing problem) and the RID of the corresponding excitation noise. 
%

\subsection*{Related works}
After the notion of information and entropy were introduced in \cite{Shannon1948},  Kolmogorov studied the rate-distortion curve for continuous measures with $\ell_{\infty}$-norm (and quadratic) distortion function in \cite{kolmogorov1956shannon} under the name $\ep$-entropy. 
Specifically, he calculated the $\ep$-entropy for Gaussian processes.
Again, Shannon returned to the rate-distortion problem in \cite{Shannon1959} and provided a general lower-bound, which is now known as the Shanon lower bound (SLB).
For the case of the difference distortion function, Lin'kov provided a set of equations in \cite{linkov1965} to evaluate the SLB. Under certain conditions, he further proved that SLB is asymptotically (vanishingly small distortion values) tight. 
%
The SLB for the more general expected $\ell_p$-norm distortion functions was investigated in \cite{yamada1980}. For these distortion functions, the tightness of the SLB was again confirmed in \cite{Zamir1994} in the low-distortion regime. 
The study of SLB for discrete-continuous probability measures appeared later in \cite{Binia1988} and \cite{gyorgy1999rate}, for the one-dimensional and multi-dimensional random vectors, respectively.

In \cite{renyi1959dimension}, instead of the previously common rate-distortion function, R\'enyi introduced the notions of information dimension (RID) and the dimensional entropy (RDE) to measure the compressibility of continuous-domain RVs (whether continuous or discrete-continuous). In simple words, RID corresponds to the asymptotic log-scaling of the entropy  when the RV is uniformly quantized with a vanishing step size. 
For bounded RVs, R\'enyi observed that \changedo{as $\ep\to 0$, the $\ep$-entropy divided by $\log \frac{1}{\ep}$ has a limiting point that coincides with the RID.}
Interestingly, the case of multi-dimensional discrete-continuous random vectors is excluded from the R\'enyi's work (R\'enyi leaves this part as an open problem). 
While the RID of $n$-dimensional continuous RVs is shown to be $n$ in \cite{renyi1959dimension}, Csisz\'ar in \cite{csiszar1961some} proved the converse that the RID is strictly below $n$ if the distribution is not continuous (e.g., contains singularities). The exact evaluation of the RID for a class of self-similar probability measures with singularity was carried out in \cite{Geronimo1989}.

In \cite{Dembo1994}, a link between the RID and rate-distortion curve was found; it was shown that for a class of distortion functions, the asymptotic (vanishingly small distortion values) log-scaling of the rate-distortion function (called rate-distortion dimension or RDD) coincides with the RID.

More recently, RID was given an operational interpretation in almost lossless compression of analog memoryless sources in \cite{wu2010renyi}. 
Specifically, it was shown that in a compressed sensing problem where the high-dimensional vector  is generated by a memoryless source,  the minimum achievable undersampling rate is given by the RID of the source. The class of memoryless sources is extended to bounded $\Psi^{*}$-mixing processes  in \cite{Jalali2016uni}.
The compressed sensing problem is a typical example of projecting a high-dimensional source onto a lower dimensional subspace. Obviously, the RID of the projected source cannot exceed the RID of the original source. Interestingly, it is shown in \cite{Hunt1997} that RID remains constant with probability $1$, if the projection is selected with Haar measure among all possible projections.

The concept of compressibility has also been generalized for discrete-domain stochastic processes. 
In particular, the notion of rate-distortion dimension (RDD) is defined in \cite{Dembo1994} using the log-scaling constant of the rate distortion function under vanishing distortion condition when asymptotically large number of samples are considered. 
The RDD notion is linked to some non-asymptotic concepts considered in the past. For instance, the results in \cite{Berger1971} imply that memoryless processes (e.g., white noise) have the maximum RDD values among stationary processes. Besides, the results in \cite{Gray1970} could be used to compute the RDD value for two special cases of discrete-domain  autoregressive Gaussian processes.
%

The extension of RID to discrete-domain stochastic processes was considered in both \cite{Jalali2016uni} and \cite{koch2019}. In \cite{Jalali2016uni}, the average RID of a block of samples with increasingly large block size is defined as  the block-average information dimension (BID). It was later shown in \cite{Rezagah2016} that BID coincides with RDD under certain conditions. The generalization in \cite{koch2019}, however, relies on the log-scaling behavior of the entropy rate of the quantized samples. Unlike the BID, the latter notion always coincides with RDD.

For continuous-domain innovation processes (i.e., continuous-domain white noise), the notions of RID and RDE are defined in \cite{ghourchian2018} by vanishingly fine quantization of the time axis and the amplitude range. 

As for the rectifiable singular measures, a new compressibility measure based on Hausdorff density is introduced in \cite{CompressibilityRectifiable}.

\subsection*{Paper Organization}
\label{sec:Paper Organization}
\changedo{The remainder of the paper is organized as follows. We provide an overview of the results and notations of the paper in Section \ref{sec: overview}. Preliminary notions related to this work are introduced in  Section \ref{sec:Preliminaries}. 
 In Section \ref{sec:Affinely Singular Random Vector} we explain the main class of random variables studied in this work, the affinely singular random variables. 
%
Our study consist of multiple compressibility measures for these random variables and associated random processes. 
We describe our main results on random vectors in Section \ref{sec:Main Contributions}. 
  %
  %
  In Section \ref{sec: mi}, we investigate a new compressibility measure (i.e., DRB) for affinely singular RVs, and explain its connections with conventional mutual information between two RVs. 
%
%
Our results for random processes, specifically samples of moving average processes, are provided in Section \ref{sec:Compressibility of stochastic processes}.
%
%
%
 Finally, Section \ref{sec:ma} concludes the paper.
 }
%

\changedo{\section{Overview of the Results and Notations}}\label{sec: overview}
Sequences of independent and identically distributed (i.i.d.) RVs (alternatively known as discrete-domain white noise) are widely used in stochastic modeling of physical phenomena. Most natural signals (such as images) have correlated structures, which enables us to compress them. It is therefore common to decorrelate a signal by means of a linear transformation so as to derive the most incompressible form of the signal without losing information. The decorrelated signal is oftentimes modeled by a discrete-domain white noise. 
As a result, the signals of interest are modeled by linear transformations of i.i.d. RVs. Because of the sparsity and compressibility properties,  discrete-continuous RVs such as the Bernoulli-Gaussian law are of special interest. This in turn implies that we are dealing with linear transformations of independent discrete-continuous RVs in such models.

To measure the compressibility of the resulting linear transformations, we focus on the RDF of the outcome as $D\to0$. While the RDF of absolutely continuous random vectors and independent collection of discrete-continuous RVs are studied in the past, there is not much to mention for the case of statistically dependent collection of discrete-continuous RVs. In this paper, we study the RDF of linear transformations of discrete-continuous RVs as a subclass of statistically dependent collection of discrete-continuous RVs.

\subsection*{Contributions}
To describe the random vectors with singular distributions, we introduce \emph{affinely singular} RVs. {\changed A central concept in this definition is a \emph{$d$-dimensional affine subset $\Acal$} of $\Rbb^m$, which is defined as a set that by a shift $\Acal-\bv$ turns into a \changedo{ $d$-dimensional} vector subspace of $\Rbb^m$.

\begin{defi}\label{def: aff_sig_measure}
    Let $\{\mu_i(\cdot)\}_i$ \changedo{be a finite or countably infinite set of} absolutely continuous probability measures on distinct $e_i$-dimensional affine subsets\footnote{\changedo{A $d$-dimensional affine set in $\Rbb^m$ with $d\leq m$ is the image of the mapping $f:\Rbb^{d}\to\Rbb^{m}$, where $f$ is an affine transformation  $f(\xv^d)=A\xv^d+\bv$.}}  $\Acal_i$ of $\Rbb^m$, \changedo{for $e_i\in \{0, 1, \ldots\}$; we interpret $e_i=0$ (for which $\Acal_i$ consists of a single point) as $\mu_i(\Bcal) = \mathds{1}_{\Acal_i\in\Bcal}$.} A measure $\mu(\cdot)$ is defined as \emph{affinely singular}, if there exists a set $\{p_i\}$, where $p_i\in \Rbb^{+}\setminus \{0\}$, such that for every \changedo{measurable set $\Bcal\subseteq\Rbb^m$} we have
    \ea{
    \mu(\Bcal)=\sum_i p_i \mu_i\big(\Bcal \cap \Acal_i\big).
    }
\end{defi}
%
In Definition \ref{def: aff_sig_measure} we employ the term ``absolute continuity'' with the following \changedo{acceptation}: 
%
define a generic transformation $f_i$ from an $e_i$-dimensional Euclidean space to the set $\Acal_i$ as a combination of a rotation and shift.
With this decomposition of $f_i$,  we say that  a measure $\mu_i(\cdot)$ is an  absolutely continuous probability measures if it is absolutely continuous  with respect to the push-forward measure $\ell_i\big(f_i^{-1}(\cdot)\big)$, where $\ell_i(\cdot)$ is the Lebesgue measure on the $e_i$-dimensional Euclidean space.
%


}

In this paper, we derive the RID and DRB of affinely singular random vectors in a closed form. 
As a special application, we study these compressibility measures for linear transformation of a sub-class of affinely singular RVs (see Section \ref{sec:Affinely Singular Random Vector}), which we refer to as \emph{orthogonally singular} RVs. 
\begin{defi}\label{def:orth_sing}
%
%
Consider the RV $\Xv^n=[X_1,\dots,X_n]^{\intercal}$ 
comprised of discrete-continuous entries 
\ea{
X_i \stackrel{\rm d}{=} \nu_i X_{\cm,i} + (1-\nu_i) X_{\dm,i},
\label{eqn: indep_disc_cont}
}
where $\nu_i$ is a Bernoulli RV ($\Pr(\nu_i=1)=\alpha_i$), $X_{\cm, i}$ is a RV with absolutely continuous distribution, and  $X_{\dm, i}$ is a discrete RV.
{\changed The RV \changedo{$\Xv^n$} is said to be  orthogonally singular if
$\{X_{\cm, i}\}_{i=1}^n$ form a set of jointly continuous RVs independent of \changedo{ $\Xv_{\dm} = [X_{\dm, 1}, \ldots, X_{\dm, 2}]$ and $\nuv = [\nu_1, \ldots, \nu_n]$, while $\Xv_{\dm}$ and $\nuv_{\dm}$} could be dependent with
the joint probability mass function $P_{\nuv, {\Xv_{\dm}}}(\cdot)$.}

%
\end{defi}

Note that if $\alpha_1=\dots=\alpha_n=1$, then, $\Xv^n$ has an absolutely continuous distribution; otherwise, the distribution of  $\Xv^n$ contains singularities that are supported on hyper-planes aligned with some of the coordinate axes, which explains the name orthogonally singular.

 Now, if we consider a linear transformation of $\Xv^n$ such as 
\ea{
\Yv^m = A_m \Xv^n,
\label{eq:affinely singular}
}
then, $\Yv^m$ is likely not to be orthogonally singular as we show in Lemma \ref{lmm: Ax_has_subsets}. In this paper, besides finding the RDF of non-orthogonally singular random vectors $\Yv^m$ for {\changed limiting distortion $D\to 0$}, we derive an expression for the RID of  $\Yv^m$ that is determined by $P_{\nuv, {\Xv_{\dm}}}(\cdot)$ and the rank of some of the submatrices of $A_m$. 
Moreover, among all Lipschitz functions $f:\Rbb^n\to \Rbb^m$ of $\Xv^n$, we prove that the RID of $f(\Xv^n)$ is maximized when $f(\Xv^n)=A_m\,\Xv^n$, where $A_m$ is a  full-rank matrix with {\changed${\spark}(A_m) = m+1$}, {\changed where SPARK quantifies the minimum number of linearly dependent columns of a matrix}. 
Interestingly, for $\Xv^n$ vectors with equal marginal (i.e. one-dimensional) distributions, the RID of $A_m\,\Xv^n$ (with $A_m$ satisfying {\changed ${\spark}(A_m) = m+1$}) is neither minimized nor maximized when $X_i$s are independent.

{\changed Further, we derive a lower-bound on the RDF of affinely singular random vectors, and for every distortion $D$. We show that the lower-bound is tight for $D\to 0$. 

Finally, we derive the formulation of mutual information between two affinely singular RVs. In fact, we show that 
\ea{
\Isf(X; Y) = b(X)+b(Y) - b(X, Y),
}
where $b(\cdot)$  denotes the DRB.
}

\subsection*{Notations}

Lower case and Boldface lower case letters denote fixed scalars and vectors, respectively; capital and boldface letters denote random variables and vectors, respectively.
Matrices are also indicated with capital letters.  Sets are indicated with calligraphic capital letters.
Measures are indicated with Greek letters. 
Logarithms are taken in base $2$ by default. 
Other notations are as follows:


\noindent
$\bullet$ \underline{\emph {Random variables and distributions:}}
the set of RVs $\{\Xv_m,\ldots \Xv_n\}$ is abbreviated as $\{\Xv_i\}_{i=m}^n$.
For brevity, we define $\{\Xv_i\} = \{\Xv_i\}_{i=-\infty}^\infty$.  
When this set of random variables is used to construct a random vector, we employ the notation  $\Xv_m^n = [X_m,X_{m+1},\ldots, X_n]$ with $n \ge m$.  
Again, when $m=1$, the subscript is omitted, i.e. $\Xv_1^n=\Xv^n$. 
By abuse of notation, for a binary vector $\sv\in\{0, 1\}^n$, $\Xv^{\sv}$ denotes a random vector formed by the elements $X_i$ of $\Xv^n$, where $s_i=1$.
The absolute $\alpha$-moment (around zero) of the RV $X$ is denoted by $M_X(\alpha)=\Ebb[|X|^{\alpha}]$ for $\alpha\in \Rbb$. 
Further, by $M_X^0(\infty)$ we mean {\changed ${\rm ess\, sup} \, |X|$}. 
Equality in distribution is indicated as $\stackrel{\rm d}{=}$.
The discrete/continuous part of the RV $X$ is indicated as $X_{\dm}/X_{\cm}$, respectively.
The Bernoulli RV with success probability $p$ is indicated as ${\rm Bern}(p)$, the Gaussian distribution with parameters $\mu$ and $\sgs$ as  $\rm {Gauss}(\mu,\sgs)$, and the Bernoulli-Gaussian distribution as ${\rm Bern-Gauss}(p,\mu,\sgs)$.

Shannon entropy function is shown by $\Hsf(\cdot)$, the differential entropy by $\hsf(\cdot)$,  the mutual information by $\Isf(\cdot;\cdot)$ and the Kullback-Leibler divergence by $\Dsf(\cdot||\cdot)$. {\changed For the sake of simplicity in expressing our results, we might define a \emph{null} or $0$-dimensional random vector $\Xv$. With an abuse of notation, we assume that $\hsf(\Xv)=0$.}

 The notation $\delta_x$ refers to the Dirac's measure, where {\changed 
 \ean{\delta_x(A) = \lcb \begin{array}{c c}
 1, & x\in A\\
 0, &\text{otherwise}.
 \end{array}\rnone
 }
 }

{\changed
\noindent
$\bullet$ \underline{\emph{Set theory}:}
%
Set subtraction is shown as $\Acal\setminus\Bcal=\Acal\cap\Bcal^{c}$.  The span of a set $\Acal$ of vectors is denoted as $\spanM(\Acal)$. Minkowski difference of two sets $\Acal$ and $\Bcal$ is indicated as $\Acal-\Bcal = \{\av-\bv:\av\in\Acal, \bv\in \Bcal\}$.
For an affine set $\Acal$, $\dim(\Acal)$ stands for its {\changed Euclidean} dimension.
The sets $\{i, i+1, \ldots, j\} \subseteq \Nbb$  and $\{1, \ldots, j\}$ are abbreviated as $[i:j]$ and $[j]$, respectively.	
{\changed A set $\Mcal$ of linearly dependent vectors} is called minimally dependent, if all proper subsets of $\Mcal$ are linearly independent. The set of all minimally dependent subsets of a set $\Acal$ of vectors is shown as $\Omega(\Acal)$.
}

\noindent
$\bullet$ \underline{\emph{Vectors and matrices:}}
Given an $m\times n$ matrix $A$, we denote the $i$-th column of $A$ by $A^{[i]}$, for $i\in[n]$.
In addition, for a binary vector $\sv\in\{0, 1\}^n$, $A^{[\sv]}$ denotes the sub-matrix of $A$ formed by columns $A^{[i]}$ of $A$ for $i\in [n]$, where $s_i=1$. 
The rank, spark and the span of the matrix $A$ are represented by $\rank(A)$, $\spark(A)$ and $\spanM(A)$, respectively. 
For an arbitrary matrix $A$ (not necessarily square), $\det^{+}{(A)}$ refers to the product of the non-zero singular values of  $A$.
The conjugate transpose of the matrix $A$ is indicated as $A^{\dagger}$.
The $\alpha$-vector-norm is represented by $\| \cdot \|_{\alpha}$.
For $\vv^n \in \{0,1\}^n$,  $\vov^n$ is the vector obtained by complementing all the elements in $\vv$.
%
%
The $n$-dimensional column vector of all zeros/ones is indicated as $\zeros_{n}$/$\ones_{n}$.
Similarly, the $n \times m$  all zeros/ones matrix is indicated as $\zeros_{n \times m}$/$\ones_{n \times m}$.

We employ the notation $I_{\text{fcr}}(A)$ as a full-rank indicator function, i.e., 
\ea{I_{\text{fcr}}(A) =\left\{\begin{array}{l l}
		1 & \rank(A) = n\\
		0 & \text{otherwise}
	\end{array}\right.
}

\noindent
$\bullet$ \underline{\emph{Other notations:}}
For $\al \in [0,1]$, define $\alb=1-\al$.
%

\medskip
\noindent
The uniform quantization of the RV $\Xv^n$ with precision $m$ is defined as
{\changed
\ea{
	[\Xv_{}^n]_m \triangleq \Big[\tfrac{ \lfloor m X_{1} \rfloor}{m}, \ldots, \tfrac{ \lfloor m X_{n} \rfloor}{m}\Big] ,
	\label{eq:uniform quantization notation}
}
}
with $[\Xv_{}^n]_m \in (\Nbb/m)^n$ and where $\lfloor x \rfloor$ is the floor of $x$.

\section{Preliminaries}
\label{sec:Preliminaries}
Our main contribution in this paper revolves around singular probability measures. Therefore, in this section, we first define various types of probability measures (including the singular ones). 
Next, we review some of the known results regarding the compressibility of a sequence of RVs and stochastic processes.

%
In the following, we shall follow standard definitions of  absolutely continuous, discrete-continuous, singular \cite[Page 121]{rudin2006real}{\changed, and \changedo{\emph{$k$-regular}} measures}.

\begin{defi}[Types of measures]
Let $\Sigma$ be an \changedo{abstract} $\sigma$-algebra of $\Rbb^n$. 
	The RV
	 $\Xv^n$ with probability measure $\mu(\cdot)$ on $\Sigma$ is called
\begin{itemize}

    \item \emph{absolutely continuous} if for every set $\Scal\in \Sigma$ with zero Lebesgue measure, we have $\mu(\Scal)=0$.
    
    \item \emph{discrete}, if there exists a finite or countable subset $\Scal\subset\Rbb^n$ such that $\mu(\Rbb^n \setminus \Scal)=0$, 

    \item  \emph{singular}, if there exists a subset $\Scal\subset\Rbb^n$ with zero Lebesgue measure 
    with
    $\mu(\Rbb^n \setminus \Scal)=0$.
\end{itemize}
\end{defi}




It is easy to verify that discrete measures are special cases of singular measures. 
The well-known Lebesgue-Radon-Nikodym theorem 
mathematically formalized the fact that absolutely continuous and singular measures are the building blocks of all probability measures.
%
%
%
%
%
More precisely,  the Lebesgue-Radon-Nikodym theorem states that, for every probability measure $\mu$ on $\Rbb^n$, there exist unique singular and absolutely continuous measures $\mu_s, \mu_a$, respectively, and  $p \in [0,1]$ such that $\mu =p\,\mu_a+\,\po\mu_s$  where $\mu_a, \mu_{s}$ are absolutely continuous, and singular probability measures, respectively.
%

%
%

It is common to call a measure \emph{discrete-continuous} if the singular component in the Lebesgue-Radon-Nikodym decomposition is purely discrete, \changedo{and} $p$ is neither one nor zero.

{\changed 
Next, we define a \emph{Lipschitz manifold}. 
\begin{defi}[\cite{naumann2011measure}, Definition 2.3] \label{def: lip_man}
    A set $\Xcal\subset \Rbb^n$ is called $k$-dimensional Lipschitz manifold, if for every $x\in \Xcal$, there exists an open set $\Ucal$ containing $x$ and a bi-Lipschitz \changedo{surjective mapping $f:\Scal\to \Ucal$, where $\Scal\subseteq \Rbb^k$ is a bounded. 
    }
\end{defi}

Using the notion of $k$-dimensional Lipschitz manifolds, we introduce $k$-regular measures. 
\begin{defi}\label{def: reg}
    A measure $\mu$ on $\Rbb^n$ is called $k$-regular, if it is absolutely continuous with respect to a finite or countably infinite union of $k$-dimensional Lipschitz sub-manifolds of $\Rbb^n$.
\end{defi}

\changedo{In Definition \ref{def: reg}, by absolute continuity of a measure $\mu$ with respect to a union of $k$-dimensional manifolds $\Xcal_i$, we mean that for any measurable set $\Acal\subseteq \Rbb^n$, there exists a function $g:\Rbb^n\to\Rbb$, such that 
    \ea{
    \mu(\Acal)=\sum_{i}\int_{\Xcal_i\cap \Acal} g(\xv) \diff {\rm Vol}_k(\xv),
    }
    where ${\rm Vol}_k$ is the natural volume of the $k$-dimensional Lipschitz manifold. The differential volume is found based on the Jacobian matrix $J\big(f(\xv)\big)$ as 
    \ea{
    \diff {\rm Vol}_k\big(f(\xv)\big)=J\big(f(\xv)\big)\diff \ell_{k}(\xv),
    }
    in which $f(\cdot)$ is defined in Definition \ref{def: lip_man} and $\ell_k(\cdot)$ is the $k$-dimensional Lebesgue measure. Note that the Jacobian matrix exists almost everywhere due to Rademacher's theorem \cite[Theorem 3.1]{heinonen2005lectures} as $f(\cdot)$ is Lipschitz.
    }
    
    The above measures are sometimes called \emph{rectifiable} in the literature 
(e.g., see \cite[Definition 4.1]{de2006lecture}). However, to avoid the confusion with other definitions of rectifiable measures (e.g., see \cite[Definition 13]{wu2010renyi}), here we refer to such measures as $k$-regular.

}

\subsection{Compressibility of random vectors}\label{sec:Compressibility of random vectors}
The classical notion of entropy is well-defined for discrete-valued RVs. For continuous-valued RVs, this notion could be defined via the limiting entropy of the quantized RV \cite{balatoni1956remarks}. 
%
%

\begin{defi}[{\cite{renyi1959dimension}}]
	\label{def:RID}
	For an RV $\Xv_{}^n$, the \emph{R\'enyi information dimension} (RID) is defined as 
	\ea{
		d(\Xv_{}^n)=\lim_{m\to\infty}\tfrac{\Hsf([\Xv_{}^n]_m)}{\log m},
	}
	if the limit exists, where $\Hsf(\cdot)$ is the Shannon entropy function. 
\end{defi}

We recall that $[\Xv^n]_m$ stands for the uniform quantization of $\Xv^n$. 

\begin{defi}[{\cite{Berger1971}}]
	\label{def:RDF}
	The \emph{quadratic rate-distortion function} (QRDF) of an RV $\Xv_{}^n$ is defined as
	\ea{
	R_2(\Xv_{}^n, D)=\inf_{\mu_{\Xhv_{}^n|\Xv_{}^n}(\cdot, \cdot): \ \Ebb\|\Xv^n-\Xhv^n\|_2^2 <D} \Isf(\Xv_{}^n; \Xhv_{}^n),
	\label{eq:QRDF}
}
	where $\mu_{\Xhv_{}^n|\Xv_{}^n}$ is the conditional probability measure of $\Xhv_{}^n$ given $\Xv_{}^n$.
\end{defi}

%

\begin{table*}
	\begin{center}
		\begin{tabular}{ |l|l| c|} 
			\hline
			Measure & Acronym &  Notation  \\
			\hline
			R\'enyi information    
			dimension  & RID & $d(\Xv_{}^n)$ \\ 
			Quadratic rate-distortion function &  QRDF & 	$R_2(\Xv_{}^n, D)$  \\
			Rate distortion dimension & RDD & $ d_{R}(\Xv_{}^n) $ \\
			Dimensional rate bias & DRB & $b(\Xv_{}^n)$ \\
			\hline 
			Block-average information dimension & BID &	$d_B\big(\{\Xv_n\}\big)$ \\ 
			Information dimension rate & IDR & 		$d_I\big(\{\Xv_n\}\big)$ \\
			\hline
		\end{tabular}
	\end{center}
	\caption{A summary of the compressibility measures in Section \ref{sec:Compressibility of random vectors} and Section \ref{sec:Compressibility of stochastic processes}.}
	\label{tab: summary}
\end{table*}
\begin{defi}[{\cite{Dembo1994}}]
	\label{def:RDD}
	For an RV $\Xv_{}^n$, the \emph{rate distortion dimension} is given by
	\ea{
		d_{R}(\Xv_{}^n)=2\lim_{D\to 0^{+}}\f {R_2(\Xv_{}^n, D)}{-\log D},
	}
	if the limit exists.
\end{defi}
The next definition follows naturally from Definition \ref{def:RDD}. {\changed To the best of our knowledge, this is the first time that such a measure is defined \changed{in this generality}.}
{\changed
\begin{defi}
	\label{def:DRB}
	 If there exists \changedo{$b(\Xv^n)\in \Rbb$} satisfying
	\ea{
		\lim_{D\to0}\big|R_2(\Xv_{}^n, D)+\tfrac{d_R(\Xv_{}^n)}{2}\log \Big(\f{2\pi e D}{d_R(\Xv^n)}\Big)-b(\Xv_{}^n)\big|=0, \label{eqn: DRB}
	}
	for $d_R(\Xv^n)\neq 0$, or 
	\ea{
		\lim_{D\to 0} \big| R_2(\Xv_{}^n, D)-b(\Xv^n) \big| =0,	
	}
	for $d_R(\Xv^n)=0$, then, it is called  the  \emph{dimensional rate bias} (DRB) of $\Xv_{}^n$. In simple words, DRB is the asymptotic value of the quadratic rate-distortion function at $D\to 0$ after removing the known diverging term (when $d_R(\Xv^n)\neq 0$).
\end{defi}
}

{\changed The above definitions imply that the \changedo{RDD} characterizes the slope of rate-distortion function with respect to $\log \tfrac{1}{D}$ as $D\to 0$, while the DRB quantifies the bias of that limiting line. As an instance, for a one-dimensional continuous RV, the \changedo{RDD} is $1$ which expresses that RDF diverges similar to $\log \tfrac{1}{D}$ as $D\to 0$, while the DRB, which in here is equal to differential entropy, determines the bias of the RDF limiting line with respect to that of uniform distribution $\Ucal_{[0, 1]}$. }

\changedo{Note that the 
term $\tfrac{d_R(\Xv_{}^n)}{2}\log \big(\f{2\pi e D}{d_R(\Xv^n)}\big)$ in
\eqref{eqn: DRB} is the differential entropy of an isotropic Gaussian RV $\nv$ with bounded norm $\Ebb[\|\nv\|^2]<D$ that is supported on a $d_R$-dimensional space. This RV is a model for 
the error of approximating $\Xv^n$ with $\Xhv$ 
that appears in the formulation of QRDF in \eqref{eq:QRDF}. }

The established link between the rate distortion {\changed dimension} and the information dimension  is as follows:
%
\begin{thm}[Prop. 3.3, \cite{Dembo1994}]
		For every RV $\Xv_{}^n$ in the metric space $(\Rbb^n, \|\cdot\|_2)$,  we have that
		\ea{
			d(\Xv_{}^n)=d_R(\Xv_{}^n).
		}
\end{thm}

\begin{thm}[\cite{Zamir1994}(Corollary 1), \cite{WuThesis}(Section 2.6)] \label{thm: abs_cont_d_b}
	For an absolutely continuous RV $\Xv_{}^n$, if $\hsf(\Xv_{}^n)>-\infty$ and $\Ebb\|\Xv_{}^n\|_2^\alpha<\infty$ for some $\alpha>0$, then
	\ean{
	d(\Xv^n) = n, ~~~ b(\Xv^n) = \hsf(\Xv^n).
		}
Further, for a discrete RV $\Yv^n$, if $\Hsf(\Yv^n)<\infty$, then
	\ean{
		d(\Yv^n) = 0, ~~~ b(\Yv^n) = \Hsf(\Yv^n).	
	}
\end{thm}

\subsection{Compressibility of stochastic processes}\label{sec:Compressibility of stochastic processes}
Next, we describe three measures of compressibility for discrete-domain stochastic processes and explain their relationships.
\changedo{A summary of acronyms and notations of these measures (as well as those introduced in Section \ref{sec:Compressibility of random vectors}) is provided in Table \ref{tab: summary}.}
\begin{defi}[{\cite{Jalali2016uni}}]
For a generic stationary process  $\{\Xv_n\}$, the \emph{block-average information dimension} (BID) is defined as
	\ea{
		d_B\big(\{\Xv_n\}\big) = \lim_{n\to\infty}\tfrac{d(\Xv^n)}{n}.\label{eqn: BID_def}
	}
\end{defi}

\begin{defi}[{\cite{koch2019}}] For a generic stochastic process  $\{\Xv_n\}$, the \emph{information dimension rate} (IDR) is defined as
	\ea{
		d_I\big(\{\Xv_n\}\big) = \lim_{m\to\infty}\lim_{n\to\infty} \tfrac{\Hsf([\Xv^n]_m)}{n\log m}.
	}
\end{defi}
One can interpret the IDR as the average number of bits needed to transmit a source normalized by the maximum average number of bits in the high-resolution regime.

\begin{defi}[{\cite{Dembo1994}}] For a generic stochastic process $\{\Xv_n\}$, the RDD is defined as
	\ea{
		d_R\big(\{\Xv_n\}\big)	= 2\lim_{D\to 0^{+}}\lim_{n\to\infty} \tfrac{R_2(\Xv^n, nD)}{-n\log D}.
	}
\end{defi}

The existence of the above {\changed double} limit for stationary processes is proved in \cite[Theorem 9.8.1]{gallager1968}. 
\begin{thm}[{\cite[Theorem 9]{koch2019}}]
	For every stochastic process $\{\Xv_t\}$ we have
	\ea{
		d_R\big(\{\Xv_t\}\big) = d_I\big(\{\Xv_t\}\big)	,
	}
	provided that $d_R\big(\{\Xv_t\}\big)$ and $d_I\big(\{\Xv_t\}\big)$ exist.
\end{thm}

	\begin{defi}{\cite{wu2010renyi}}
%
Let $\{\Xv_n\}$ be a random process. 
For a given $n\in\mathbb{N}$, we call 	$(f_n,g_n)$ an $\ep$-encode-decode pair with rate $R_n$, if  $f_n:\, \mathbb{R}^n\,\to \Rbb^{\lfloor n R_n \rfloor}$, $g_n:\, \Rbb^{\lfloor n R_n \rfloor}\to \Rbb^{n}$, and
	\ea{
			\Pr\Big(g_n\big(f_n (\Xv^n)\big)\neq \Xv^n\Big)\leq \ep.
		}
Given the set of all achievable rates $R_n$ (i.e., all values of $R_n$ such an $\ep$-encode-decode pair with rate $R_n$ exists), we call $\liminf_{n\to\infty} R_n$  the minimum $\ep$-achievable rate \changedo{(or in short, $\ep$-compression rate)}. If $f_n$ is further restricted to be linear, we call the result of $\liminf$ the minimum linear-encode $\ep$-achievable  rate and denote it by $R^*(\ep)$. If $g_n$ is restricted to be Lipschitz, then, we call the result of $\liminf$ the minimum Lipschitz-decode $\ep$-achievable  rate and denote it by $R(\ep)$.
%
	\end{defi}

	\begin{lem}{\cite[Lemma $10$]{wu2010renyi}} \label{lmm: min_R}
	The minimum linear-encode $\ep$-achievable  rate $R^{*}(\ep)$ coincides with the  $\liminf_{n\to\infty} R_n$ given that there exists a Borel set $\Scal^n\subset \Rbb^n$ and a subspace $\Hcal^n\subset \Rbb^n$ of the dimension at least $\lceil (1-R_n)\, n \rceil$ such that
		\ea{
			\Pr(\Xv^n\in \Scal^n)\geq 1-\ep,	 \label{eqn: low_bound_s_n}
		}
		and 
		\ea{
			\big(\Scal^n - \Scal^n\big) \cap \Hcal^n =\{0\}.
		}
	\end{lem}

{\changed

\section{Affinely Singular Random Vectors}
\label{sec:Affinely Singular Random Vector}

The definition of affinely singular RV in Definition \ref{def: aff_sig_measure} provides a measure-theoretical description of this random object. 
As we proceed, we  often find it more convenient to describe an affinely singular RV through a \emph{constructive} approach which sees the RV as a result of an affine transformation. 
This approach also shows  that orthogonally singular RVs, as introduced in Definition \ref{def:orth_sing},  form a subclass of affinely singular RVs.
%
In addition, we show that linear transformations of orthogonally singular RVs form affinely singular RVs.

\begin{lem} \label{lem: aff_sing}
	 For $i\in\mathbb{N}$ let  $e_i\in\{1, \ldots, m\}$,   and define the affine function $\fv_i(\cdot):\Rbb^{e_i}\to\Rbb^m$ as
	 \ea{
	    \fv_i(\xv) = U_i \Big[ \xv^{\intercal}, \underbrace{0, \ldots, 0}_{m-e_i}\Big]^{\intercal}+\bv_i, \label{eqn: f_i}
	 }
	 where $U_i$ is an $m\times m$ unitary matrix and $\bv_i\in\mathbb{R}^m$ is a fixed vector. 
	 We denote the \changedo{image} of $\fv_i$ by $\Kcal_i $  and assume that $\Kcal_i$s are distinct. For an $e_i$-dimensional random column vector $\Cv_i$ with an absolutely continuous distribution, define the RV $\Zv^{(i)}$  as 
	\ea{
		\Zv^{(i)} =  \fv_i(\Cv_i). \label{eqn: Zv(i)}
	}
	%
	%
	Further, let $V_m$  be any RV supported on $\mathbb{N}$, \changedo{and assume it is}  independent of $\Zv^{(i)}$s.	
	Then, 
	\ea{
		\Zv^m =  \Zv^{(V_m)},\label{eqn: singular_subsets}
	}
	has an affinely singular probability measure as in Definition \ref{def: aff_sig_measure} with
	%
	%
	\eas{
	p_i & =\Pr(V_m = i) \\
	\mu_i(\cdot) & =\mu_{\Cv_i}\big(\fv_i^{-1}(\cdot)\big) \\
	\Acal_i &= \Kcal_i, 
	}{\label{eq:cond lem 2}}
	for all $i$.
	Conversely, every affinely singular probability measure can be constructed as in \eqref{eqn: singular_subsets} with a proper choice of $(\{\fv_i\}, \{\Cv_i\}, V_m)$ that satisfy the  conditions in \eqref{eq:cond lem 2}.
\end{lem}

\begin{IEEEproof}
First, we prove that \eqref{eqn: singular_subsets} has an affinely singular probability measure. 
By the conditioning rule, we have
\ea{
\changedo{\mu_{\Zv^m}(\Ccal)} = \sum_{i} \Pr(V_m=i)\mu_{\Zv^{(i)}}(\Ccal), \label{eqn: conditioning}
}
for any set $\Ccal\subseteq \Rbb^m$. Since $\Zv^{(i)}$ is only supported on $\Kcal_i$, we can rewrite \eqref{eqn: conditioning} as
\ea{
\changedo{\mu_{\Zv^m}(\Ccal)}=  \sum_{i} \Pr(V_m=i)\mu_{\Zv^{(i)}}(\Ccal\cap \Kcal_i). \label{eqn: new_aff}
}
Because of the $1$-$1$ mapping from $\Cv_i$ to $\Zv^{(i)}$ and absolute continuity of $\Cv_i$ with respect to the Lebesgue measure on $\Rbb^{e_i}$,  $\Zv^{(i)}$ is also absolutely continuous with respect to the push-forward Lebesgue measure to $\Kcal_i$. 
Therefore, we conclude that 
\eqref{eqn: new_aff} is in accordance with the formulation of affinely singular RVs in Definition \ref{def: aff_sig_measure}, as $\Kcal_i$s are affine sets.

Conversely, for an affinely singular RV, we provide a representation as in \eqref{eqn: singular_subsets}. To do so, for every affine set $\Acal_i$ in Definition \ref{def: aff_sig_measure}, we find $U_i$ and $\bv_i$ such that with $\fv_i(\xv)$ as in \eqref{eqn: f_i}, 
we have $\Acal_i  =  \{\fv_i(\xv)|\xv\in \Rbb^{e_i}\}$. Note that this is always possible, since $\Acal_i$ is an affine set. Next, we generate RVs $\Cv_i$ with push-forward measure $\mu_i\big(\fv_i^{-1}(\cdot)\big)$, which is by definition absolutely continuous. We also define $V_m$ as a RV with probability $p_i$. Finally, we recall that if $\Zv^{(i)}$ is defined as in \eqref{eqn: Zv(i)}, we have
\ea{
\mu_{\Zv^{(V_m)}}(\Ccal)=\sum_i p_i\mu_i (\Ccal\cap \Acal_i).
}
\end{IEEEproof}

The next remark  shows that orthogonally singular RVs form a subclass of affinely singular RVs.

\begin{rem} \label{rem: ortho_aff}
    According to Definition \ref{def:orth_sing} of orthogonally singular RVs, one can rewrite \eqref{eqn: indep_disc_cont} in the form of \eqref{eqn: singular_subsets} as
    \ea{
    \Xv^n = \Xv^{(\nuv, \xv_{\dm}^{\nuov})},
    }
    where 
    \ea{
    \Xv^{(\sv, \xv_{\dm}^{\sov})} := I_n^{[\sv]}\Xv_c^{\sv} + I_n^{[\sov]} \xv_{\dm}^{\sov}, \label{eqn:orth_sing}
    }
     $\nuv= [\nu_1,\ldots \nu_n]$ for $\nu_i$ in  
    \eqref{eqn: indep_disc_cont}, and 
    the realizations of this RV are indicated by $\nuv=\sv$. 
    Since $\Xv_{\cm}$ is absolutely continuous, its marginal probability measure  $\Xv_c^{\sv}$ 
    is also absolutely continuous. Now, by comparing \eqref{eqn:orth_sing} and \eqref{eqn: f_i}, we conclude that orthogonally singular RVs are also affinely singular. 
    Moreover, since the affine subsets are formed by the functions 
    \ean{
    \fv_i(\xv)=(I_n^{[\sv]}, I_n^{[\sov]})\lsb \xv^{\intercal}, \underbrace{0, \ldots, 0}_{|\sov|}\rsb^{\intercal}+ I_n^{[\sov]}\xv_{\dm}^{\sov},
    }
    we see that such sets are parallel to the Euclidean axes.
\end{rem}
\begin{figure*}
    \centering
    
\begin{tikzpicture}
[
roundnode/.style={circle, draw=green!60, fill=green!5, very thick, minimum size=7mm},
]

\coordinate (A) at (0, 0);
\coordinate (B) at (1, 0.5);
\coordinate (C) at (1, -2.5);
\coordinate (D) at (0, -3);

\coordinate (E) at (-1.5, -1.5);
\coordinate (F) at (-0.5, -1);
\coordinate (G) at (1, -1);
\coordinate (H) at (0, -1.5);

\coordinate (I) at (2.5, -1);
\coordinate (J) at (1.5, -1.5);


\coordinate (l11) at (-1,-1.25);
\coordinate (l12) at (2.,-1.25);

\coordinate (l111) at (-1.75,-1.25);
\coordinate (l121) at (2.75,-1.25);

\fill[left color=red!50!black, right color=red!25!black!25]
  (E) -- (F) -- (G) -- (H);

\draw[color=white , line width=1mm,] (l11)--(l12);
\draw[color=black!60!cyan , line width=0.75mm,] (l11)--(l12);

\fill[bottom color=red!50!black, top color=white!50]
  (A) -- (B) -- (C) -- (D);

\fill[left color=red!25!black!25, right color=white!50]
  (G) -- (I) -- (J) -- (H);

\draw[color=white , line width=1mm,] (0.5,-1.25)--(l12);
\draw[color=black!60!cyan , line width=0.75mm,] (0.5,-1.25)--(l12);

\draw[color=black!60!cyan , line width=0.5mm, dotted] (l11)--(l111);
\draw[color=black!60!cyan , line width=0.5mm, dotted] (l12)--(l121);

\coordinate (r1) at (8,-2.5);
\coordinate (r2) at (10,-1);
\coordinate (r4) at (9,-0.5);
\coordinate (r3) at (11,1);

\fill[left color=blue!50!black, right color=white!25!black!25]
  (r1) -- (r2) -- (r3) -- (r4);

\coordinate (r11) at (8.5,-1.5);
\coordinate (r12) at (10.5,0);
\coordinate (r111) at (8.5-.5,-1.5-0.5*1.5/2);
\coordinate (r121) at (10.5+0.5,0+0.5*1.5/2);

\draw[color=white , line width=1mm,] (r11)--(r12);
\draw[color=black!60!green , line width=0.75mm,] (r11)--(r12);
\draw[color=black!60!green , line width=0.5mm,dotted] (r11)--(r111);
\draw[color=black!60!green , line width=0.5mm,dotted] (r12)--(r121);

\node(K2) at      (11.25,0.25+.25*1.5/2) {$\Kcal_2$};
\node[] (K1)     at  (9.25,-.6) {$\Kcal_1$};

\begin{scope}[node distance=2.5mm and 1mm]

\node [above= of B] (A1) {};
\node [above= of I] (A11) {};
\node [left= .1cm of r4] (A2) {};

\node [below = of l12] (C1) {};
\node [left = of r111] (C2) {};
\end{scope}

\path[->] (A1) edge [out=60, in=100, line width=2pt, color=gray ]  node[above, color=black]  {$\Wv_{[0, 1, 1]}$ w.p. $\Pr(\Upsilon_3^{(1)}=[0, 1, 1])$}   (A2);
\path[->] (A11) edge [out=60, in=100, line width=2pt, color=gray ]
node[above, color=black ]  {\hspace{-2cm}$\Tcal_1^{(3)}= \{[0, 1, 1], [1, 1, 0]\}$} node[below=0.5cm, color=black] {$\Wv_{[1, 1, 0]}$ w.p. $\Pr(\Upsilon_3^{(1)}=[1, 1, 0])$} (A2);

\path[->] (C1) edge [out=-60, in=220, line width=2pt , color=gray ]
node[below, color=black ]  {$\Tcal_2^{(3)}=\{[0, 1, 0]\}$} node[above, color=black] {$\Wv_{[0, 1, 0]}$ w.p. $1$} (C2);

\node [ draw, rounded rectangle, fill=blue!20] at (-1,2)  {\Large \bf{$\Xv^3$}};
\node [ draw, rounded rectangle , fill=blue!20] at (11,2)  {\Large \bf{$\Yv^3$}};

\node[] (nu1) at (0,0.5) {$\nuv=[0, 1, 1]$};
\node[] (nu1) at (-1.75,-1) {$\nuv=[0, 1, 0]$};
\node[] (nu1) at (-1,-1.8) {$\nuv=[1, 1, 0]$};

\end{tikzpicture}
\caption{A conceptual representation of a linear transformation of orthogonally singular RV into an affinely singular RV.}
    \label{fig: ortho}
\end{figure*}
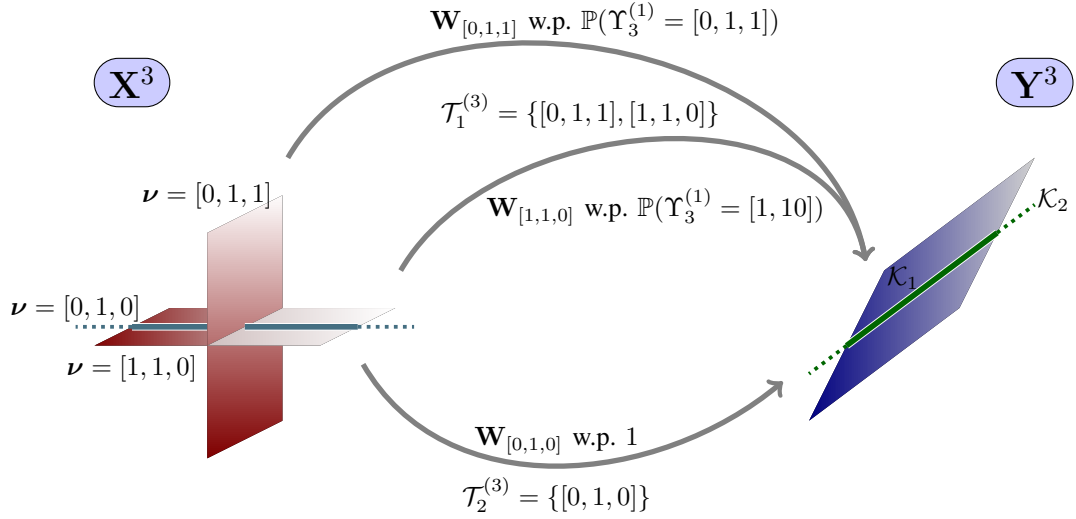

\changedo{Next, we investigate affinely singular RVs generated by linear transformations of   orthogonally singular RVs. 

\subsection{Linear Transformation of  Orthogonally Singular Random Vectors}}}
{\changed 
Let us consider the linear transformation $\Yv^m = A_m\Xv^n$ where $\Xv^n$ is an orthogonally singular RV. 
Following  Remark \ref{rem: ortho_aff}, $\Xv^n$ is composed of absolutely continuous measures on  affine sets that are specified by pairs $(\sv, \xv_{\dm}^{\sov})$. The linear transformation $A_m\xv^n$ of such affine sets forms a collection $\{\Kcal_i\}_i$ of distinct $e_i$-dimensional affine sets.}
%
\begin{table}
    \centering
    \begin{tabular}{|c|c|}\hline
         \rowcolor{red!20}$\sv$ & $\digamma_3(\sv)$ \\\hline
         $[0, 1, 1]$ & $1$\\\hline
         $[1, 1, 0]$ & $1$\\\hline
         $[0, 1, 0]$ & $2$ \\\hline
    \end{tabular}
    \medskip
    \caption{The definition of $\digamma_3(\cdot)$ in example of Figure \ref{fig: ortho}}
    \label{tab: ortho_dig}
\begin{tabular}{|c|c|}\hline
    \rowcolor{red!20}$i$ & $\Pr(\Upsilon_3=i)$ \\\hline
    $1$ & $\Pr(\nuv = [0, 1, 1])+\Pr(\nuv = [1, 1, 0])$\\\hline
    $2$ & $\Pr(\nuv = [0, 1, 0])$\\\hline
\end{tabular}
\medskip
    \caption{The probability mass function of $\Upsilon_3$  in  example of Figure \ref{fig: ortho}}
    \label{tab: ortho_ups}
\begin{tabular}{|c|c|c|}\hline
    \rowcolor{red!20}$\sv$ & $\Pr(\Upsilon_3^{(1)}=\sv)$ &  $\Pr(\Upsilon_3^{(2)}=\sv)$ \\\hline
    $[0, 1, 1]$ & $\frac{\Pr(\nuv = [0, 1, 1])}{\Pr(\nuv = [0, 1, 1])+\Pr(\nuv = [1, 1, 0])}$ & 0\\\hline
    $[1, 0, 1]$ & $\frac{\Pr(\nuv = [1, 1, 0])}{\Pr(\nuv = [0, 1, 1])+\Pr(\nuv = [1, 1, 0])}$ & 0\\\hline
    $[0, 1, 0]$ & $0$ & $1$\\\hline
\end{tabular}
\medskip
    \caption{The probability mass function of {$\Upsilon_3^{(1)}$ }
    in example of Figure \ref{fig: ortho}}
    \label{tab: ortho_ups_3}
\end{table}
{\changed 
Further, let $\Tcal_i^{(m)}$ be the set of all pairs of $(\sv\,,\, \xv_{\dm}^{\sov})$ that generate the same affine subset $\Kcal_i$ (it is possible that after the linear transformation, some of the singular components of $\Xv^n$ are mapped to the same affine sets). Besides, define
$\digamma_m:\, \bigcup \Tcal_i^{(m)}\to \Nbb$ as the function that identifies the index $i$ for each pair $(\sv, \xv_{\dm}^{\sov})$, i.e., if $(\sv, \xv_{\dm}^{\sov}) \in \Tcal_i^{(m)}$, then $\digamma_m\big((\sv, \xv_{\dm}^{\sov})\big)=i$. 
For this indices, we define the RV $\Upsilon_m$ as 
\ea{
\Upsilon_m=\digamma_m\big((\nuv, \Xv_{\dm}^{\nuov})\big)\label{eqn: upsilon_m}} 
%
which implies that the total probability measure of $\Yv^m$  corresponding to $\Tcal_i^{(m)}$ is ${\Pr}(\Upsilon_m=i)$. 

Moreover, for each $i\in\Nbb$ we define the RV $\Upsilon_m^{(i)}$ over $\Tcal_i^{(m)}$ independent of $\Upsilon_m^{(j)}$ for $j\neq i$ such that
%
\ea{
	\forall\, (\sv, \xv_{\dm}^{\sov})\in \Tcal_i^{(m)}:~~~{\Pr}\big(\Upsilon_m^{(i)}=(\sv, \xv_{\dm}^{\sov})\big) = \frac{{\Pr}_{\nuv,\Xv_{\dm}^{\nuov}}(\sv,\xv_{\dm}^{\sov})}{{\Pr}(\Upsilon_m=i)}.
	\label{eqn: v_prime_t}
}

Finally, for each pair $(\sv, \xv_{\dm}^{\sov})\in \bigcup\limits_{i}\Tcal_i^{(m)}$, we define a continuous random vector $\Wv_{\sv, \xv_{\dm}^{\sov}}$ by
\ea{
	\Wv_{\sv, \xv_{\dm}^{\sov}} &= \widetilde{D}_{\sv}
	\underset{\hspace{-1mm}{\rightharpoonup}}{U_{\sv}}^{\dagger} 
	\Xv_{\cm}^{\sv} + 
	\underset{{\hspace{-1mm}\leftharpoonup}}{\widetilde{U}_{\sv}\;\;}^{\hspace{-3mm}\dagger} A_m^{[\overline{\sv}]} \xv_{\dm}^{\sov},\label{eqn: W_s_x}
}
 where $\underset{\hspace{-1mm}\leftharpoonup}{U_{\sv}}$, $\widetilde{D}_{\sv}$, and $ \underset{\hspace{-1mm}\rightharpoonup}{\widetilde{U}_{\sv}}$ stand for the matrices of left-singular vectors of $A_m^{[\sv]}$, 
 the upper $r_{\sv}^{(m)} \times m$ part of the diagonal matrix of the singular values of $A_m^{[\sv]}$ with $r_{\sv}^{(m)}=\rank (A_m^{[\sv]})$, 
  and the $m\times r_{\sv}^{(m)}$ matrix formed by the first $r_{\sv}^{(m)}$ columns of the right-singular vectors of $A_m^{[\sv]}$, respectively. 
  Note that $\underset{{\hspace{-1mm}\rightharpoonup}}{\widetilde{U}_{\sv}\;\;}^{\hspace{-3mm}\dagger} A_m^{[\overline{\sv}]}$ is a fixed term in the definition of $\Wv_{\sv, \xv_{\dm}^{\sov}}$ in \eqref{eqn: W_s_x}. 
  The RVs $\Wv_{\sv, \xv_{\dm}^{\sov}}$ essentially identify the continuous part of the $\Yv^m$ over lower-dimensional affine subsets for each specific discrete part of $\Xv^n$; we should highlight that because of the rank of $A_m^{[\sv]}$ submatrices, the continuous part of $\Xv^n$ does not necessarily generate a continuous component in $\Yv^m$.
  %


We are now equipped to state the result on the linear transformation of  orthogonally singular RVs.
\begin{lem} \label{lmm: Ax_has_subsets}
For a matrix $A_m \in \Rbb^{m \times n}$ and a vector $\sv \in \{0,1\}^n$, let 
	%
	 $\underset{\hspace{-1mm}\leftharpoonup}{U_{\sv}}$ stand for 
	 the $m\times m$ matrix of left-singular vectors of $A_m^{[\sv]}$.
We form functions $\fv_i(\cdot)$ as 
	\ea{
		\fv_i(\xv)= \underset{\hspace{-1mm}\leftharpoonup}{U_{i}} \Big[ \xv^{\intercal}, \underbrace{0, \ldots, 0}_{m-r_i}\Big]^{\intercal}+\bv_i, \label{eqn: ytv_i}
	}
where $\underset{\hspace{-1mm}\leftharpoonup}{U_{i}}=\underset{\hspace{-1mm}\leftharpoonup}{U_{\sv}}$, $r_i=\rank(A_m^{[\sv]})$, and $\bv_i = \underset{\hspace{-1mm}\leftharpoonup}{U_{\sv}} \underset{\hspace{-1mm}\leftharpoonup}{\widehat{U}_{\sv}}^{\dagger} A_m^{[\sov]} \xv_{\dm}^{\sov}$ \changedo{in which $\underset{\hspace{-1mm}\leftharpoonup}{\widehat{U}_{\sv}}$ is constructed by zeroing off the first $\rank(A_m^{[\sov]})$ of left-singular vectors of $A_m^{[\sv]}$} and for $(\sv,\xv_{\dm}^{\sov})\in\Tcal_i^{(m)}$ (we show that for all pairs in $\Tcal_i^{(m)}$, these values are fixed and are therefore, indexed with $i$). Then, $\Yv^m$ in \eqref{eq:affinely singular} is affinely singular as in Lemma \ref{lem: aff_sing} with the triplet $\big(\{\fv_i\}, \{\Ytv_i\}, \Upsilon_m\big)$, where $\Ytv_i$ is defined as
\ea{
	\Ytv_{i} =  \Wv_{\Upsilon_m^{(i)}}. \label{eqn: y_tilde_i}
}
\end{lem}
\begin{IEEEproof}
 See Appendix \ref{app: proof_Ax_has_subsets}.
\end{IEEEproof}

	For illustration of the above definitions, we have depicted a toy example of linear transformation of an orthogonally singular RV in Figure \ref{fig: ortho}. In this example, the definition of  $\digamma_3(\cdot)$,  the probability mass function of $\Upsilon_3$, and the RVs $\Upsilon_3^{(1)}$ and $\Upsilon_3^{(2)}$ are specified in Tables \ref{tab: ortho_dig}, \ref{tab: ortho_ups}, and \ref{tab: ortho_ups_3}, respectively. In this example, since the discrete component of the probability measure is $\Xv_{\dm} = [0, 0, 0]$ (see Definition \ref{def:orth_sing}), we omit its symbol in the notation of other parameters.
}


\changedo{\section{Results on Random Variables}
\label{sec:Main Contributions}}

Our main contribution in this paper is the derivation of the RDF for affinely singular RVs
for $D\to 0$ through evaluating DRB and RID of such RVs. 
Through  this result we are able to address
the RDF of affinely transformed orthogonally singular RVs.
%

\changedo{
We begin by deriving the RID and the DRB of affinely singular RVs and the RID of linear transformations of orthogonally singular RVs. 
In Section \ref{sec:The gap between our upper bound and the known upper bound}, using a simple example, we show how the dependence among elements of an orthogonally singular RV can maximize the information transfer in terms of RID, during the course of  dimensionality reduction.  Finally, in Section \ref{sec:An illustrative example} we illustrate three  examples of transforming orthogonally singular RVs. 

}
%


\subsection{RID {\changed and DRB} of affinely singular RV}\label{sec: RID_affine}

In this section, we derive an expression for the RID {\changed and DRB} of the affinely singular RV $\Zv^m$ through its representation as in Lemma \ref{lem: aff_sing}.
As a reminder to the reader, the probability measure of $\Zv^m$ is formed by a union of absolutely continuous measures indexed by $i\in \mathbb{X} \subset \Nbb$, where the $i$th measure  is supported on an $e_i$-dimensional affine subset of $\Rbb^m$ and consists of $P(V_m=i)$ of the total probability measure.

\begin{thm}\label{thm: singular_subset}
 Let $\Zv^m$ be an affinely singular RV as per Definition \ref{def: aff_sig_measure}, then, the RID of $\Zv^m$ 
	is obtained as 
	\ea{
	d(\Zv^m) = \Ebb_{V_m}[e_{V_m}], \label{eqn: aff_sing_RID}
	}
	for $V_m$  as in Lemma \ref{lem: aff_sing}, provided that $\Hsf(V_m)<\infty$. 
{\changed 
Further, if $\hsf(\Cv_i)>-\infty$ and $\Ebb[\|\Zv^m\|_{\alpha}^{\beta}]<\infty$ for some $\alpha, \beta\in \Rbb^{+}$,
 	then the DRB of $\Zv^m$ is obtained as
 \ea{
 		b(\Zv^m) &= \sum_{i=1}^{\infty} p_i \hsf(\Cv_i)+\Hsf(V_m),
 	}
 	where $p_i = \Pr(V_m =i)$. 
}
 \end{thm}
 
\begin{IEEEproof}
	See Appendix \ref{prf: singular_subset}.
\end{IEEEproof}

{\changed 
The proof of Theorem \ref{thm: singular_subset} heavily relies on the following lemma that is proved for a mixture of $k_i$-regular measures. Such measures, as stated in \cite[Thm. 3.1 ]{de2006lecture} form the largest class of measures on which one could define probability density functions.

\begin{lem}\label{lem: reg}
    Let $\mu$ be a probability measure on $\Rbb^m$ that can be written as a mixture of $k_i$-regular probability measures $\mu_i$ for $i\in \Nbb$, as
    \ea{
    \mu(\Acal)= \sum_{i} p_i \mu_i(\Acal),
    }
    for every set $\Acal\in \Rbb^m$, and $p_i\in (0, 1]$, where $\sum_i - p_i \log{p_i}<\infty$. Then, the RID of a random variable $\Zv^m$ with probability measure $\mu(\cdot)$ is given as
    \ea{
    d(\Zv^m) = \sum_{i} p_i k_i.
    }
\end{lem}
\begin{IEEEproof}
	First, we show that the RID of a RV $\Xv^m$ with $k$-regular measure $\tilde{\mu}(\cdot)$ is equal to $k$.
	
	Based on \cite[Equation 17]{wu2010renyi}, the RID of this $\Xv^m$ can be obtained as
	\ea{
		d(\Xv^m) = \lim_{\ep\to 0}\Ebb_{\xv^m\sim \Xv^m}\Big[ \frac{\log \tilde{\mu}\big(B_{\ep}(\xv^m)\big)}{\log \ep}\Big], \label{eqn: rid_exp}
	}
	where $B_{\ep}(\xv^m)$ is a ball with radius $\ep$ around $\xv^m$. 
	Further, using \cite[Theorem 1.1]{de2006lecture} and \cite[2.86]{ambrosio2000functions}, we know that if $\xv^m$ follows a $k$-regular measure $\tilde{\mu}(\cdot)$,  the following limit is finite and non-zero 
	almost surely
	\ea{
		\lim_{\ep\to0}\frac{\tilde{\mu}\big(B_{\ep}(\xv^m)\big)}{\ep^{k}}.
	}
	This is equivalent to the fact that
	\ea{
		\lim_{\ep\to0}\log{\tilde{\mu}\big(B_{\ep}(\xv^m)\big)}-k\log{\ep},
	}
	is finite almost surely, that leads to the fact that
	\ea{
		\lim_{\ep}\frac{\log{\tilde{\mu}\big(B_{\ep}(\xv^m)\big)}}{\log{\ep}}=k,
	}
	almost surely. This, together with \eqref{eqn: rid_exp} shows that $d(\Xv^m)=k$.
	
	We complete the proof by recalling \cite[Theorem 5]{WuThesis} to obtain $d(\Zv^m)=\sum_i p_i k_i$; note that we assumed $\sum_i p_i\log \frac{1}{p_i}<\infty$.
\end{IEEEproof}

}
{\changed 
So far, we have expressed  the DRB as summation of the average differential entropy of   absolutely continuous components of the measure and the uncertainty in the choice of affine subsets. 
%
Since $V_m$ covers all possible choices of the discrete components, we can express the DRB  as
\ea{
& b(\Zv^m) = \sum_{i=1}^{\infty} p_i \hsf(\Cv_i)+\Hsf(\Zv^m_{D}) \Pr(\Zv^m~ \text{is \ discrete})\\
&  \quad  + \Hsf(V_m | \Zv^m ~\text{is \ not \  discrete}) \Pr(\Zv^m ~\text{is \ not \ discrete}), \nonumber
}
\changedo{where $\Zv_D^{m}$ is the discrete component of $\Zv^m$
and $\Pr(\Zv^m ~\text{is \ not \ discrete})$ refers to the probability  that the dimension of the corresponding set is $e_{V_m}=0$. }
}
%

{\changed
The latter form is useful in studying the 
compressibility of discrete and continuous \emph{mixed-pairs} as defined in \cite{nair2006entropy}.

\begin{example}
For a discrete random variable $X$ taking values from $\Scal = \{x_1, x_2, \ldots\}$ each with non-zero probability, and a continuous random variable $Y$ supported on $\Rbb$ \changedo{that have a joint probability measure $X, Y\sim\mu_{X, Y}$}, the pair $\Zv = (X, Y)$ is called a mixed-pair. We observe that the pair $\Zv$ has the probability measure
\changedo{
\ea{
\mu_{\Zv}(\Ccal)= \sum_i \mu_i(\Ccal)\cdot \Pr(X=x_i),
}
where $\mu_i(\cdot)$ denotes the conditional probability 
\ea{
\mu_i(\Ccal)=\Pr\big((X, Y)\in \Ccal|X=x_i\big).
}
}
Note that $\mu_i(\cdot)$ is absolutely continuous on the $1$-dimensional affine set \changedo{$\Acal'_i=\{x_i\}\times \Rbb$}; otherwise, if for some  $x=x_i$ and a zero-Lebesgue measure $\Scal\in \Rbb$, we have \changedo{$\mu_i\big(\{x_i\} \times \Scal\big)>0$, then, which implies that $\mu_Y(\Scal)=\sum_i \Pr(X=x_i)\mu_i(\{x_i\}\times \Scal)>0$} that contradicts the absolute continuity of $Y$. 
\changedo{As a result, $\Zv$ is affinely singular and using Lemma \ref{lem: aff_sing} it is described by the triplet $\big(\{f_i\}, \{C_i\}, V_m\big)$ where $C_i$ is a RV with probability measure $\mu_{C_i}=\mu_{Y|X=x_i}$, $f_i(y)=(x_i, y)$, and $\Pr(V_m=i)=\Pr(X=x_i)$. Hence, using Theorem \ref{thm: singular_subset}, we have
\ean{
d(\Zv)&=\Ebb_{X}[1]=1,\\
b(\Zv)&={\sum_i \Pr(X=x_i) \hsf(Y|X=x_i)+\Hsf(X)}={\mathbb{H}(\Zv)},
}
where 
$\mathbb{H}(\Zv)$ is the \emph{entropy of the mixed-pair}  defined in \cite{nair2006entropy}.}

This example reveals that the DRB coincides with the notion of entropy of mixed-pairs in this specific setting. In Section \ref{sec: mi}, we further show that $b(\Zv)$ for an affinely singular RV $\Zv$ coincides with the generalized entropy of its probability measure with respect to the mixture of Lebesgue measures on affine subsets.
\end{example}

}

{
\changed
Next, we evaluate the RID and DRB of linearly transformed orthogonally singular RVs.

}

\begin{thm}\label{thm: disc_cont_fat}
	Let  $\Xv^n$ be an orthogonally singular RV as in Definition \ref{def:orth_sing} 
	such that $\Hsf(\nuv, \Xv_{\dm})<\infty$. 
	Then, the RID of $\Yv^m=A_m \Xv^n$ satisfies
	\ea{
		d(\Yv^m) = \Ebb_{\nuv}[{\rank} (A_m^{[\nuv]})].\label{eqn: lin_RID}
	}
	{\changed 
	Further, if  $\hsf(\Xv_{\cm})>-\infty$ and $\Ebb[\|\Xv_{\cm}\|^{\beta}_{\alpha}], \Ebb[\|\Xv_{\dm}\|^{\beta}_{\alpha}]<\infty$ for some $\alpha, \beta\in \Rbb^{+}$, then, the DRB of $\Yv^m$ is obtained as
	\ea{
	b(\Yv^m) = \Hsf(\Upsilon_m) + \sum_{i=1}^{\infty} p_i\hsf(\Ytv_{i})	,\label{eqn: b_y^m}
}
where $\Upsilon_m$ is the discrete RV introduced in \eqref{eqn: upsilon_m}, and $\{\Ytv_i\}$ is the set of absolutely continuous RVs mentioned in \eqref{eqn: y_tilde_i}.
	}
\end{thm}

\begin{IEEEproof}
	See Appendix \ref{app: disc_cont_fat}.
\end{IEEEproof}

%
As a special case, Theorem \ref{thm: disc_cont_fat} implies that if $\Xv^n$ is absolutely continuous (i.e. $\alpha_i=1$ for all $i$), then, $d(\Yv^m)=\rank (A_m)$. {\changed Further, in this case \eqref{eqn: b_y^m} reduces to
\ea{
b(\Yv^m)= \log \det^{+}(A_m)+\hsf( \underset{\rightharpoonup}{\widetilde{U}}^{\dagger}\Xv^n),\label{eqn: abs_cont_DRB}
}
where $\underset{\rightharpoonup}{\widetilde{U}}$ is composed of the first $r=\rank(A_m)$ right singular vectors of $A_m$.
}

Another special case is when $A_m$ is of full column-rank, i.e. $\rank (A_m) =n$. In this case, ${\rank} \lb A_m^{[\sv]}\rb =\sum_{i=1}^{n}s_i$, and the expectation in \eqref{eqn: lin_RID} simplifies to $d(\Yv^m) = \sum_{i=1}^n \alpha_i$. This result is in agreement with a similar result for bi-Lipschitz transformations in \cite[Theorem 2]{WuThesis}.
%
{\changed 
One can verify that in this case, $\digamma_m(\cdot)$ is a one-to-one mapping. This fact, together with \eqref{eqn: upsilon_m}, \eqref{eqn: y_tilde_i}, and \eqref{eqn: b_y^m}  implies that
%
\begin{align}
	b(\Yv^m)=& \Hsf(\nuv, \Xv_{\dm}^{\overline{\nuv}})+m \psi_{m,\boldsymbol{\nu}}(A) \nonumber \\
	& +\sum_{\sv \in \{0, 1\}^n}\Pr(\nuv = \sv)\hsf(\Xv_{\cm}^{\sv}), \label{eqn: sing_drb}
\end{align}
where $\hsf(\Xv_{\cm}^{\nuv})$ is the differential entropy of $\Xv_{\cm}^{\nuv}$, the sub-vector of $\Xv_{\cm}$ associated with non-zero elements in $\nuv$, and $\psi_{m,\boldsymbol{\nu}}(A)$ is defined as
%
\begin{align}\label{eq:psi_def}
\psi_{m,\boldsymbol{\nu}}(A_m) := \tfrac{1}{m}\Ebb_{\nuv}\Big[\log \det^{+} \big(A_m^{[\nuv]}\big)\Big].
\end{align}
}
{\changed 
\changedo{The subscript $\nuv$ in $\psi_{m, \nuv}(A_m)$ is to highlight that this value is a function of the probability distribution of $\nuv$ (but not its random value). }

In the special case that $\Yv^m$ contains independent discrete-continuous elements (i.e., $m=n$, $A_m=I_n$ and $\Xv^n$ is element-wise independent), formulation of RID and DRB in Theorem \ref{thm: disc_cont_fat} reduces to
	\ea{
	d(\Yv^m) = \sum_{i=1}^m \Pr(\nu_i = 1),
	}
	and
	\ea{
		b(\Yv^m)= \sum_{i=1}^m	\Hsf(\nu_i) + \Pr(\nu_i=0)\Hsf(X_{\dm, i}) + \Pr(\nu_i=1) \hsf(X_{\cm, i}). \label{eqn: b_decomp}
		}
The latter identity is an extension of \cite[Theorem 1]{Binia1988} to RVs with bounded moment (instead of bounded variance).
}


\begin{rem}\label{rem:ContWeight}
A closer look at \eqref{eqn: lin_RID} reveals that $d(\Yv^m)$, besides $A_m$, also depends on $\{\alpha_i=\Pr(\nu_i=1)\}$. 
In other words, the choice of the continuous and discrete distributions of the elements of $\Xv^n$ does
not affect the RID: this quantity  only depends on how much weight is assigned to the continuous components.
Additionally, $d(\Yv^m)$ is an increasing function of each $\alpha_i$.
\end{rem}

For fixed $\Xv^n$, the maximum value of  $d(\Yv^m)$ is achieved when
\ea{
	\rank(A_m^{[\sv]}) &= \min\lcb m, \sum_{i=1}^n s_i \rcb,~~~ \forall \  \sv \ \in \{0, 1\}^n. \label{eqn: rank_vand}
} 
Using the compressed-sensing terminology, the above condition is equivalent to {\changed $\spark(A_m) =m+1$}. 
{\changed One can prove two directions for such equivalence as follows: (i) first, if we assume $\mathrm{SPARK}(A_m) = m +1$, it means that 
every $d$  columns for $d\leq m$ are linearly independent. Hence, if $\sum_{i=1}^n s_i\leq m$, then, $\mathrm{rank}( A_m^{[\sv]}) = \sum_{i=1}^n s_i = \min \{m, \sum_{i=1}^n s_i\}$. Further, (ii) if $\sum_{i=1}^n s_i\geq m$, then, there is at least $m$ columns in $A_m^{[\sv]}$ which based on our assumption are linearly independent. 
As a result, $\mathrm{rank} (A_m^{[\sv]})\geq m$.  Moreover, since the number of rows in $A_m^{[\sv]}$ is $m$, then we have $\mathrm{rank} (A_m^{[\sv]})\leq m$ which proves that $\mathrm{rank} (A_m^{[\sv]})= m =\min \{m, \sum_{i=1}^n s_i\}$. 
On the other hand, if $\mathrm{rank} (A_m^{[\sv]}) =\min \{m, \sum_{i=1}^n s_i\}$, it means that if $\sum_{i=1}^n s_i\leq m$, then $\mathrm{rank}(A_m^{[\sv]})$ is equal to its number of columns, which shows that the columns are linearly independent. As a result, $\mathrm{SPARK}(A_m^{[\sv]})\geq m+1$ and using the property $\mathrm{SPARK}(A_m^{[\sv]})\leq m+1$, one shows that $\mathrm{SPARK}(A_m^{[\sv]}) = m+1$. }

The Vandermonde matrices are among the examples that satisfy {\changed the condition \eqref{eqn: rank_vand}}. Therefore, Vandermonde matrices can be considered among the transformations that maximally preserve the information \changedo{measured by the RID.}
%
%

\begin{figure*}\centering
	\subfigure[$d\big(T\Xv^n\big)$ vs. input dimension]{
		\includegraphics[width=0.48\linewidth]{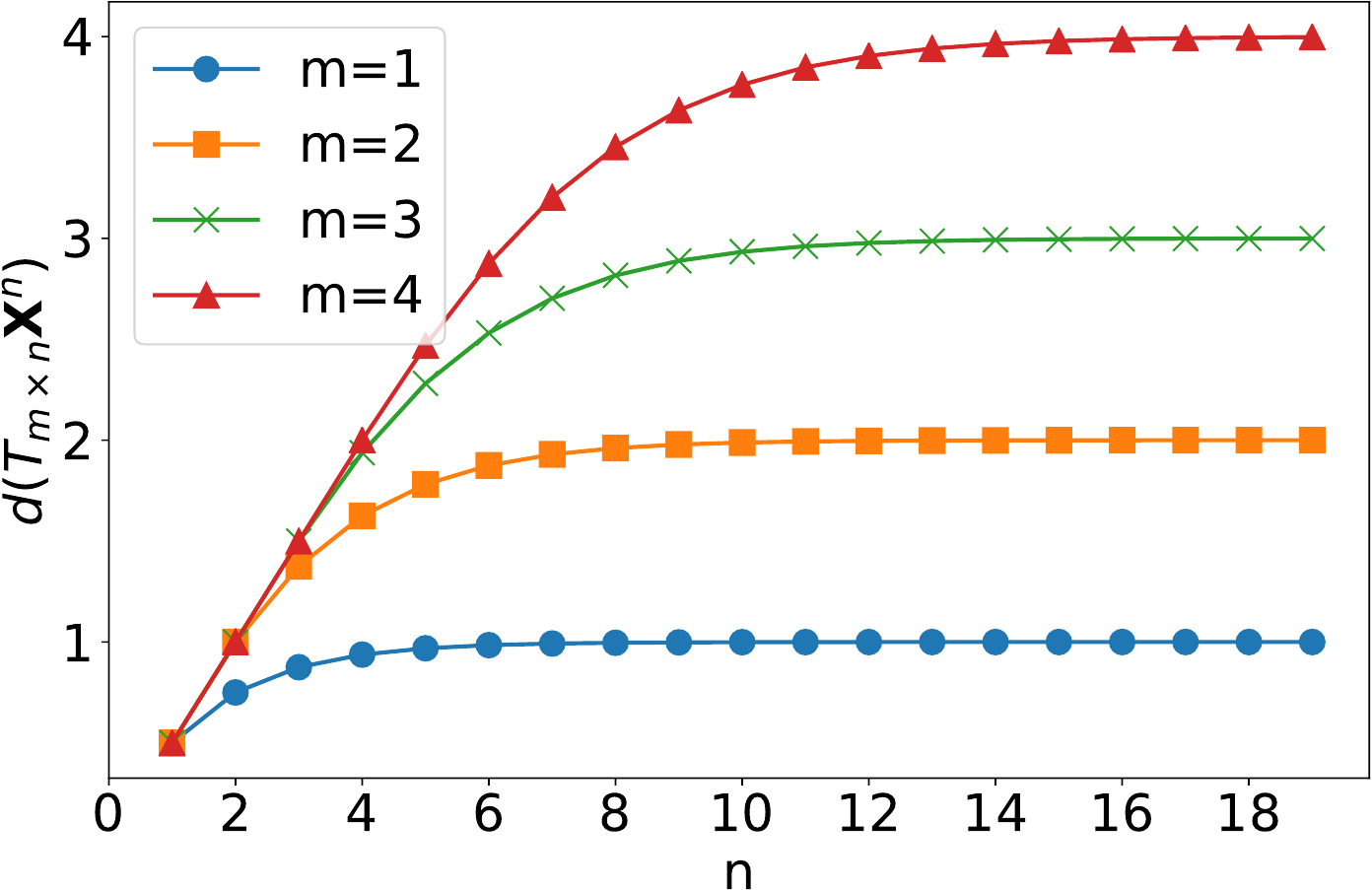}					
	}
	\subfigure[$d\big(T\Xv^5\big)$ vs. input RID]{
		\includegraphics[width=0.48\linewidth]{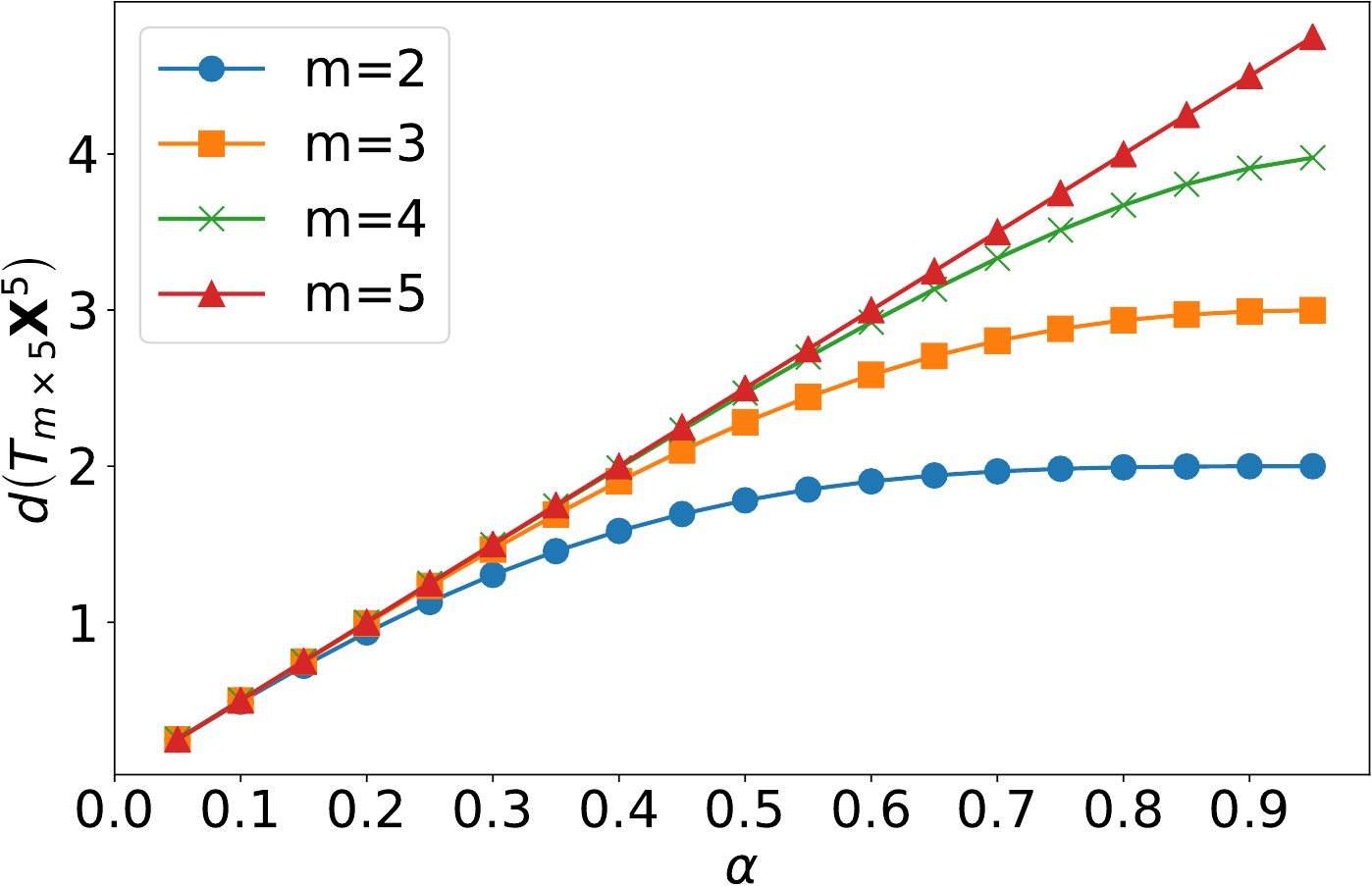}	
	}
	\caption{(a) The RID of $T_{m\times n}\,\Xv^n$ where {\changed $\spark(T)=m+1$}, and $d(X_i) =1/2$. One can see that for $n\leq m$ the curve is equal to $Y=n/2$ (linear) because $T_{m\times n}$ is full column rank and preserves the RID. Moreover, for large $n$'s the RID saturates to $m$. 
		(b)\changedo{ Information dimension of $T_{m\times 5} \,\Xv^5$ for various values of $d(X_i)=\alpha$, where $T$ is a Vandermonde matrix and the elements of $\Xv^5$ are independent discrete-continuous RVs. As one could conclude from Theorem \ref{thm: disc_cont_fat}, for $m>5$, $d(T_{m\times 5}\Xv^{5})$ has the same value as that of $m=5$.}
	}\label{fig:vand_rid}
\end{figure*}

As a special case of the above, the output RID of such linear transformations for independent discrete-continuous RVs is plotted in Figure \ref{fig:vand_rid}.

This maximal behavior is even valid among Lipschitz functions, as shown by the next lemma.
%
 
%
\begin{lem}\label{thm: lip_opt}
Let $\Xv^n$ be an orthogonally singular RV as in Definition \ref{def:orth_sing} with $\Hsf(X_{\dm,i})<\infty,~i\in[n]$, and $A_{m}$ is an $m\times n$ matrix that satisfies  {\changed$\spark(A_m)=m+1$}. Then, for any Lipschitz function $f:\Rbb^n\to \Rbb^m$ we have that
\begin{align}\label{eq:Lipschitz Maximal}
\overline{d}\big( f(\Xv^n) \big) \leq d( A_m \, \Xv^n ),
\end{align}
where $\overline{d}\big( f(\Xv^n) \big)$ denotes the upper RID of $f(\Xv^n)$ (the RID itself might not exist for functions of $\Xv^n$).
%

%
\end{lem}

\begin{IEEEproof}
	See Appendix \ref{proof: lip_opt}.
\end{IEEEproof}

{\changed As another application of Theorem \ref{thm: disc_cont_fat}, below we study the effect of a random matrix on the information dimension.
	
	\begin{prop} \label{prop: random_matrix}
		Let the elements of the matrix $A=[a_{ij}]_{i, j=1}^{m, n}$ be distributed as $a_{ij}\sim f_{\mathrm{a}}(a)$, where $f_{\mathrm{a}}(\cdot)$ is a valid density function. For an $n$-dimensional truncation $\Xv^n$ of an orthogonally singular random process with $\Pr(\nu_i=1)=p$, and $m=\lceil rn \rceil$ where $r$ is a fixed sampling rate, we have \changedo{$\lim_{n\to\infty, m=\lceil rn\rceil} \frac{1}{m} d(A \Xv^n)= \min\{1, \frac{p}{r}\}$}.
	\end{prop}
	\begin{IEEEproof}
	    See Appendix \ref{app: proof_random_matrix}.
	\end{IEEEproof}
	\begin{IEEEproof}[Sketch of proof]
	    Since the set of roots of the determinant polynomial has zero Lebesgue measure, every sub-matrix of $A$ is of full-rank with probability $1$. Using \eqref{eqn: rank_vand}, this results in $\frac{1}{m}d(A\Xv^n) = \Ebb\big[\min\{1, \frac{\sum\nu_i}{m}\}\big]$. Finally, we make  use of the concentration of  $\frac{\sum \nu_i}{n}$ around $\frac{p}{r}$ to complete the proof.
	\end{IEEEproof}
}

\changedo{\subsection{A study of RID in RVs with dependent components}
\label{sec:The gap between our upper bound and the known upper bound}}
In this part, we present a simple example to examine the RID of a function {\changed of an} orthogonally singular RV; in particular, we study  the effect of  dependence among the elements of an RV. 
If $\Xv^n$ is an orthogonally singular RV, we know from Lemma \ref{thm: lip_opt} that among all Lipschitz functions of $\Xv^n$, linear matrix operators $A_m$ with {\changed $\spark(A_m)=m+1$} maximize the RID. 
%
However, the RID of $A_m \Xv^n$  is not the same if the elements of $\Xv^n$ are dependent or independent.
 As we show here, 
 the independent case corresponds to neither the maximum RID case nor the minimum RID case.
\begin{table}
	\caption{Joint probability of two selection variables $\nu_1, \nu_2$  for the example in Section \ref{sec:The gap between our upper bound and the known upper bound}.}\label{tbl: dist_nus}
	\centering
	\newcolumntype{a}{>{\columncolor{purple!10}}c}
	\setlength\arrayrulewidth{1pt}
	\newcolumntype{b}{>{\columncolor{yellow!20}}c}
	\begin{tabular}{|b ||c |c|}
		\hline
		\rowcolor{yellow!20}
		\backslashbox{$\boldsymbol{\nu_1}$}{$\boldsymbol{\nu_2}$} & \bf 0 & \bf 1\\
		\hline
		\hline
		\bf0 & 0.18 & 0.12\\
		\hline
		\bf1 & 0.42 & 0.28\\
		\hline
	\end{tabular}
	\begin{tabular}{|a ||c |c|}
		\hline
		\rowcolor{purple!10}
		{ \backslashbox{$\boldsymbol{\nu_1}$}{$\boldsymbol{\nu_2}$}} &  \bf0 &  \bf1\\
		\hline
		\hline
		{\bf 0} & 0 & 0.3\\
		\hline
		{\bf 1} & 0.6 & 0.1\\
		\hline
	\end{tabular}
	\newcolumntype{d}{>{\columncolor{blue!20}}c}
	\begin{tabular}{|d ||c |c|}
		\hline
		\rowcolor{blue!20}
		\backslashbox{$\boldsymbol{\nu_1}$}{$\boldsymbol{\nu_2}$} & \bf 0 & \bf 1\\
		\hline
		\hline
		\bf0 & 0.3 & 0\\
		\hline
		\bf1 & 0.3 & 0.4\\
		\hline
	\end{tabular}
\end{table}

We consider a 2D RV with three  dependency cases for $\nu_1$ and $\nu_2$ (defined in \eqref{eqn: indep_disc_cont}) as shown in Table \ref{tbl: dist_nus}:
\begin{enumerate}
\item the $Q$ distribution corresponds to the case of independent $\nu_1$ and $\nu_2$ (equivalently, $X_1$ and $X_2$), 

\item the $Q'$ distribution in which  $X_1$ and $X_2$ cannot take discrete values at the same time,

\item and the $Q''$ distribution in which a discrete value of $X_1$ never coincides with a continuous value of $X_2$.
\end{enumerate}
We need to highlight that the marginal distribution of $\nu_i$s are the same in all the three cases. Next, we evaluate the RID of $Y=[1, 2]\cdot \Xv^2$ (a linear transformation onto the one-dimensional space) for these cases. Obviously,  $A_1=[1, 2]$ is a Vandermonde matrix. 

%
%
We first recall that the sum 
of a continuous RV and another (arbitrary) RV is always a continuous RV \cite[Lemma 11]{ghourchian2018}; hence, the discrete component in the distribution of $Y$ is a result of both $X_1$ and $X_2$ taking discrete values. Furthermore, as $Y$ is a scalar, the RID of $Y$ equals to the total probability of the continuous component; i.e., $0.82$, $1$, and $0.7$ for $Q$, $Q'$, and $Q''$, respectively.
%
%
%
%
%
%
 %

Since $A_1$ is a Vandermonde matrix, Lemma \ref{thm: lip_opt} guarantees that  $Y$ has the maximum possible RID among all Lipschitz functions $f:\Rbb^2\to\Rbb$ on $\Xv^2$. Nevertheless, as the distribution of $\Xv^2$ changes, the RID of $Y$ also changes. Surprisingly, the above example reveals that the RID of $Y$ for the element-wise independent $\Xv^2$ and $A_1$ being a Vandermonde matrix does not necessarily provide a universal upper-bound. 
 
%
%

\begin{figure}
	\centering
		\includegraphics[width=\linewidth]{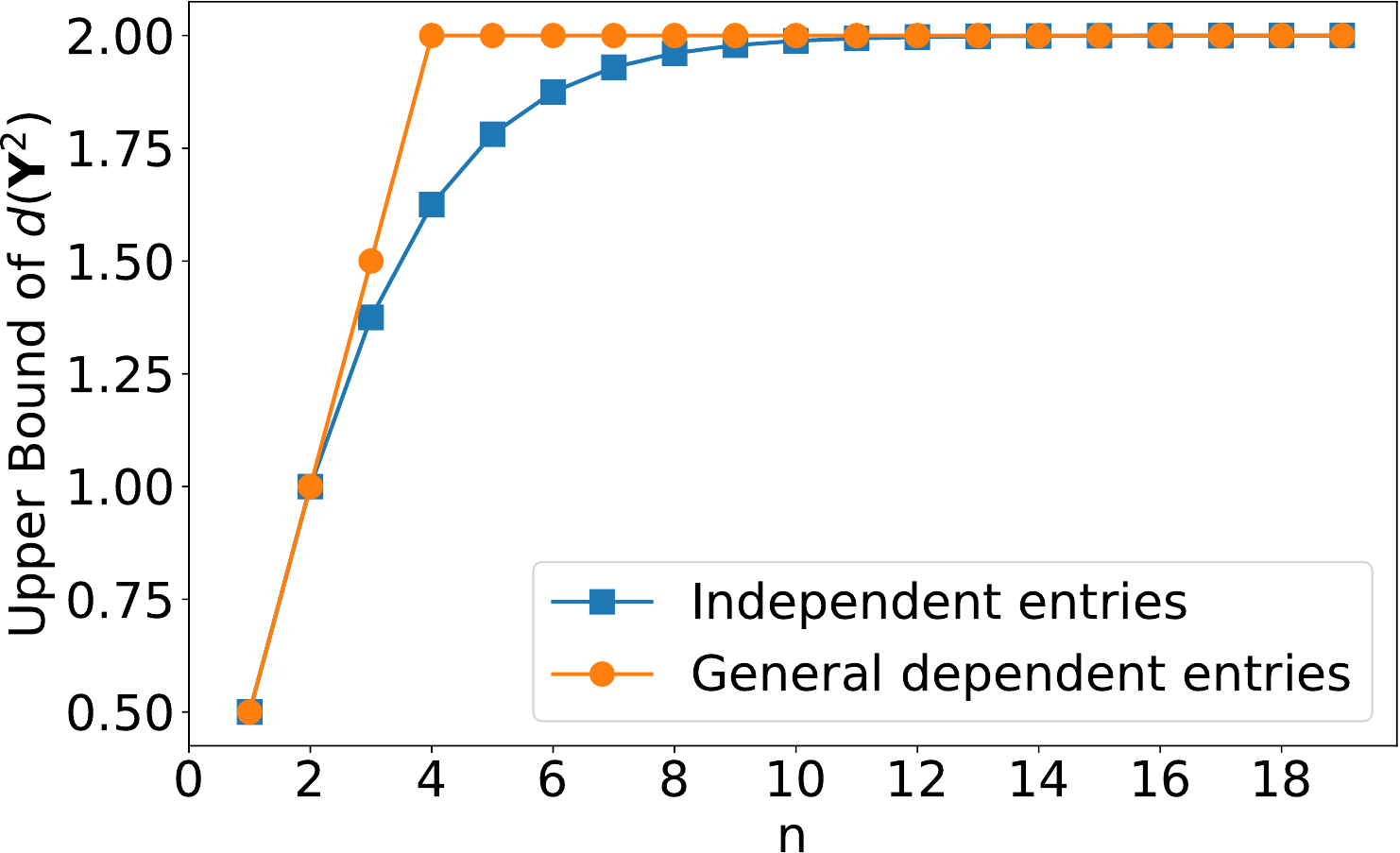}

	\caption{Upper bounds of a two-dimensional Lipschitz function of element-wise dependent and independent discrete-continuous identically distributed RVs.{\changed Here, we portray the case in which $d(X_1)=0.5$.}}
	\label{fig: gap}
\end{figure}

For Lipschitz functions of a general RV (not necessarily orthogonally singular), the upper-bound
\ea{
d\big(f(\Xv^n)\big)\leq  \min \lcb d(\Xv^n), m\rcb.
\label{eq:upp dependent}
}
is proved in \cite[Theorem 2]{WuThesis} and \cite[Eq. (80)]{renyi1959dimension}. To compare this bound with our result in Theorem \ref{thm: disc_cont_fat}, we consider an input  $\Xv^n$ with i.i.d. elements that is mapped into $\Yv^2$ (i.e., $m=2$) using a Vandermonde matrix. Figure \ref{fig: gap} depicts the upper-bound on the RID using both the general bound in \eqref{eq:upp dependent} and our bound {\changed of the RID that is calculated in} Theorem \ref{thm: disc_cont_fat} {\changed by setting $\rank(A_m^{[\sv]}) = \min\lcb m, \sum_{i=1}^n s_i \rcb$}. This figure indicates that our result provides a tighter bound by restricting the input type.

%
%

%

\subsection{An illustrative example}
\label{sec:An illustrative example}

\begin{figure*}
	\centering
	\subfigure[Orthogonally singular RV]{
		\includegraphics[width=0.25\linewidth]{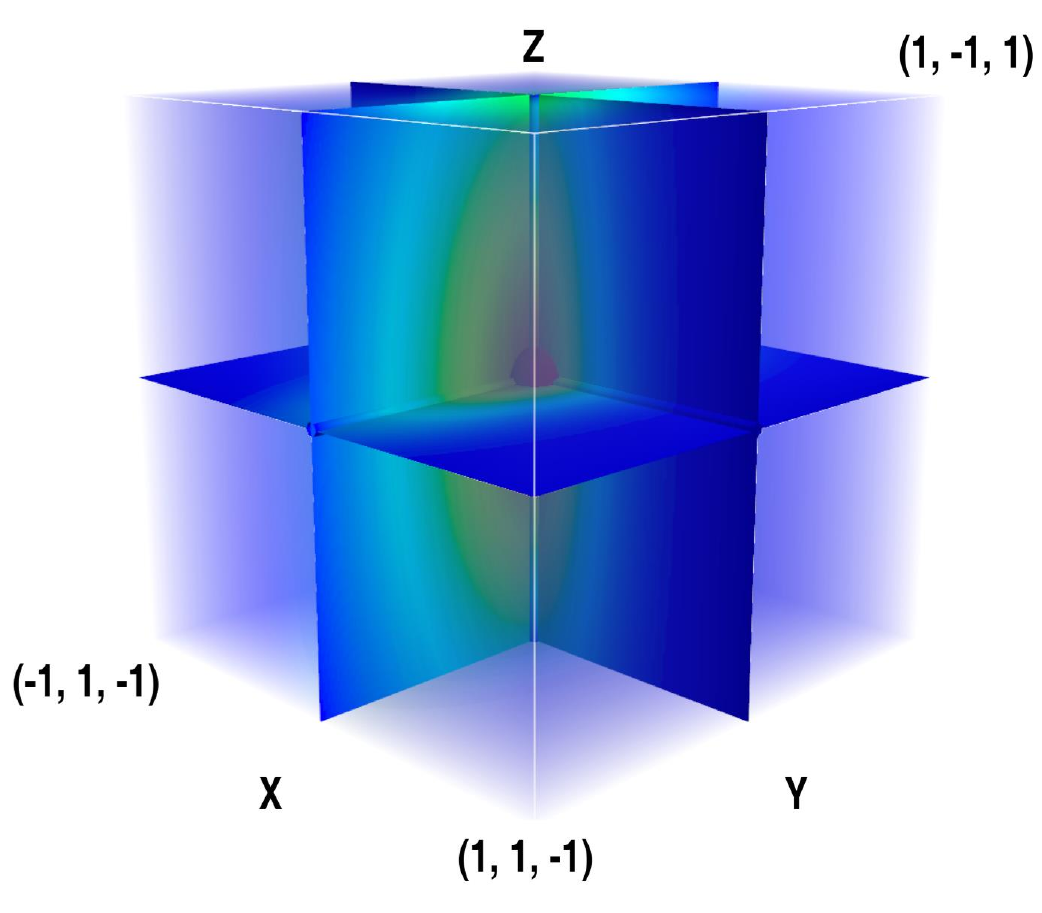}
		\label{fig:Ortogonally singular distribution}
	}	
	\subfigure[Affinely singular RV]{\centering
		\includegraphics[width=0.27\linewidth]{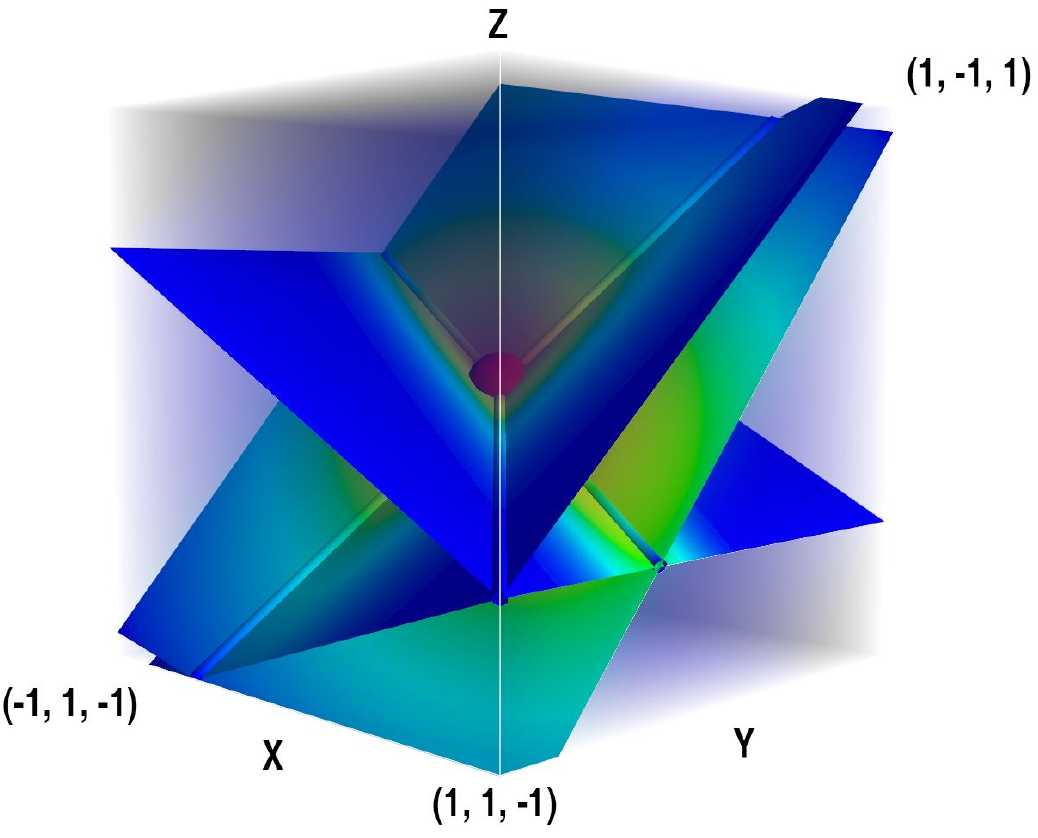}
		\label{fig:An affinely singular distribution}
	}
	\subfigure[Degraded affinely singular RV]{\centering
		\includegraphics[width=0.25\linewidth]{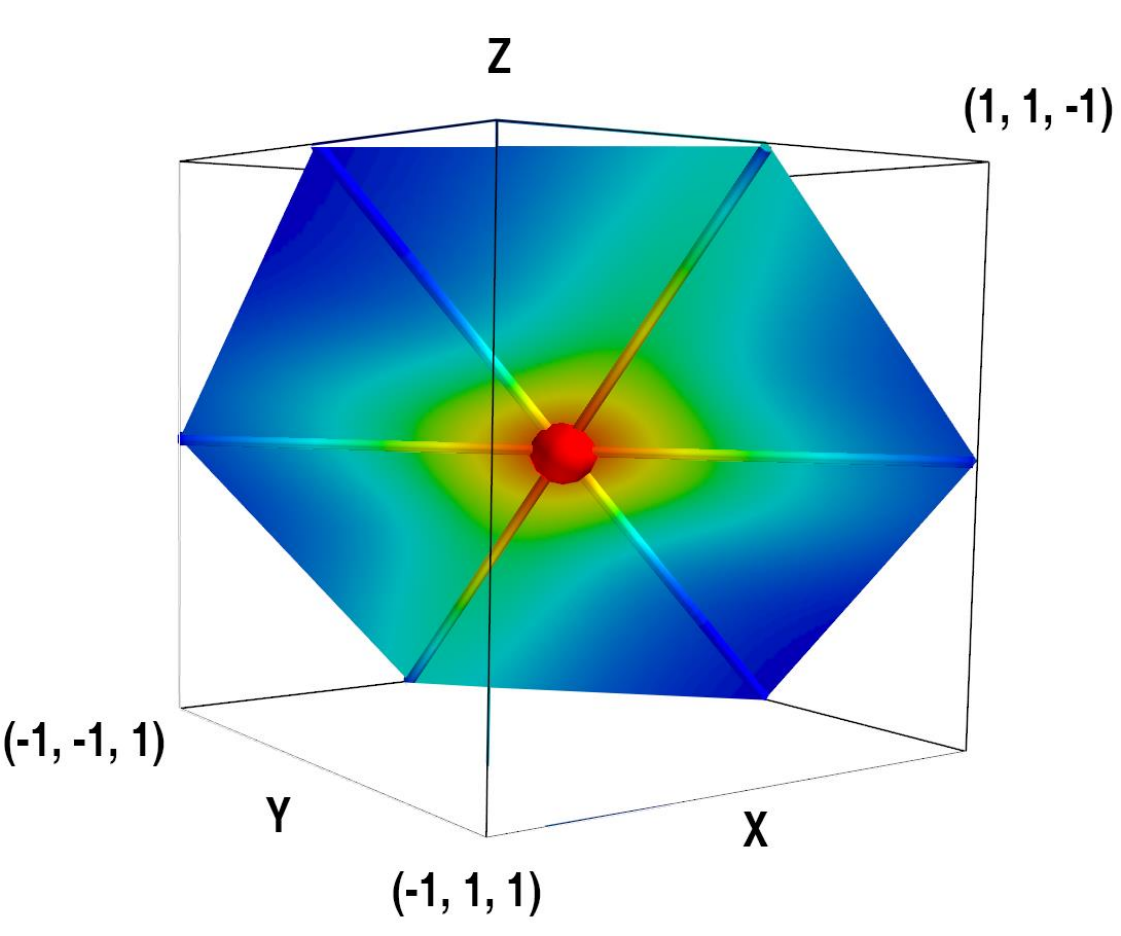}
		\label{fig:A degraded affinely singular distribution}
	}
	%
	\label{fig:singular distribution}
	\caption{A representation of orthogonally, affinely and {\changed degraded} affinely singular random vectors (RVs). }
\end{figure*}

As mentioned earlier, our goal in this paper is to derive the RID and DRB for linearly transformed random vectors in which entries have independent discrete-continuous distributions.  To better clarify the problem formulation and our contributions, let us present in this section a  rather simple but illustrative example. 
This example is illustrated in Fig. \ref{fig:singular distribution}.

\noindent
{\bf Fig. \ref{fig:Ortogonally singular distribution} -- Orthogonally singular RV:}
Let $\Xv^3$ be a three-dimensional random vector in which each entry has an independent Bernoulli-Gaussian distribution, that is 
\ea{
	\mu_{i}^{\rm BG} (\sigma_i, \alpha)=\alpha \mu_{\rm Gauss} (0, \sigma_i^2), +\overline{\alpha}\delta_0,
	\label{eq:BG}
} 
for $i\in[3]$, with $\alpha=1/2$, $\boldsymbol{\sigma}=[0.2, 0.4, 0.8]^{\intercal}$ and where $\delta_x$ indicates the Dirac's measure distributed around $x$, and $\mu_{\rm Gauss}$ is the Gaussian (normal) measure. 
Since each coordinate has a mass probability (of $\overline{\alpha}=1/2$) at zero, the three-dimensional product measure has a mass probability in the origin.
The singularity in zero in each dimension also yield one-dimensional singularities along the axes and two-dimensional singularities on principal planes (i.e., $x-y$, $y-z$ and $x-z$ planes). 
Because of the structure of the singularities, we call $\Xv^3$   an \emph{orthogonally singular RV}. %
As the singular parts of $\Xv^3$ belong to orthogonal planes, it is straightforward to check that $d(\Xv^3)=d(X_1)+d(X_2)+d(X_3)=3/2$.

\noindent
{\bf Fig. \ref{fig:An affinely singular distribution} -- Affinely singular RV:
	}
Next, let us consider the RV $\Yv^3$ obtained as the linear transformation $A\, \Xv^3$ for $A$ full-rank, %
The full-rank linear transformation causes three-dimensional rotation on the singularity patterns. Therefore, it is no longer possible to decompose the singularities and derive the RID according to one-dimensional distributions.
Our result -- see Theorem \ref{thm: disc_cont_fat} --  characterizes the  RID of this class of vectors, which we term \emph{affinely singular RVs}.
{\changed This theorem, asserting previous results on the RID of bi-Lipschitz functions of a RV, shows that for every invertible matrix $A$, we have $d(\Yv^3)=3/2$.}
%
%
%
%
%

\noindent
{\bf Fig. \ref{fig:A degraded affinely singular distribution} -- Degraded affinely singular RV:}
The case of most interest is the case of 
rank-deficient transformations.
Let us consider the case in which  $\Ytv^3=\At \, \Xv^3$, where
\ea{
	\At=\left[\begin{array}{ccc}1&1&0\\0&1&1\\1&0&-1\end{array}\right].
	\label{eq:degraded A}
}
Since $\rank(\At)=2$, $\Ytv^3$ has a two-dimensional structure; i.e., the distribution of $\Ytv^3$ as a 3-dimensional vector is purely singular. 
See Figure \ref{fig:A degraded affinely singular distribution}.
Besides,  it consists of one dimensional singularities as well as a point mass.
Quite interestingly, our results in this paper are general enough to even include this case -- see Theorem \ref{thm: disc_cont_fat}.
In particular, for the transformation in  \eqref{eq:degraded A}, we have  that $d(\Ytv^3)=11/8$.

	\section{Results on Mutual Information} \label{sec: mi}
	In this section, we apply the notion of DRB to evaluate the mutual information between two affinely singular random vectors. 
	We first study the Kullback-Leibler (KL) divergence of two affinely singular random vectors $\Xv$ and $\Yv$,  
	since their mutual information could be stated as the KL-divergence divergence between the joint probability measure $\mu_{\Xv, \Yv}(\cdot)$ and the product measure $\mu_{\Xv}\times \mu_{\Yv}(\cdot)$. 
	
	Next, we show that if $\Xv$ and $\Yv$ are affinely singular, such product and joint probability measures, in case of existence of the mutual information, must also be affinely singular. 
	Employing these results, we show that 
	\ea{
		\Isf(\Xv;\Yv)= b(\Xv)+b(\Yv)-b(\Xv, \Yv). \label{eqn: mutual_info}
	}
	Indeed, this is an extension of the classical results when $\Xv$ and $\Yv$ are either both discrete or both continuous, where   $b(\Xv)=\Hsf(\Xv)$, $b(\Yv)=\Hsf(\Yv)$ and $b(\Xv)=\hsf(\Xv )$, $b(\Yv)=\hsf(\Yv)$, respectively.
	
	We start by the following lemma that states absolutely continuous probability measures with respect to an affinely singular probability measure are affinely singular themselves with the same singularity subsets.
	
	
	\begin{lem} \label{lem: abs_cont_affine}
		Let  $\mu_{\Xv}(\cdot)$ be an affinely singular probability measure according to Definition  \ref{def: aff_sig_measure} on a collection of affine subsets $\Scal_{\Xv}$, and with measures $\mu_{\Xv}^{\Acal}(\cdot)$ on each set $\Acal\in \Scal_{\Xv}$. Then, $\mu_{\Yv}(\cdot)$ is absolutely continuous with respect to $\mu_{\Xv}(\cdot)$, if and only if $\mu_{\Yv}(\cdot)$ is  affinely singular on a collection of affine subsets $\Scal_{\Yv}\subseteq\Scal_{\Xv}$ and $\mu_{\Yv}^{\Acal}(\cdot)$ is absolutely continuous with respect to $\mu_{\Xv}^{\Acal}(\cdot)$ for every $\Acal\in \Scal_{\Xv}$. 
	\end{lem}
	
	\begin{IEEEproof}
		See Appendix \ref{app: prf_abs_cont_aff_sing}.
	\end{IEEEproof}
	
	Using Radon-Nikodym theorem and the above lemma, we can evaluate the KL-divergence for a pair of affinely singular RVs.
	
	\begin{lem} \label{lem: kl}
		Let $\mu_{\Xv}(\cdot)$ and $\mu_{\Yv}(\cdot)$ be two affinely singular probability measures according to Definition \ref{def: aff_sig_measure} on affine subsets $\Scal_{\Xv}$ and $\Scal_{\Yv}$ with total probability $p_{\Xv}^{\Acal}$, $p_{\Yv}^{\Bcal}$ on each affine set $\Acal\in \Scal_{\Xv}$ and $\Bcal\in \Scal_{\Yv}$, and with measures $\mu_{\Xv}^{\Acal}(\cdot)$ and $\mu_{\Yv}^{\Bcal}(\cdot)$ on such sets, respectively.  The KL-divergence $\Dsf(\mu_{\Xv}\|\mu_{\Yv})$ is defined if and only if $\Scal_{\Xv}\subseteq \Scal_{\Yv}$ and for every $\Acal\in \Scal_{\Xv}$, the measure $\mu_{\Xv}^{\Acal}$ is absolutely continuous with respect to the measure $\mu_{\Yv}^{\Acal}$.  In that case, we can evaluate the KL-divergence as
		\ea{
			\Dsf(\mu_{\Xv}\|\mu_{\Yv}) = \sum_{\Acal\in \Scal_{\Xv}} p_{\Xv}^{\Acal}\Dsf(\mu_{\Xv}^{\Acal}\|\mu_{\Yv}^{\Acal})+\Dsf(p_\Xv\|p_{\Yv}),
		}
		where  $\Dsf(p_{\Xv}\|p_{\Yv}):=\sum_{\Acal\in \Scal_{\Xv}} p_{\Xv}^{\Acal} \log \frac{p_{\Xv}^{\Acal}}{p_{\Yv}^{\Acal}}$. 
	\end{lem}
	
	\begin{IEEEproof}
		See Appendix \ref{app: prf_kl}.
	\end{IEEEproof}
	
	\begin{rem}
		One can check that the proof of Lemma \ref{lem: kl} is not restricted to the case that $\mu_{\Yv}$ is a probability measure. 
		Using Lemma \ref{lem: aff_sing}, we know that there exists a triplet $\big(\{\fv_i\}, \{\Cv_i\}, V_m\big)$, such that
		\ea{
			\mu_{\Xv}(\Ccal) =\sum_i \Pr(V_m = i) \mu_{\Cv_i}\big(\fv_i^{-1}(\Ccal\cap \Acal_i)\big).
		}
		Then, we can define $\mu_{\Yv}(\cdot)$ as
		\ea{
			\mu_{\Yv}(\Ccal) = \sum_{i} \ell_{e_i}\big(\fv_i^{-1}(\Ccal\cap \Acal_i)\big),
		}
		where $e_i$ is the dimension of $\Cv_i$, and $\ell_{e_i}$ is the Lebesgue measure on $\Rbb^{e_i}$. In that case, since $\mu_{\Yv}(\cdot)$ is $\sigma$-finite, with the same line of proof as above, the generalized entropy of $\mu_{\Xv}$ with respect to $\mu_{\Yv}$ can be obtained as
		\ea{
			H_{\mu_{\Yv}}&(\mu_{\Xv}) \nonumber\\&:= -\int \diff \mu_{\Xv} \log \frac{\diff \mu_{\Xv}}{\diff \mu_{\Yv}} \\&= -{\textstyle\sum_i} \Pr(V_m = i) \int_{\Acal_i} \tfrac{\diff \fv_i^*(\mu_{\Cv_i})}{\diff \fv_i^*(\ell_{e_i})}\log \tfrac{\diff \fv_i^*(\mu_{\Cv_i})}{\diff \fv_i^*(\ell_{e_i})} {\diff \fv_i^*(\ell_{e_i})} \nonumber \\&\quad- {\textstyle\sum_{i}} \Pr(V_m = i) \log \Pr(V_m = i), \label{eqn: gen_entropy}
		}
		where $\fv_i^*(\mu)$ for a measure $\mu$ denotes the push-forward measure $\mu(\fv_i^{-1}(\cdot))$. 
		
		With a change of variables, it is possible to 
		rewrite \eqref{eqn: gen_entropy} as
		\ea{
			H_{\mu_{\Yv}}&(\mu_{\Xv}) \nonumber\\&= -{\textstyle\sum_i} \Pr(V_m = i) \int_{\Rbb^{e_i}} \tfrac{\diff \mu_{\Cv_i}}{\diff \ell_{e_i}}\log \tfrac{\diff \mu_{\Cv_i}}{\diff \ell_{e_i}} {\diff \ell_{e_i}} \nonumber + \Hsf(V_m)\\
			& = \sum_i \Pr(V_m = i) \hsf(\Cv_i) + \Hsf(V_m) \overset{(a)}{=} b(\Xv),
		}
		in which $(a)$ holds because of Theorem \ref{thm: singular_subset}. %
		This expresses the DRB of an affinely singular RV in terms of the generalized entropy of the probability measure with respect to a mixture of Lebesgue measures.
		
		
	\end{rem}
	
	We now state the main result of this section which connects the notion of DRB to mutual information between two affinely singular RVs.
	
	\begin{thm}
		Let $\mu_{\Xv}(\cdot)$ and $\mu_{\Yv}(\cdot)$ be probability measures of two affinely singular RVs as in Lemma \ref{lem: kl}. The mutual information $\Isf(\Xv ;\Yv)$ exists only if the joint probability measure $\mu_{\Xv, \Yv}(\cdot)$ is affinely singular on the collection of affine sets $\Scal_{\Xv} \otimes \Scal_{\Yv}:=\{ \Acal\times \Bcal \, : \Acal\in \Scal_{\Xv}\, , \Bcal\in \Scal_{\Yv}\}$. In this case, $\Isf(\Xv; \Yv)$ can be calculated as \eqref{eqn: mutual_info}. 
	\end{thm}
	
	\begin{IEEEproof}
		\changedo{Assume $\Xv$ and $\Yv$ are $m$ and $n$-dimensional  RVs, respectively.} Lemma \ref{lem: aff_sing} implies the existence of the triplets  $\big(\{\fv_i\}, \{\Cv_i\}, U_m\big)$ and $(\{\gv_i\}, \{\Dv_i\}, V_n)$,  such that
		\ea{
			\mu_{\Xv}(\Ccal) = \sum_i \Pr(U_m = i) \mu_{\Cv_i}\big(\fv_i^{-1}(\Ccal\cap \Kcal_i)\big), \label{eqn: mux}
		}
		and 
		\ea{
			\mu_{\Yv}(\Dcal) = \sum \Pr(V_n = i) \mu_{\Dv_i}\big(\gv_i^{-1}(\Dcal\cap \Lcal_i)\big), \label{eqn: muy}
		}
		for $\Ccal\subseteq \Rbb^m$ and $\Dcal\subseteq\Rbb^n$, where $\Kcal_i$ and $\Lcal_i$ are the \changedo{images} of $\fv_i(\cdot)$ and $\gv_i(\cdot)$, respectively, and \changed{where $\Cv_i$s and $\Dv_i$s are $e_i$- and $e_i'$-dimensional, respectively.}
		
		We construct the product measure 
		\ea{
			\mu_{\Xv}\times \mu_{\Yv}(\Ecal) = \int_{\yv\in \Rbb^n} \mu_{\Xv}(\Ecal^{\yv})\diff \mu_{\Yv}(\yv), \label{eqn: prod}
		}
		where $\Ecal^y = \{(\xv', \yv')\in \Ecal:\, \yv'=\yv\}$. 
		Using Fubini's theorem and  \eqref{eqn: mux}-\eqref{eqn: prod}, we conclude that
		\ea{
			\mu_{\Xv}&\times \mu_{\Yv}(\Ecal) \nonumber\\&=\sum_{i, j}p_{ij} \int_{\yv \in \Lcal_i} \fv_i^*(\mu_{\Cv_i})(\Ecal^y \cap \Kcal_i) \diff \gv_i^*(\mu_{\Dv_i})(\yv),
		}
		where $\fv_i^*(\mu_{\Cv_i})(\cdot)$ denotes the push-forward measure $\mu\big(\fv_i^{-1}(\cdot))$, and $p_{ij}$ indicates the product mass probability ${p_{ij}=\Pr(U_m = i)\Pr(V_n = j)}$. 
		Hence, by the definition of the product measure, we have
		\ea{
			\mu_{\Xv}&\times \mu_{\Yv}(\Ecal)=\sum_{i, j}p_{ij}\fv_i^*(\mu_{\Cv_i})\times \gv_i^*(\mu_{\Dv_j})(\Ecal \cap \Kcal_i\times \Lcal_j).
		}
		Since the product of absolutely continuous measures is also absolutely continuous, $\mu_{\Xv}\times \mu_{\Yv}$ is an affinely singular probability measure on the collection of affine sets $\Scal_{\Xv}\otimes \Scal_{\Yv}$, as \changedo{$\fv_i^*(\mu_{\Cv_i})$ and $\gv_j^*(\mu_{\Dv_j})$} are absolutely continuous on $\Kcal_i$ and $\Lcal_j$.
		One can also see that $\mu_{\Xv}\times \mu_{\Yv}$ is absolutely continuous with respect to the mixture measure
		\ea{
			\ell_{\Ccal} = \sum_{i, j} \changedo{\ell_{e_i+e_j'}}\big(h_{ij}^{-1}(\Ccal\cap \Kcal_i\times \Lcal_j)\big),
		}
		where $h_{ij}\big([\xv_1, \xv_2]\big)=\big[\fv_i(\xv_1), \gv_j(\xv)\big]$, and $e_i$ is the dimension of the domain of $h_{ij}(\cdot)$, and \changedo{ $\ell_{e_i+e_j'}(\cdot)$ is the Lebesgue measure on $\Rbb^{e_i+e_j'}$}. Since $\fv_i(\cdot)$ and $\gv_j(\cdot)$ are linear functions \changedo{with images} $\Kcal_i$ and $\Lcal_j$, the inverse of the function $h_{ij}(\cdot)$ always exists.
		
		Now, we recall the well-known identity
		$$\Isf(\Xv; \Yv) = \Dsf(\mu_{\Xv, \Yv}\|\mu_{\Xv}\times \mu_{\Yv}).$$ 
		With the help of Lemma \ref{lem: kl}, 
		we conclude that $\Isf(\Xv;\Yv)$ exists only if $\mu_{\Xv, \Yv}$ is affinely singular on the collection of affine sets $\Scal_{\Xv}\otimes \Scal_{\Yv}$. Hence, we can define $\mu_{\Xv, \Yv}$ as
		\ea{
			\mu_{\Xv, \Yv} (\Ecal)= \sum_{i, j} p^{\Xv, \Yv}_{i,j} h_{ij}^*(\mu_{\Ev_{ij}})(\Ecal\cap \Kcal_i\times \Lcal_j),
		}
		for $\Ecal\subseteq \Rbb^{m+n}$, where $\Ev_{ij}$ is an absolutely continuous RV on $\Rbb^{e_{ij}}$, in which $e_{ij}$ is the dimension of the domain of $h_{ij}(\cdot)$. Further, since $\mu_{\Xv, \Yv}(\Ccal\times \Rbb^{n})=\mu_{\Xv}(\Ccal)$ for every $\Ccal\subseteq \Rbb^m$,  \eqref{eqn: mux}, implies that
		\ea{
			\sum_{j} p_{i, j}^{\Xv, \Yv} \mu_{\Ev_{ij}}(\Ccal\times \Rbb^n) =\Pr(U_m = i) \mu_{\Cv_i}(\Ccal), \label{eqn: sum_prob_e_1}
		}
		which yields
		\ea{
			\sum_{j} p_{i, j}^{\Xv, \Yv} = \Pr(U_m = i).\label{eqn: sum1}
		}
		Similarly, one could prove that
		\ea{
			\sum_{j} p_{i, j}^{\Xv, \Yv} \mu_{\Ev_{ij}}(\Rbb^m\times \Ccal) =\Pr(V_n = j) \mu_{\Dv_j}(\Ccal),\label{eqn: sum_prob_e_2}
		}
		and
		\ea{
			\sum_{i} p_{i, j}^{\Xv, \Yv} = \Pr( V_n = j).\label{eqn: sum2}
		}
		
		Again by using Lemma \ref{lem: kl}, we have that
		\ea{
			\Isf(&\Xv;\Yv)\nonumber\\&=\sum_{i, j}p^{\Xv, \Yv}_{i, j}\Dsf(h_{ij}^*(\mu_{ij})\|\fv_i^*(\mu_{\Cv_i})\times \gv_i^*(\mu_{\Dv_j})) \nonumber\\&\quad+\sum_{i,j} p^{\Xv, \Yv}_{i, j}\log \frac{p^{\Xv, \Yv}_{i, j}}{p_{i j}}\\
			&\overset{(a)}{=} \sum_{i, j}p^{\Xv, \Yv}_{i, j} \int \diff h_{ij}^*(\mu_{\Ev_{ij}})\log \frac{ \diff h_{ij}^*(\mu_{\Ev_{ij}}) }{\changedo{\diff\big( \fv_i^*(\mu_{\Cv_i})\times \gv_i^*(\mu_{\Dv_j})\big)}}\nonumber\\
			&\quad +\Hsf(U_m)+\Hsf(V_n)+\sum_{i, j} p_{i, j}^{\Xv, \Yv}\log p_{i, j}^{\Xv, \Yv}\\
			&\overset{(b)}{=}\sum_{i, j} p_{i, j}^{\Xv, \Yv}\int \diff \mu_{\Ev_{ij}}\log \changedo{\frac{\diff\mu_{\Ev_{ij}}}{\diff\big(\mu_{\Cv_i}\times \mu_{\Dv_j}\big)}}\nonumber\\
			&\quad +\Hsf(U_m)+\Hsf(V_n)+\sum_{i, j} p_{i, j}^{\Xv, \Yv}\log p_{i, j}^{\Xv, \Yv}\\
			&\overset{(c)}{=} \sum_i \Pr(U_m = i) \hsf(\Cv_i) + \sum_j \Pr(V_n = j) \hsf(\Dv_j) \nonumber\\&\quad-\sum_{i, j}p_{i, j}^{\Xv, \Yv}\hsf(\Ev_{ij}) +\Hsf(U_m)+\Hsf(V_n) \nonumber\\
			&\quad+\sum_{i, j} p_{i, j}^{\Xv, \Yv}\log p_{i, j}^{\Xv, \Yv}\\
			&= b(\Xv)+b(\Yv)-b(\Xv, \Yv),
		}
		where $(a)$ holds because of \eqref{eqn: sum1} and \eqref{eqn: sum2},  $(b)$ follows from a change of variables, and $(c)$ is validated through  \eqref{eqn: sum_prob_e_1} and \eqref{eqn: sum_prob_e_2},  \changedo{and by chain rule $\frac{\diff\mu_{\Ev_{ij}}}{\diff\big(\mu_{\Cv_i}\times \mu_{\Dv_j}\big)} = \frac{\diff\mu_{\Ev_{ij}}}{\diff\hv_{ij}^*(\ell_{e_{i}+e_j'})}\frac{\diff\hv_{ij}^*(\ell_{e_{i}+e_j'})}{\diff\big(\mu_{\Cv_i}\times \mu_{\Dv_j}\big)}$}.
	\end{IEEEproof}

\section{Results on Random Processes} \label{sec:ma}

\changedo{In this section, we study the role of the DRB, BID, and IDR introduced in Section \ref{sec:Compressibility of random vectors} and \ref{sec:Compressibility of stochastic processes} in evaluating compressibility of stochastic processes. More specifically, in the following, we show that under certain conditions, the BID coincides with the $\ep$-compression rates. 
In Section \ref{sec: bid_ma}, we show that moving-average processes satisfy such conditions. Next, by finding the IDR and BID of moving-average processes, we evaluate the $\ep$-compression rates for these cases. Finally, in Section \ref{sec:DRB}, we quantify the DRB of a class of moving-average processes.}

\subsection{BID and $\epsilon$-compression rates} \label{sec: eps_comp}

{\changed 
For processes that have samples with affinely singular probability measures, 
we show that the $\epsilon$-compression rate    coincides with the BID
under some conditions. 
This is our first result that makes a connection between information-theoretic and compressed-sensing notions of compressibility. 

\begin{thm}\label{thm: concent_Rs}
\changedo{Let $\{\Zv_t\}_{t=-\infty}^{\infty}$ be a discrete-domain stochastic process for which the distribution of all finite subset of samples is
affinely singular as in Lemma \ref{lem: aff_sing}. Then,
}
	if the BID exists, it is achievable as
	\ea{
		d_B\big(\{\Zv_t\}\big) = \lim_{m\to\infty} \Ebb\big[\tfrac{e_{V_m}}{m}\big].\label{eqn: BID_di_m}
	}
	Moreover, if $V_m$ has a finite sample space for all $m$ and \changedo{if for all $\ep \in (0,1]$, $\delta\in \Rbb^{+}$, there exists a finite integer $m(\ep, \delta)\in \Nbb$ such that for all $m\geq m(\ep, \delta)$ we have}
		\ea{
		\Pr_V\bigg(i: \Big|\tfrac{e_i}{m}-d_B\big(\{\Zv_t\}\big)\Big|<\delta\bigg)>1-\ep,	\label{eqn:Concentration}
	}  
%
	then, for the process $\{\Zv_t\}$ 
	we know that
	\ea{
		R^*(\ep)=R_B(\ep)=R(\ep)=d_B\big(\{\Zv_t\}\big),\label{eqn: Rs_equal}
	}
	where $R_B(\ep)$ is the Minkowski dimension compression rate defined in \cite[Definition 10]{wu2010renyi}.
\end{thm}

\begin{IEEEproof}
	See Appendix \ref{app: prf_concent_Rs}.
\end{IEEEproof}

{\changed
\begin{IEEEproof}[Sketch of proof]
The proof consists of three steps. In the first step, we show that $R(\ep)\leq d_{B}(\{\Zv_t\})+\delta$, for any value of $\delta\in(0, 1)$. For this purpose, we find a high-probability $\lfloor m(d_{B}(\{\Zv_t\})+\delta)\rfloor$-rectifiable set $\Rcal_m$ by ignoring high-dimensional affine subsets and bounding $\Zv^m$ to a compact set. 

In the second step, we prove $R^*(\ep)\geq d_B(\{\Zv_t\})$ 
by contradiction; we show that if $d_B(\{\Zv_t\})-R^*(\ep)=\delta\in(0, 1)$, then, for every $m\geq M$, every $\lceil (1-d_B(\{\Zv_t\})+\delta/2)m\rceil$-dimensional subspace $\Hcal^m$, and every high-probability set $\Scal^m$, the set $(\Scal_m - \Scal^m)\cap \Hcal^m$ contains at least one non-zero vector. This coupled with Lemma \ref{lmm: min_R} shows that $R^*(\ep)\geq d_B(\{\Zv_t\})-\delta/2$, which is a contradiction.

Finally, in the third step, using inequality $R^*(\ep)\leq R_B(\ep)\leq R(\ep)$ (see \cite[Eqn. 75]{wu2010renyi}), we complete the proof.
\end{IEEEproof}
}

As an application of the above result, we study compression-rates of moving-average processes in the following subsection.
}
\subsection{Compression rates and BID of moving-average processes} \label{sec: bid_ma}
Using Theorem \ref{thm: disc_cont_fat} {\changed and Theorem \ref{thm: concent_Rs}}, we study a compressed sensing problem below: 
consider the moving average (MA) process 
\ea{
Y_i = \sum_{j=-l_1}^{l_2} c_{j} W_{i-j},	
\label{eq:MA def}
}
where the excitation noise $\{W_j\}_{j \in \Zbb}$ is a set of i.i.d. RVs with discrete-continuous distribution.
%
We further assume that  $d(W_j)=\alpha$, and $\{c_j\}_{j \in [-l_1,l_2]}$  are constants with $c_{-l_1}, c_{l_2}\neq 0$. 
Let us consider the problem of recovering a truncated version of the process $\{Y_t\}$ (e.g., $Y_1,\dots, Y_m$) from noisy random projections onto lower dimensional sub-spaces. The goal is to determine the minimum dimension of such sub-spaces in terms of the truncated length (i.e., $m$). More formally, we would like to find the minimum value of $k$ such that $\Yv^{m}=(Y_1,\dots,Y_m)^{\intercal}$ could be fairly recovered from $H_{k\times m} \Yv^{m} + \Nv^k$ where $H_{k\times m}$ is a random projection matrix and $\Nv^k$ stands for the noise vector. 


%
%

{\changed 
Theorem \ref{thm: concent_Rs} suggests that  $k=m\alpha$ if \eqref{eqn:Concentration} is fulfilled;}
{\changed below we show that,} 
if the sample space of the discrete part of $W_i$s is finite in the moving-average process, then, this inequality holds. Leveraging this result, we are able to evaluate $R(\ep)$ in such cases. 

{\changed
\begin{lem}\label{lmm: concent_MA}
	If we generate an $m$-dimensional realization of a moving-average process as in \eqref{eq:MA def}, the probability of that realization being on an at least $(k+1)$-dimensional affine singularity for $\f{k+l_1+l_2}{m+l_1+l_2}<\alpha$, can be bounded as
	\ea{
		\Pr_{V_m}({e_{V_m}}> k) \geq 1- {\rm exp}\Big\{-(m+l_1+l_2)\Dsf\big(\tfrac{k+l_2+l_1}{m+l_1+l_2}\|\alpha\big)\Big\},\label{eqn: up_bound_d_ma}
	}
	and the probability of being on an at most $(k-1)$-dimensional affine singularity in the case $\f{k}{m}> \alpha$, is bounded as
	\ea{
		\Pr_{V_m}(e_{V_m}< k) \geq 1- {\rm exp}\Big\{-m\Dsf\big(\tfrac{k}{m}\|\alpha\big)\Big\},\label{eqn: low_bound_d_ma}
	}
	where $\Dsf(p\|q)$ is the Kullback-Leibler divergence between $\text{Bern}(p)$ and $\text{Bern}(q)$.
\end{lem}

\begin{IEEEproof}
See Appendix \ref{app: prf_concent_MA}.
\end{IEEEproof}

\begin{Cor}\label{cor: ma_concentrated}
For each pair $(\ep, \delta)\in (0, 1]^2$, there exists a large enough $m$ such that $\Pr_{V_m}\big(|\frac{e_{V_m}}{m}-\alpha|>\delta\big)<\ep$.
\end{Cor}

\begin{IEEEproof}
To show this, let $\ep, \delta\in \Rbb^+$ and define $k_m=m(\alpha-\delta)$. If 
	\changedo{\ea{
		m\geq \max \Big\{\tfrac{2(l_1+l_2)(1-\alpha+\delta)}{\delta}, \tfrac{-\log \ep}{\Dsf(\alpha-\delta/2\|\alpha)}-l_1-l_2\Big\},\label{eqn: low_bound_n}
	}}
	then, we have that
	\eas{
		\tfrac{k_m+l_1+l_2}{m+l_1+l_2} &= \alpha-\delta+ \tfrac{(l_1+l_2)(1-\alpha+\delta)}{m+l_1+l_2}\label{eqn: up_bound_k'_n_first}\leq \alpha-\tfrac{\delta}{2}\\&<\alpha.	\label{eqn: up_bound_k'_n}
	}
	Now, using Lemma \ref{lmm: concent_MA} we conclude that
	\ea{
		\Pr_{V_m}(e_{V_m}&>k_m)\nonumber\\& \geq 1- {\rm exp} \Big\{-(m+l_1+l_2)\Dsf\big(\tfrac{k_m+l_2+l_1}{m+l_1+l_2}\|\alpha\big)\Big\} \nonumber \\
		&\overset{(a)}{\geq} 	1-{\rm exp}\big\{-(m+l_1+l_2)\Dsf(\alpha-\delta/2\|\alpha)\big\} \nonumber \\
		&\overset{(b)}{\geq} 	1-\ep,\label{eqn: low_bound_d}
	}
	where 	$(a)$ is due to \eqref{eqn: up_bound_k'_n_first} and $(b)$ is because of \eqref{eqn: low_bound_n}.
	
	To upper-bound  $e_{V_m}$,  let $k'_m=m(\alpha+\delta)$. Thus,
	\ea{
		\f{k'_m}{m}= \alpha+\delta>\alpha.\label{eqn: kprime_low_bound}
	} 	
	Now, if 
	\ea{
		m\geq  \tfrac{-\log\ep}{\Dsf(\alpha+\delta\|\alpha)},\label{eqn: low_bound_n_ma2}
	}
	we can conclude 
	\changedo{\ea{
		\Pr_{V_m}(e_{V_m}<k'_m)&\geq 1-\exp\big\{-m\Dsf(\tfrac{k'_m}{m}\|\alpha)\big\}\nonumber\\
		&\overset{(a)}{\geq} 1- \exp(-m\Dsf(\alpha+\delta\|\alpha))\nonumber\\
		&\overset{(b)}{\geq} 1- \ep,
	}
	} using \eqref{eqn: low_bound_d_ma}. Here, $(a)$ is because of \eqref{eqn: kprime_low_bound} and $(b)$ is due to \eqref{eqn: low_bound_n_ma2}.
\end{IEEEproof}
}


Next, we find the IDR and BID of MA processes in the general case; here, the alphabet of the discrete component is not necessarily restricted to be finite \changedo{as opposed to the result in Theorem \ref{thm: concent_Rs}}. 

{\changed
\begin{thm}\label{thm:MA_IDR_BID}
Let $\{Y_i\}$ be a MA process as in \eqref{eq:MA def} with $d(W_j) = \alpha$. Then, we have
\ea{
d_{B}\big(\{\Yv_t\}\big) = d_{I}\big(\{\Yv_t\}\big) 
=\alpha.
}
\end{thm}
}

{\changed
\begin{IEEEproof}
For a better explanation of the linear transformation in \eqref{eq:MA def}, we can express the truncated process $\{Y_t\}$ as
%
\ea{
\Yv^m = \underbrace{\left[\begin{array}{c c c c c c }
	c_{-l_1}&  \ldots & c_{l_2} & 0 & \ldots & 0 \\
	0 & c_{-l_1}  & \ldots & c_{l_2} & \ldots & 0 \\
	\vdots & \ldots  & \ddots & \ddots &\ddots &\vdots \\
	0& \ldots & \ldots & c_{-l_1}  & \ldots & c_{l_2}
\end{array}\right]}_{A_m} \Wv^{m+l_1+l_2}, \label{mat: MA}
}
where \changedo{$c_j$s} are the constants introduced in \eqref{eq:MA def}.
 Due to the Lipschitz dependence of $\Yv^m$ on $\Wv^{m+l_1+l_2}$, we know that
\ea{
d(\Yv^m) \leq (m+l_1+l_2) \alpha, \label{eq:Ym_Upper}
}
where $\alpha$ stands for the probability of $W_i$s being drawn from the continuous component ($\alpha_i$ in \eqref{eqn: indep_disc_cont}). To lowerbound $d(\Yv^m)$, let us consider $W_i$s separately and assume each has its own $\alpha_i$ (i.e., $\alpha_i$s could be different).  
From Remark \ref{rem:ContWeight}, we know that $d(\Yv^m)$ does not depend on the choice of the discrete component of $W_i$s; thus, if we assume $W_i$ takes the value $0$ with probability $1-\alpha_i$ (and is drawn from a continuous distribution with probability $\alpha_i$), $d(\Yv^m)$ remains unchanged. Furthermore, if we decrease any $\alpha_i$, then, $d(\Yv^m)$ decreases or remains unchanged. 
 Therefore, a lower-bound for $d(\Yv^m)$ can be achieved by setting the first $l_1+l_2$ terms $W_1=\dots=W_{l_1+l_2}\equiv 0$ which corresponds to $\alpha_1=\dots=\alpha_{l_1+l_2}=0$. This implies that
%
%
\ea{
d(\Yv^m) &\geq {\changed d\Big(Q_m \big[W_{l_1+l_2+1} \,,\, \dots \,,\, W_{l_1+l_2+m}\big]^{\intercal}\Big)} \nonumber\\
&=
\sum_{\sv \in \{0, 1\}^m} \rank(Q_m^{[\sv]}) \alpha^{\|\sv\|_1}(1-\alpha)^{(1-\|\sv\|_1)} ,
}
where $Q_m$ is the matrix formed by columns with index $[l_1+l_2+1: l_1+l_2+m]$ of $A_m$. 
Since $Q_m$ is a lower triangular matrix with non-zero diagonal elements, we conclude that it is  full-rank. Hence, we have
\ea{
d(\Yv^m) &\geq \sum_{\sv \in \{0, 1\}^m} \Big(\sum_{i=1}^{m} s_i\Big) \alpha^{\lb \sum_{i=1}^{m} s_i\rb }(1-\alpha)^{(1-\sum_{i=1}^{m} s_i)} \nonumber\\
&= \sum_{i=0}^m i \, \alpha^{i} (1-\alpha)^{1-i}
 = m\alpha. \label{eq:Ym_lower}
}
By combining \eqref{eq:Ym_Upper} and \eqref{eq:Ym_lower} as $m\to\infty$, we have that
%
\ea{
d_{B}\big(\{\Yv_t\}\big) = \lim_{m\to\infty} \f {d(\Yv^m)}{m} =\alpha. \label{eqn: Block_MA}
}
 Moreover, since MA processes are $\psi^*$-mixing according to the definition in \cite{Jalali2016uni}, due to \cite[Corollary 15]{koch2019} we know that the IDR and BID for such processes are equal, which together with \eqref{eqn: Block_MA} concludes
\ea{
d_I\big(\{\Yv_t\}\big)=\alpha.
}
\end{IEEEproof}
}

\begin{figure}
	\centering
	\includegraphics[width=.9\linewidth]{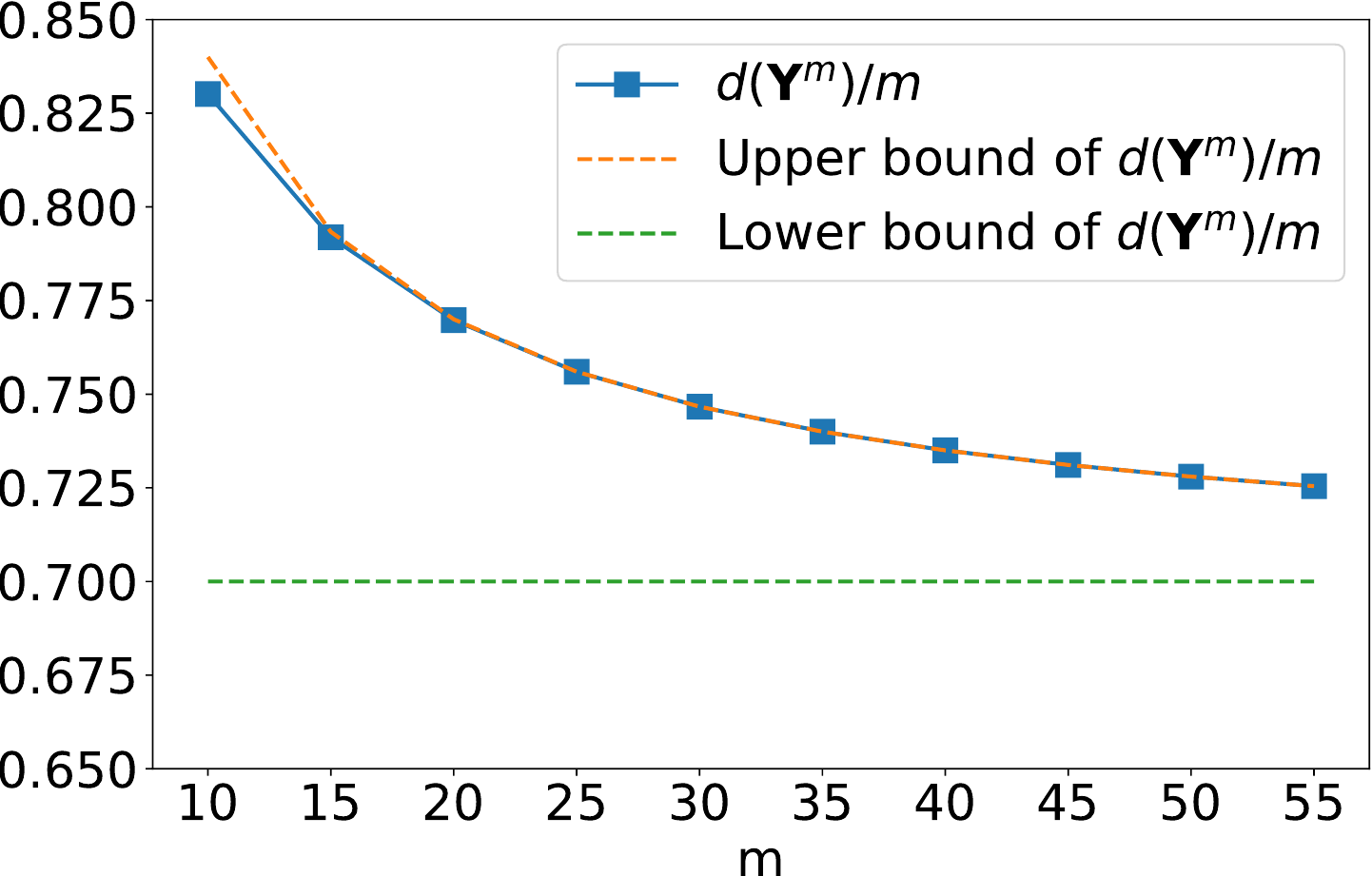}
	\caption{Block-average information dimension of a moving-average process samples and its bounds as in \eqref{eq:Ym_Upper} and \eqref{eq:Ym_lower}, in the special case in which $\boldsymbol{a} = [-2, 0.5, 1]$ and $\alpha=0.7$.}
	\label{Fig: BID}
\end{figure}

{\changed
To better visualize Theorem \ref{thm:MA_IDR_BID}, we consider the MA process associated with $\boldsymbol{c} = [-2, 0.5, 1]$ and $\alpha=0.7$; Figure \ref{Fig: BID} depicts the evolution of $\f{d(\Yv^m)}{m}$, as well as the upper and  lower-bounds in 
\eqref{eq:Ym_Upper} and \eqref{eq:Ym_lower}.

Note that Theorem \ref{thm:MA_IDR_BID}
is an improvement over \cite[Theorem 5 ]{Jalali2016uni} wherein the proved equality in this work is presented only as an upper-bound. Nevertheless, 
}
 \cite[Theorem 9]{Jalali2016uni} implies that if a bounded MA process is sampled using an $m\times n$ random matrix with independent standard normal entries, then, 
the Lagrangian-minimum entropy pursuit (MEP) algorithm reconstructs the realization of MA process with no asymptotic loss as $n$ grows to infinity.

\subsection{DRB of moving-average processes}\label{sec:DRB}
As an application of the result in Theorem \ref{thm: disc_cont_fat} and the evaluation of the value of DRB therein, we consider the samples of a moving-average process with discrete-continuous excitation. Indeed, we truncate the samples into an $m$-dimensional vector and evaluate the DRB of the vector normalized by $m$ as $m\to \infty$. 
This measure is useful for comparing the compressibility of processes with the same BID (e.g., two MA processes with discrete-continuous excitation that have the same information dimension). 

\changedo{To express our results on the DRB of a process, we introduce the following notations 
\ea{
\Psi_{{\alpha}}(c_{-l_1}, \ldots, c_{l_2})
=\Psi_{\alpha}(A_m):= \lim_{m\to\infty}\psi_{m,\nuv}(A_m)
\label{eqn: psi_alpha_2},
}
where {\changed $\psi_{m,\nuv}(\cdot)$ is defined in \eqref{eq:psi_def}}.
Note that $A_m$ is determined by moving-average process coefficients $(c_{-l_1}, \ldots, c_{l_2})$.
}
\begin{prop}\label{prop:DRB_MA}
For the MA process in \eqref{eq:MA def}, if \changedo{ $\hsf(W_{\cm, 1})>-\infty$, $M_{W_{\cm, 1}}(\beta)<\infty$ for some $\beta\in \Rbb^{+}$, and the support of $W_{\dm, 1}$} is a finite set, then, with $\Psi_{\alpha}$ defined as \eqref{eqn: psi_alpha_2} we have that
\begin{itemize}
	\item $
	\lim_{m\to\infty} \frac{b(\Yv^m)}{m} = b(W_1)+\Psi_{\alpha}(c_{-l_1}, \ldots, c_{l_2})$, 
 given $\alpha\in (0, \tfrac{1}{l_1+l_2})$, and
	
	\item $
	\lim_{m\to\infty} \frac{b(\Yv^m)}{m} = b(W_1) $, 
	given $\alpha=0$, and \changedo{$\Pr(W_{\dm, 1}\neq 0)\in (0, \tfrac{1}{l_1+l_2})$}.
\end{itemize}

\end{prop}

{\changed
Before we proceed to the proof of Proposition \ref{prop:DRB_MA}, we provide the below lemma \changedo{that considers a general linear transformation of discrete-continuous RVs (i.e., not necessarily samples of an MA process)}.
}

{\changed
\begin{lem}\label{lem: rho}
	Let $\{\Xv_t\}$ be an i.i.d. process with discrete-continuous elements as \eqref{eqn: indep_disc_cont}. \changedo{We denote the continuous and discrete components of $X_i$ by $X_{\cm, i}$ and $X_{\dm, i}$, respectively; $\alpha_i$ also stands for the probability that $X_i$ takes a value from the corresponding continuous measure.
	} Further, assume \changedo{$\hsf(X_{\cm, 1})>-\infty$, $M_{X_{\cm, 1}}(\beta)<\infty$} for some $ \beta\in \Rbb^{+}$, and $X_{\dm, i}$ takes values in a finite set. 	
	Let $g:\, \Nbb\to\Nbb$ be a function  that satisfies $g(m)>\beta e$ for large enough $m$, and let $G(m) = \max\big( g(1),\dots , g(m) \big)$. 
	For each $m=1, 2, \dots$, we define $Y_m=\cv_m^{T} \, \Xv^{g(m)}$ where $\cv_m\in \Rbb^{g(m)}$ is an arbitrary but known vector. Hence, $\{\Yv_t\}$ is a random process that linearly depends on $\{\Xv_t\}$. Besides, we form $m\times G(m)$ matrices $A_m$ by zero-padding each $\cv_i^{\intercal}$ in the range $i=1,\dots,m$ with enough zeros to increase the length of the vector to $G(m)$ and stacking all of them as the rows of the matrix $A_m$.
	Now, there exist \changedo{fixed values} $C_i\in \Rbb$ (or $\Ct_i\in \Rbb$) for $i\in [2]$ such that
%
%
	\begin{itemize}
		\item if $\alpha_1\neq 0$, then, for sufficiently large $m$, 
		\ea{
			&\big|\tfrac{b(\Yv^m)}{m}-\tfrac{G(m)}{m}b(X_1)-\psi_{m,\nuv}(A_m)\big| \label{eqn: estimate_DRB} \\&\leq  \tfrac{G(m)}{m}\rhom \big(C_1 G(m)+C_2\log \rhom\big), \nonumber 
		}
		where
		\ea{
			\rhom :=  \Pr_{\nuv} \big(I_{\text{fcr}}(A_m^{[\nuv]}) = 0\big), \label{eqn: rhom}
		}
		provided that $G(m)\rhom\leq \frac{1}{2e}$ for such $m$,
		\item if $X_1$ is discrete (i.e., \changedo{$\alpha_1=0$}), then, for sufficiently large $m$, 
		\ea{
			&\big|\tfrac{b(\Yv^m)}{m}-\tfrac{G(m)}{m}b(X_1)\big|  \label{eqn: disc_b} \\&\leq  \tfrac{G(m)}{m}\rhotm \big( \Ct_1 G(m)+\Ct_2\log \rhotm\big), \nonumber
		}
		where
		\ea{
			\rhotm =  \Pr_{\xiv} \big(I_{\text{fcr}}(A_m^{\xiv}) = 0\big), \label{eqn: rhotm}
		}
		provided that $G(m)\rhotm\leq \frac{1}{2e}$ for such $m$.
		Here, $\xiv =[\xi_1, \ldots, \xi_{G(m)}]$ and $\{\xi_i\}$  is a set of i.i.d. Bernoulli RVs with $\Pr(\xi_i=0)= \Pr(X_{\dm, 1}=0)$,
	where  $\psi_{m, \nuv}(A_m)$ is defined as  in \eqref{eq:psi_def}	and $\nuv$ is the selection RV defined in \eqref{eqn: indep_disc_cont}.
		\end{itemize}
\end{lem}

\begin{IEEEproof}
	See Appendix \ref{app: prf_lem_rho}.
\end{IEEEproof}

\begin{Cor} \label{Cor: rhom}
	Provided that the conditions in Lemma \ref{lem: rho} are satisfied, one can now verify that 
	\begin{itemize}
		\item if $\alpha_1\neq 0$, $G(m)-m\sim o(m)$, and $\rhom\sim o(1/m)$, then
		\ea{
			\lim_{m\to\infty} \tfrac{b(\Yv^m)}{m} = b(X_1)+\Psi_{\alpha}(A_m),	
		}
		\changedo{where
		$\Psi_{\alpha}(A_m)$ in \eqref{eqn: psi_alpha_2}
		is well-defined.}
		
		\item if $\alpha_1=0$, $G(m)-m\sim o(m)$, and $\rhotm\sim o(1/m)$, then
		\ea{
			\lim_{m\to\infty} \tfrac{b(\Yv^m)}{m} = b(X_1),
		}
	\end{itemize}
\end{Cor}

}

\begin{IEEEproof}[Proof of Proposition \ref{prop:DRB_MA}]
Since $c_{-l_1}$ and $c_{l_2}$ are non-zero, for $m\geq l_1+l_2$, $A_m$ cannot have a zero column (each column includes at least one of $c_{-l_1}$ and $c_{l_2}$). We further check when $I_{\text{fcr}}(A_m^{[\sv]})$ can be zero (linear dependence among columns of $A_m$ for which $s_i\neq 0$). Let $1\leq i_1<\dots<i_t\leq n$ be such that $s_{i_j}=1$ and a linear combination of  columns $\{i_1,\dots,i_t\}$ of $A_m$ with non-zero coefficients is zero. We claim that $i_{j+1}-i_{j} \leq l_1+l_2$ for all $1\leq j\leq n-1$. Indeed, if $i_{j+1}-i_{j} > l_1+l_2$, then, the $(i_{j+1}-l_1-l_2)$th element in any linear combination of $A_{m}^{[i_1]},\dots,A_m^{[i_t]}$ with non-zero coefficients, is non-zero. Note that the $(i_{j+1}-l_1-l_2)$th element of all $A_m^{[i_k]}$s is zero for all $k\in[t]$ except for $k=j+1$. A similar argument shows that $i_1\leq l_1+l_2$; otherwise, the $(i_1-l_1-l_2)$th element in any linear combination with non-zero coefficients shall be non-zero. Thus, we conclude that $t\geq \lfloor\tfrac{n-1}{l_1+l_2}\rfloor$, or equivalently, $\sum_{i=1}^n s_i\geq \lfloor\tfrac{n-1}{l_1+l_2}\rfloor$.
%
	 Using this fact, \cite[Theorem 1]{arratia1989}, {\changed and the definition of $\rhom$} in \eqref{eqn: rhom}, we have that
	\ea{
	 \rhom &\leq \exp\big(-n(\Dsf\tfrac{n-1}{n(l_1+l_2)}\|\alpha)\big) \nonumber\\
	 &<\exp\big(-n\Dsf(\tfrac{1}{l_1+l_2}\|\alpha)\big), \label{eqn: bnd_rhom}
	}
	provided that $\alpha<\tfrac{n-1}{n(l_1+l_2)}$. {\changed Note that $\Dsf(p\|q)$ is the KL-divergence between ${ \rm Bern} (p)$ and ${\rm Bern}(q)$. } Since we assumed $\alpha\in (0, \tfrac{1}{l_1+l_2})$, this condition is satisfied for large enough $n$.
	
	Similarly, \changedo{if we let $\alpha' = \Pr(W_{\dm, 1}\neq 0)$,} we can bound $\rhotm$ as
	\changedo{\ea{
		\rhotm <\exp\Big(-n\Dsf\big(\tfrac{1}{l_1+l_2}\|\alpha'\big)\Big). \label{eqn: bnd_rhotm}
	}}

	Finally, \eqref{eqn: bnd_rhom} and \eqref{eqn: bnd_rhotm} together with Corollary \ref{Cor: rhom} yield the results in Proposition \ref{prop:DRB_MA}.
\end{IEEEproof}

\changedo{\begin{prop}\label{Prop: Psi}
	\changedo{$\Psi_{\alpha}(c_1, c_2)$ as defined in \eqref{eqn: psi_alpha_2}} exists, is bounded and is equal to
	\ea{
		\Psi_{\alpha}(c_1, c_2) = \frac{1-\alpha}{2}\Ebb_Y[\log E_Y],
	}
	in which $Y$ is an $\alpha$-geometric RV (defined in \eqref{eq:Geometric}), and $E_k$ is defined as
	\ea{
	E_k := \left\{\begin{array}{c c}
		\frac{c_1^{2k+2}-c_2^{2k+2}}{c_1^2-c_2^2}, & c_1^2\neq c_2^2\\
		c_1^{2k}(1+k) & c_1^2=c_2^2
	\end{array} \right. .	
	}
\end{prop}}
\begin{IEEEproof}
	See Appendix \ref{app: prf_prop_psi}.
\end{IEEEproof}

\begin{figure*}
	\centering
	\vspace{-0.65cm}
	\includegraphics[width=\linewidth]{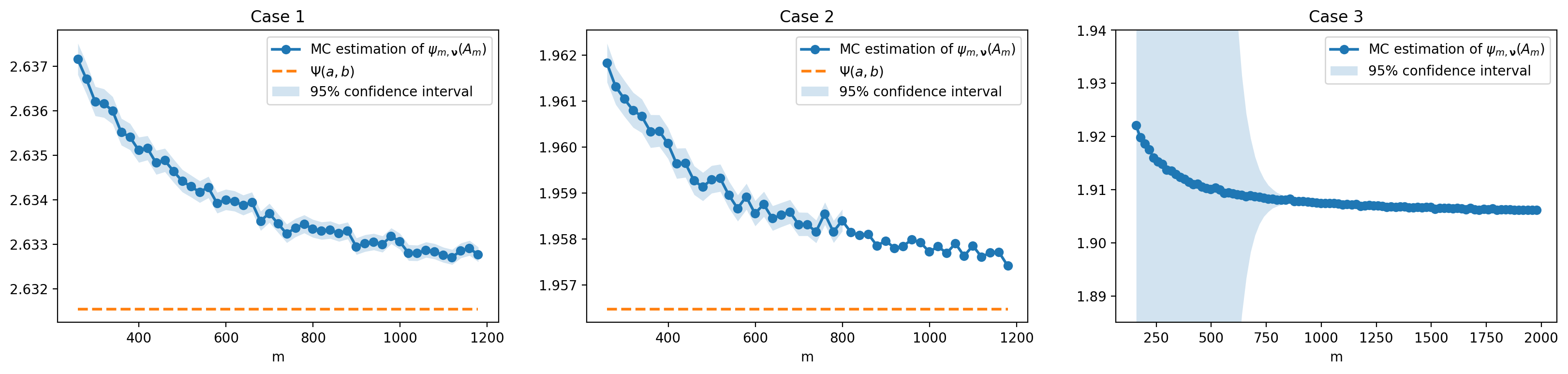}
	\caption{Monte-Carlo estimation of $\psi_{m, \nuv}(A_m)$, the $95\%$ confidence interval of the difference $\tfrac{b(\Yv^m)}{m}-b(X_1)$, and the value of $\Psi$ for the three cases mentioned in Section \ref{sec:DRB}.}
	\label{Fig:DRB}
\end{figure*}

	For the visual illustration of the results, we consider three examples in Figure \ref{Fig:DRB}, namely
	\begin{itemize}
		\item[Case] \hspace{-1.5mm}1) a Bernoulli-Gaussian excitation noise with $\alpha = 0.6$, and Gaussian variance $\sigma^2 = 1$, passed through the MA system $Y_m = 1.5 X_{m-1} + X_m$,
		\item[Case] \hspace{-1.5mm}2) a Bernoulli-Laplace excitation noise with $\alpha = 0.6$, and Laplace parameter $b = 1$, passed through the MA system $Y_m = 1.5 X_{m-1} + X_m$, and
		\item[Case] \hspace{-1.5mm}3) a Bernoulli-Laplace excitation noise with $\alpha = 0.4$, and Laplace parameter $b = 1$, passed through the MA system $Y_m = X_{m-2} + 1.2X_{m-1}+X_{m}$.	
	\end{itemize}
	For these cases, we have plotted the Monte-Carlo evaluation of $\psi_{m, \nuv}(\cdot)$ via Prop. \ref{prop:DRB_MA}, and its convergence to the \changedo{average} information loss \changedo{$\lim_{m\to\infty} \frac{b(\Yv^m)}{m}-\lim_{n\to\infty} \frac{b(\Xv^n)}{n}$} based on Proposition  \ref{Prop: Psi}.
	{\changed The Monte-Carlo simulation of $\psi_{m, \nuv}(\cdot)$ is done by forming a set of randomly generated binary selection variables $\Scal_M$ from the distribution ${\rm Bern}(\alpha)$, and finding $\widehat{\psi}_{m} = \frac{1}{mM}\sum_{\sv\in \Scal_M} \log \det^{+}(A_{m}^{[\sv]})$. Next, we can estimate the variance of $\log \det^{+}(A_{m}^{[\sv]})$ via an unbiased variuance estimator. Since the terms $\log \det^{+}(A_{m}^{[\sv]})$ are bounded, as $M$ grows, $\widehat{\psi}_M$ converges in distribution to a Gaussian random variable. Hence, using error function (erf) we can approximate the interval \changedo{in which $\widehat{\psi}_M$ falls} with $95\%$ probability. 
	Further, using \eqref{eqn: estimate_DRB}, we can find an upper- and lower-bound on $\frac{b(\Yv^m)}{m}-b(X_1)$ by $\psi_{m, \nuv}(A_m)$ and a diminishing term. This, coupled with the previous Monte-Carlo analysis, could give us a $95\%$ confidence interval on  $\frac{b(\Yv^m)}{m}-b(X_1)$.
	In all the three cases, we observe that the $95\%$ confidence interval shrinks as $n$ increases.}

{\changed

}

\section{Conclusion}
\label{sec:Conclusion}
\changedo{In this paper, we defined a new class of probability measures that have singularities over affine subsets.
This is a high-dimensional 
extension of a sparse random variable that includes a mass probability at the origin. 
We studied the compressibility of this new class using conventional notions such as RID and also the new notion of DRB.
Specifically, we found the analytical value of the RID and DRB of a linear functions of i.i.d. vectors with element-wise discrete-continuous RV. 
Furthermore, we provided an upper-bound for Lipschitz functions of this class of RVs. Next, we established a connection between the notions of DRB and mutual information between two affinely singular RVs. 
 Studying this connection for   two general RVs (not necessarily affinely singular) is a potential research direction.
We evaluated the BID, IDR, and DRB of a discrete-domain moving-average process with discrete-continuous excitation noise.
The RID was shown to be  closely related to the notion of $\ep$-compression rates in such cases. 
Overall, the results 
show that the RID plays a fundamental role in evaluating the compressibility of sequences with dependent components in compressed-sensing scenarios. 
Furthermore, our results show that instead of sparsity, it is affine singularity that determine the extent to which we could smoothly compress and robustly reconstruct a sequence. Indeed, although the samples of a moving-average process are less sparse than their excitation noise, they could be equally compressed. Such phenomenon could be further investigated for more general processes, such as auto-regressive moving-average processes. 
Our results on the DRB, its application in evaluating mutual information, and its equality with Shannon's entropy and differential entropy in discrete and continuous cases nominate DRB  as a useful replacement for  compressibility measures such as RDE in comparing the compressibility of sources with the same RID.
 }

%
%

\bibliographystyle{IEEEtran}
\bibliography{v31}
\appendices 

\section{Proof of Lemma \ref{lmm: Ax_has_subsets}} \label{app: proof_Ax_has_subsets}

{\changed In this section, we first prove the following lemma, which expresses that a full-row-rank matrix preserves absolute continuity of a random vector. By means of this lemma, then we prove Lemma \ref{lmm: Ax_has_subsets}.
	
	\begin{lem}\label{lmm: abs_cont_rank_preserve}
		Let $\Xv^n$ be an absolutely continuous RV and 
		let $\Yv^m=A_m \, \Xv^n$. If  $\rank(A_m)=m$, then, $\Yv^m$ is also an absolutely continuous RV.
	\end{lem}
	
	\begin{IEEEproof}
		Since $A_{m}$ has full row rank, the rows of $A_m$ are linearly independent. Therefore, it is possible to extend these $m$ rows to a set of $n$ linearly independent row vector (\cite[Theorem 2.12]{axler1997linear}). We can now form the $n\times n$ matrix $A_{\rm aug}$ by using this linearly independent set as the rows. Obviously, $A_{\rm aug}$ is invertible. Since $\Xv^n$ is absolutely continuous and $A_{\rm aug}$ is invertible, we conclude that $\Yv_{\rm aug}=A_{\rm aug} \Xv^n$ is also absolutely continuous ($A_{\rm aug}^{-1}$ is a linear operator and maps sets with zero-Lebesgue-measure  into sets with zero-Lebesgue-measure \cite[Theorem 2.20]{rudin2006real}). Finally,  note that the probability distribution of $\Yv = A_m\Xv^n$ is a marginal distribution of $\Yv_{\rm aug}$; hence, it is also absolutely continuous.
	\end{IEEEproof}
	
	%
	%
	Now we get back to the proof of Lemma \ref{lmm: Ax_has_subsets}.} We can represent $A_m\Xv^n$ as
\ea{
	A_m\Xv^n &= \sum_{\sv\in \{0, 1\}^n} \mathds{1}_{\nuv=\sv} A_m\Xv^n \nonumber \\
	&= \sum_{\sv\in \{0, 1\}^n} \mathds{1}_{\nuv=\sv} (A_m^{[\sv]} \Xv_{\cm}^{\sv}+A_m^{[\sov]}\Xv_{\dm}^{\sov}) \nonumber  \\
	&=\sum_{\sv \in \{0, 1\}^n}\sum_{\xv_{\dm}^{\sov}\in \Xcal_{\sov}} \mathds{1}_{\nuv=\sv, \Xv_{\dm}^{\sov} =\xv} (A_m^{[\sv]}\Xv_{\cm}^{\sv} +A_m^{[\sov]}\xv_{\dm}^{\sov}).\label{eqn: first_decomp_ax}
}
The definition of $\Tcal_i^{(m)}$ for $i\in \Nbb$ (before \eqref{eqn: W_s_x}) implies that $\bigcup\limits_{i\in \Nbb} \Tcal_i^{(m)} = \big\{(\sv, \xv_{\dm}^{\sov})\,:\,\sv\in \{0, 1\}^n, \xv_{\dm}^{\sov}\in \Xcal_{\dm}^{\sov}\big\}$. As a result, one can rewrite \eqref{eqn: first_decomp_ax} as
\ea{
	A_m\Xv^n=\sum_{i\in \Nbb}\sum_{(\sv, \xv_{\dm}^{\sov})\in \Tcal_i^{(m)}} \mathds{1}_{\nuv=\sv, \Xv_{\dm}^{\sov} =\xv} (A_m^{[\sv]}\Xv_{\cm}^{\sv} +A_m^{[\sov]}\xv_{\dm}^{\sov}). \label{eqn: AXn_first}
}
%
Next, we deploy the singular value decomposition for $A_m^{[\sv]}$ as
\ea{
	A_m^{[\sv]} = \underset{\hspace{-1mm}\leftharpoonup}{U_{\sv}} D_{\sv} \underset{\hspace{-1mm}\rightharpoonup}{U_{\sv}}^{\dagger}, \label{eqn: svd_as}
}
where 	$D_{\sv}$,  	$\underset{\hspace{-1mm}{\rightharpoonup}}{U_{\sv}}$ and
$\underset{\hspace{-1mm}\leftharpoonup}{U_{\sv}}$  indicate
the diagonal $m\times k$ matrix of the singular values, the $k\times k$ matrix of right-singular vectors and 
the $m\times m$ matrix of left-singular vectors of $A_m^{[\sv]}$; here, $k$  represents the number of non-zero elements in $\sv$. Hence, we see that
\ea{
	A_m^{[\sv]}\Xv_{\cm}^{\sv} +A_m^{[\sov]}\xv_{\dm}^{\sov} = \underset{\hspace{-1mm}\leftharpoonup}{U_{\sv}} (\Yv_{\sv}^{\intercal},\underbrace{0, \ldots, 0}_{m-\rank (A^{[\sv]})} )^{\intercal} + A_m^{[\sov]}\xv_{\dm}^{\sov}, \label{eqn: as_xs}
}
where using Lemma \ref{lmm: abs_cont_rank_preserve}, $\Yv_{\sv}$ is an absolutely continuous RV defined as
\ea{
	\Yv_{\sv} = \widetilde{D}_{\sv} \underset{\hspace{-1mm}\rightharpoonup}{U_{\sv}}^{\dagger}\Xv_{\cm}^{\sv},
}
with $\widetilde{D}_{\sv}$ being a $\rank(A_m^{[\sv]})\times k$ matrix formed by the first $\rank(A_m^{[\sv]})$ rows of $D_{\sv}$.
If the columns of $\underset{\hspace{-1mm}\leftharpoonup}{U_{\sv}}$ are shown by $\{\uv_i^{\sv}\}_{i\in [n]}$, \eqref{eqn: as_xs} reveals that $A_m^{[\sv]}\Xv_{\cm}^{\sv} +A_m^{[\sov]}\xv_{\dm}^{\sov}$ is supported on the affine subset $\Acal(\sv, \xv_{\dm}^{\sov})$ of $\Rbb^m$, defined via
\ea{
	\Acal(\sv, \xv_{\dm}^{\sov}) = \bigg\{\sum_{i=1}^{\rank(A^{[\sv]})}a_i\uv_i^{\sv}+\bv^{\perp}(\sv, \xv_{\dm}^{\sov}): \av\in \Rbb^{\rank(A_m^{[\sv]})}\bigg\},\label{eqn: A_s_xds}
}
where 
\ea{
	\bv^{\perp}(\sv, \xv_{\dm}^{\sov}) = \underset{\hspace{-1mm}\leftharpoonup}{U_{\sv}} \underset{\hspace{-1mm}\leftharpoonup}{\widehat{U}_{\sv}}^{\dagger} A_m^{[\sov]} \xv_{\dm}^{\sov},
}
is perpendicular to $\spanM\big(\{\uv_i^{\sv}\}_{i\in [\rank(A_m^{[\sv]})]}\big)$;
$\underset{\hspace{-1mm}\leftharpoonup}{\widehat{U}_{\sv}}$ is the $m\times m$ matrix formed by zeroing off the first $\rank(A_m^{[\sv]})$ columns of $\underset{\hspace{-1mm}\leftharpoonup}{U_{\sv}}$. 
We should highlight that for all $(\sv, \xv_{\dm}^{\sov})\in \Tcal_i^{(m)}$, $A_m^{[\sv]}$ shall span the same space; therefore, without loss of generality, we use $\rank(A_m^{[\sv]}) = r_i$, $\underset{\hspace{-1mm}\leftharpoonup}{U_{\sv}} = \underset{\hspace{-1mm}\leftharpoonup}{U_{i}}$, and $\bv^{\perp}(\sv, \xv_{\dm}^{\sov}) = \bv_i$; i.e., only the index $i$ for $\Tcal_i^{(m)}$ is important, and not the specific vector of $\sv$ (as long as $(\sv, \xv_{\dm}^{\sov})\in \Tcal_i^{(m)}$). We can now rewrite \eqref{eqn: as_xs} as
%
\ea{
	A_m^{[\sv]}\Xv_{\cm}^{\sv} +A_m^{[\sov]}\xv_{\dm}^{\sov} = \underset{\hspace{-1mm}\leftharpoonup}{U_{i}} \big(\Wv^{\intercal}_{\sv, \xv_{\dm}^{\sov}}, \underbrace{0, \ldots, 0}_{m-r_i}\big)^{\intercal}+\bv_i, \label{eqn: AXn_elements}
}
in which $\Wv_{\sv, \xv_{\dm}^{\sov}}$ is defined in \eqref{eqn: W_s_x}. 
%
%
\changedo{For the sake of simplicity, we define the RV $\Upsilon_m^{(i)}$ for $i\in \Nbb$ as 
{\changed
	\ea{
		\Upsilon_m^{(i)} = \left\{\begin{array}{c c}
			(\nuv, \Xv_{\dm}^{\sov}) & \Upsilon_m=i,\\
			T_i & \text{otherwise},
		\end{array}\right.
	}
}
where $T_i$ is a RV independent of $\Upsilon_m$ and supported on $\Tcal_i^{(m)}$, 
with probability mass function
\ea{
	\forall(\sv, \xv_{\dm}^{\sov}) \in \Tcal_i^{(m)}: ~~~ \Pr\big(T_i=(\sv, \xv_{\dm}^{\sov})\big) = \frac{{\Pr}_{\nuv,\Xv_{\dm}^{\nuov}}(\sv,\xv_{\dm}^{\sov})}{{\Pr}(\Upsilon_m=i)}. 
}
The definition of $T_i$ is such that $\Pr(\Upsilon_m^{(i)}  \,|\, \Upsilon_m=j)$ for all $j$ is the same (whether $j=i$ or $j\neq i$); consequently, $\Upsilon_m^{(i)}$ and $\Upsilon_m$ are independent. Furthermore, the probability mass function of $\Upsilon_m^{(i)}$ coincides with \eqref{eqn: v_prime_t}.
%

With these notations in hand, we combine \eqref{eqn: AXn_first} and \eqref{eqn: AXn_elements} to rewrite $A_m\Xv^n$ as
\ea{
	A_m\Xv^n = \sum_{i\in \Nbb} \mathds{1}_{\Upsilon_m=i} \Big[\underset{\hspace{-1mm}\leftharpoonup}{U_{i}} \big(\Ytv_i^{\intercal}, \underbrace{0, \ldots, 0}_{m-r_i}\big)^{\intercal}+\bv_i\Big],
}
where $\Ytv_i$ is defined in \eqref{eqn: y_tilde_i}. This representation coupled with \eqref{eqn: ytv_i} proves 
\ea{\Yv^m = \fv_{\Upsilon_m}(\Ytv_{\Upsilon_m}).\label{eqn: fv_yv}}

The definitions of $\Xv^n$ and $\Wv_{\sv, \xv_{\dm}^{\sov}}$ in \eqref{eqn: indep_disc_cont} and \eqref{eqn: W_s_x}, respectively, show that $\Wv_{\sv, \xv_{\dm}^{\sov}}$ is independent of $(\nuv, \Xv_{\dm}^{\sov})$. 
Since $\Ytv_i$ is a function of $\Wv_{\sv, \xv_{\dm}^{\sov}}$ and $\Upsilon_m^{(i)}$, it is independent of $\Upsilon_m$. 
This point implies that \eqref{eqn: fv_yv} has a similar representation as \eqref{eqn: singular_subsets}.}

\section{Proof of Theorem \ref{thm: singular_subset}}\label{prf: singular_subset}

{\changed We calculate RID and DRB of the affinely singular random vectors as follows.

\subsection{RID of affinely singular RVs}}
By comparing Definition \ref{def: aff_sig_measure} and \ref{def: reg}, and making use of Lemma \ref{lem: aff_sing}, one can see that the probability measure of the affinely singular RV $\Zv^m$ is a mixture of $e_i$-regular measures with set of probabilities $\{\Pr(V_m = i)\}_i$. Since $\Hsf(V_m)<\infty$ and using Lemma \ref{lem: reg}, one can show that
\ea{
d(\Zv^m) = \Ebb_{V_m}[e_{V_m}].
}

{\changed \subsection{DRB of affinely singular RVs}

The proof of this part of the theorem is an extension of \cite[Theorem 1]{gyorgy1999rate} to the affinely singular RVs with a bounded moment and differential entropy. 
{\changed To prove this part, we first show in the following Lemma that differential entropy is continuous in terms of additive Gaussian random vectors with infinitesimal Frobenious norm of covariance matrix.

\begin{lem}\label{lem: frob_diff_ent}
	Let $\Xv$ be an absolutely continuous RV with finite moment $\Ebb[\|\Xv\|_{\alpha}^{\beta}]$ for some $\alpha, \beta\in \Rbb^{+}$, and differential entropy $\hsf(\Xv)>-\infty$. Also let $\{\Nv_i\}_i\in\mathbb{N}$ be a sequence of normal RVs with covariance matrices $\{\Sigma_i\}_i$. If
	\ea{
	\lim_{i\to\infty} \|\Sigma_i\|_F = 0,
	}
	we have that
	\ea{
	\lim_{i\to\infty}\hsf(\Xv+\Nv_i) = \hsf(\Xv).	 \label{eqn: diff_ent_continuity}
	}
\end{lem}
\begin{IEEEproof}
	Let $q=\min\{1, \beta\}$. One can bound $\Ebb\big[\|\Xv\|_{\alpha}^{q}\big]$ as
\ea{
\Ebb\big[\|\Xv\|^q_{\alpha}\big] &= \Ebb\big[\|\Xv\|^q_{\alpha}\mid \|\Xv\|_{\alpha}\leq 1\big]\Pr(\|\Xv\|_{\alpha}\leq 1) \nonumber\\&\qquad+ \Ebb\big[\|\Xv\|^q_{\alpha}\mid \|\Xv\|_{\alpha}> 1\big] \Pr(\|\Xv\|_{\alpha}> 1)\\
&\leq 1+ \Ebb\big[\|\Xv\|^{\beta}_{\alpha}\mid \|\Xv\|_{\alpha}> 1\big] \Pr(\|\Xv\|_{\alpha}> 1)\\
&\leq 1+ \Ebb\big[\|\Xv\|^{\beta}_{\alpha}\big]<\infty,
}
where the last inequality holds because $\Ebb[\|\Xv\|_{\alpha}^{\beta}]<\infty$.
By setting $\dbar = \max\{1, \alpha\}$, the norm inequality in \cite[Lemma 2.1]{kirmaci2008some} shows that
\ea{
\|\vv\|_{\dbar}\leq \|\vv\|_{\alpha},
}
for every vector $\vv$, which concludes that
\ea{
\Ebb[\|\Xv\|_{\dbar}^{q}]\leq \Ebb[\|\Xv\|_{\alpha}^{q}] <\infty.\label{eqn: boundedness_moment}
}


Further, for two-dimensional vectors $\vv_1=\big(\|\Xv+\Nv(D)\|_{\dbar}, \|\Nv(D)\|_{\dbar}\big)^{\intercal}$, and $\vv_2=\big(\|\Xv\|_{\dbar}, \|\Nv(D)\|_{\dbar}\big)^{\intercal}$, we have
\ea{
\|\Xv\|_{\dbar}&\overset{(a)}{\leq}\|\Xv+\Nv(D)\|_{\dbar}+\|\Nv(D)\|_{\dbar}\nonumber\\&\leq \big(\|\Xv+\Nv(D)\|_{\dbar}^q+\|\Nv(D)\|_{\dbar}^q\big)^{\tfrac{1}{q}},\label{ineq: XND1}
}
and
\ea{
	\|\Xv+\Nv(D)\|_{\dbar}&\overset{(b)}{\leq}\|\Xv\|_{\dbar}+\|\Nv(D)\|_{\dbar}\nonumber\\&\leq \big(\|\Xv\|_{\dbar}^q+\|\Nv(D)\|_{\dbar}^q\big)^{\tfrac{1}{q}}, \label{ineq: XND2}
}
where $(a)$ and $(b)$ are followed by the triangle inequality for $\dbar\geq 1${\changed, and the fact that $\|\cdot\|_1\leq \|\cdot\|_q$ for $q=\min\{1, \beta\}\leq 1$ (again, see  \cite[Lemma 2.1]{kirmaci2008some})}. Combining \eqref{ineq: XND1} and \eqref{ineq: XND2}, we arrive at
\ea{
-\|\Nv(D)\|_{\dbar}^q \leq \|\Xv+\Nv(D)\|_{\dbar}^q - \|\Xv\|_{\dbar}^q \leq \|\Nv(D)\|_{\dbar}^q ,
}
and
\ea{
\big|\Ebb[\|\Xv+\Nv(D)\|_{\dbar}^q]& -\Ebb[\|\Xv\|_{\dbar}^q]\big|\leq \Ebb[\|\Nv(D)\|_{\dbar}^q] \nonumber\\
\leq \Ebb[\|\Nv(D)\|_{1}^q] & \leq \Ebb[\|\Nv(D)\|_{q}^q] = \sum_{i=1}^{n} \Ebb[|N^i(D)|^q].  \label{eqn: difference_normal}
}
As $\Nv(D)$ is a normal random vector, each $N^i(D)$ is also 
  a normal RV with standard deviation $\sigma_i$ that satisfy
\ea{
\sigma_i \leq \Big(\sum_{i}\sigma_i^2\Big)^{1/2}  \leq \|\Sigma(D)\|_F.
}
Using the bound in \cite[Theorem 3]{Shanbhag77} for the moments of a normal RV, we have
\ea{
		\Ebb[|N^i(D)|^q]\leq \sigma_i^q \tfrac{2^{q/2}\Gamma(\f{q+1}{2})}{\sqrt{\pi}} \leq \|\Sigma(D)\|_F^q \tfrac{2^{q/2}\Gamma(\f{q+1}{2})}{\sqrt{\pi}},\label{eqn: normal_gamma}
} 
where $\Gamma(.)$ is the Gamma function.

As a result of \eqref{eqn: difference_normal} and \eqref{eqn: normal_gamma}, and the assumption that $\lim_{D\to0}\|\Sigma(D)\|_F= 0$, we have that
\ea{
	\lim_{D\to 0}\Ebb[\|\Xv+\Nv(D)\|_{\dbar}^q] = \Ebb[\|\Xv\|_{\dbar}^q].\label{eqn: continuity_moment}
}
Finally, \cite[Theorem 1]{Zamir1994} results in the desired claim as  $\hsf(\Xv)>-\infty$ (other requirements of this theorem are fullfilled due to \eqref{eqn: boundedness_moment} and \eqref{eqn: continuity_moment}).
\end{IEEEproof}
}

Now, we prove the second part of Theorem \ref{thm: singular_subset} in two parts: (i) first we find a lower bound for the $\liminf$ and then (ii) we find an upper bound for the $\limsup$. The proof of the second part of Theorem \ref{thm: singular_subset} follows from the equality of the two bounds.
\begin{noindlist}
	\item {\bf Step (i)}:
	Using the definition of rate-distortion function, there exists a sequence $\{\Zhv^m_{k, D}\}_{k\in\Nbb}$ of $m$-dimensional RVs that
	\ea{
		\lim_{k\to\infty}\Isf(\Zv^m; \Zhv^m_{k, D})=R_2(\Zv^m, D), \label{eqn: rdf_seq}
	} 
	and 
	\ea{
		\Ebb[\|\Zv^m-\Zhv^m_{k, D}\|_2^2]\leq D. \label{eqn: err_D}
	}
	Next, because $\Zv^{(i)}$'s are mutually singular,  $V_m$ is a function of $\Zv^m$ with probability $1$. As a result, one has
	\ea{
		\Isf(\Zv^m; \Zhv^m_{k, D})&= \Isf(\Zv^m, V_m; \Zhv^m_{k, D})  \\
		&= \Isf(V_m;\Zhv^m_{k, D})+\Isf(\Zv^m; \Zhv^m_{k, D}|V_m)   \\
		&=\Isf(V_m; \Zhv^m_{k, D})+ \sum_{i=1}^{\infty} p_i \Isf(\Zv^{(i)}; \Zhv^{(i)}_{k, D}), \label{eqn: decomp_mutu_rate_dist}
	}
	where $\Zhv^{(i)}_{k, D}$ is a RV with the probability measure equal to the conditional probability measure of $\Zhv^m_{k, D}$ given $V_m=i$. 
	
	\changedo{If $e_i=0$ for a selection $V_m=i$, then we have
	\ea{
	0\leq \Isf(\Zv^{(i)} ;\, \Zhv^{(i)}_{k, D})\leq  \Hsf(\Zv^{(i)})=0.
	}
	}
	
	\changedo{Next, by assuming $e_i\neq 0$} we bound each expression in \eqref{eqn: decomp_mutu_rate_dist}. First, we claim that
	\ea{
		\liminf_{D\to0}\liminf_{k\to\infty} \Isf(V_m; \Zhv^m_{k, D}) \geq \Hsf(V_m). \label{eqn: mutual_est_func}
	}
	The reason is that for every $\ep\in \Rbb^{+}$, there exists $\tilde{k}(D, \ep)\in \Nbb$ such that
	\ea{ \label{eqn: mutual_est_func1}
	\liminf_{k\to\infty} \Isf(V_m; \Zhv^m_{k, D})\geq \Isf(V_m; \Zhv_{\tilde{k}(D, \ep), D}^m)-\ep.	
	}
	As for any $D>0$ we know that $\Ebb[\|\Zv^m-\Zhv_{\tilde{k}(D, \ep), D}^m\|^2]\leq D$, we conclude that $\big(V_m; \Zhv_{\tilde{k}(D_t, \ep), D_t}^m\big)$ converges to $\big(V_m; \Zv^m\big)$ in mean (also in distribution) for any sequence $\{D_t\}_{t\in\mathbb{N}}$ with $D_t\stackrel{t\to \infty}{\longrightarrow}0$. Recalling the  lower semi-continuity property of the mutual information (see \cite[Eqn. 3.13]{polyansky2017}), we can write
	\begin{align}\label{eq:semi-continuity}
	\liminf_{t\to\infty} \Isf(V_m; \Zhv_{\tilde{k}(D_t, \ep), D_t}^m) \geq \Isf(V_m; \Zv^m).
	\end{align}
	Now, by combining \eqref{eqn: mutual_est_func1} and \eqref{eq:semi-continuity}, we conclude that
%
%
	\ea{
	\liminf_{D\to0}\liminf_{k\to\infty} \Isf(V_m; \Zhv^m_{k, D})&\geq \Isf(V_m; \Zv^m)-\ep\\
	&= \Hsf(V_m) -\ep. \label{eq:eps_bound}
	}
	As \eqref{eq:eps_bound} holds for all $\ep>0$, we can use $\ep\to 0$ and \eqref{eqn: mutual_est_func} is immediate.
	
	Second, using the fact that an invertible function preserves the mutual information, we have
	\ea{
		\Isf(\Zv^{(i)}; \Zhv^{(i)}_{k, D})= \Isf\big(\Cv_i; U^{\dagger}(\Zhv^{(i)}_{k, D}-\bv_i)\big).\label{eqn: mut_preserve}
	}
	By defining  the first $e_i$ elements of $U^{\dagger}(\Zhv^{(i)}_{k, D}-\bv_i)$ as $\Ztv^{(i)}_{k, D}$ and, we can rewrite \eqref{eqn: mut_preserve} as
	\ea{
		\Isf(\Zv^{(i)}; \Zhv^{(i)}_{k, D})&\overset{(a)}{\geq} \Isf(\Cv_i; \Ztv^{(i)}_{k, D}) \nonumber \\
		&= \hsf(\Cv_i)- \hsf(\Cv_i|\Ztv^{(i)}_{k, D}) \nonumber \\
		&=\hsf(\Cv_i)- \hsf(\Cv_i-\Ztv^{(i)}_{k, D}|\Ztv^{(i)}_{k, D})\nonumber  \\
		&\overset{(b)}{\geq} \hsf(\Cv_i)- \f{e_i}{2}\log \f {2\pi e \Ebb[\|\Cv_i-\Ztv^{(i)}_{k, D}\|^2_2]}{e_i}, \label{eqn: decomp_single_lateral}
	}
	where $(a)$ holds due to the fact that dropping a RV from one side of the mutual information, decrease its value. The validity of $(b)$ is also because of the decreasing property of differential entropy with conditioning; besides,  the Gaussian RV with i.i.d. components maximizes the differential entropy among all RVs with the second-order moment constraint.

	Moreover, we know that
	\ea{
		\|\Cv_i-\Ztv^{(i)}_{k, D}\|_2^2&\leq \| (\Cv_i^{\intercal}, \underbrace{0, \ldots, 0}_{m-e_i})^{\intercal}- U^{\dagger}(\Zhv^{(i)}_{k, D}-\bv_i)\|_2^2 \nonumber \\
		&\overset{(a)}{=} \|\Zv^{(i)}-\Zhv^{(i)}_{k, D}\|_2^2, \label{eqn: less_dist}
	}
	where $(a)$ is correct because multiplication by a unitary matrix preserves the Euclidean norm of a vector. Using this inequality, we have
	\ea{
		 \sum_{i=1}^{\infty} p_i \Ebb[\|\Cv_i-\Ztv^{(i)}_{k, D}\|_2^2]\leq \sum_{i=1}^{\infty} p_i \Ebb [ \|\Zv^{(i)}-\Zhv^{(i)}_{k, D}\|^2_2]\leq D, \label{eqn: less_D}
	}
	where the last inequality follows from \eqref{eqn: err_D}, and the definition of $\Zv^{(i)}$ and $\Zhv^{(i)}_{k, D}$.
	We recall the log-sum inequality as
\begin{align}\label{eq:log-sum}
\sum_i a_i \log\tfrac{b_i}{a_i} \leq a\log\tfrac{b}{a},
\end{align}	
where $a_i,\,b_i$ are arbitrary non-negative reals,  $a=\sum_i a_i$ and $b=\sum_i b_i$. Thus, if we set $a_i= p_i e_i$ and $b_i=2\pi e \, p_i \Ebb[\|\Cv_i-\Ztv^{(i)}_{k, D}\|_2^2$, the log-sum inequality \eqref{eq:log-sum} implies that
%
	\changedo{\ea{
		&\sum_{i=1\, e_i\neq 0}^{\infty} \f {p_ie_i}{2} \log \f{2\pi e \Ebb[\|\Cv_i-\Ztv^{(i)}_{k, D}\|_2^2]}{e_i} \nonumber \\
		&\leq \f{\sum_{i=1\, e_i\neq 0}^{\infty}p_ie_i}{2} \log \f{2\pi e \sum_{i=1}^{\infty} p_i\Ebb [\|\Cv_i-\Ztv^{(i)}_{k, D}\|_2^2]}{\sum_{i=1\, e_i\neq 0}^{\infty}p_ie_i} \nonumber  \\
		&\leq \f{d(\Zv^m)}{2} \log \f{2\pi e \sum_{i=1}^{\infty} p_i\Ebb [\|\Cv_i-\Ztv^{(i)}_{k, D}\|_2^2]}{d(\Zv^m)} \nonumber  \\
		&\overset{(a)}{\leq} \f {d(\Zv^m)}{2} \log \f {2\pi e D}{d(\Zv^m)}, \label{eqn: log_less_jensen}
	}}
	where $d(\Zv^m) = \sum_i p_i e_i$ and $(a)$ is due to \eqref{eqn: less_D}. 
	As a result of \eqref{eqn: rdf_seq}, \eqref{eqn: decomp_mutu_rate_dist}, \eqref{eqn: decomp_single_lateral}, and \eqref{eqn: log_less_jensen} one has
	\ea{
	R_2&(\Zv^m, D)\nonumber\\\geq& \liminf_{k\to\infty} \Isf(V_m; \Zhv_{k, D}^m) +\sum_{i=1}^{\infty} p_i \hsf(\Cv_i)\nonumber\\& -	\f {d(\Zv^m)}{2}\log \f {2\pi e D}{d(\Zv^m)}. \label{eqn: rdf_lb}
	}
	Finally, by combining \eqref{eqn: mutual_est_func} and \eqref{eqn: rdf_lb} we conclude that
	\ea{
		\liminf_{D\to 0} \bigg(&R_2(\Yv^m, D)+\f {d(\Zv^m)}{2}\log \f {2\pi e D}{d(\Zv^m)}\bigg) \nonumber\\&\geq \Hsf(V_m)+\sum_{i=1}^{\infty} p_i \hsf(\Cv_i).
		\label{eq: liminf}
	}
	
	\item {\bf Step (ii)}: To upper-bound  the QRDF of $\Zv^m$, \changedo{if $e_i\neq 0$}, we perturb the RV with a Gaussian noise of (vector) variance  $D$.  Then, we upper-bound the mutual information between the original and the perturbed versions. 
	More specifically, assume that
		\ea{
		\Zhv^{(i)}= \Zv^{(i)} + U_i (\Nv_i^{\intercal}, \underbrace{0, \ldots, 0}_{m-e_i})^{\intercal},
	}	
	where $\Nv_i$ is an $e_i$-dimensional Gaussian RV with zero-mean i.i.d. elements and element-wise variance $\f{D}{d(\Zv^m)}$. Besides, $\Nv_i$ and $\Nv_j$ are independent of each other for $i\neq j$.
	%
	%
	We further define $\Zhv^m$ as 
	\ea{
		\Zhv^m=\sum_{i=1}^{\infty} \mathds{1}_{V_m=i} \Zhv^{(i)}.
	}
	Now, we have 
	\ea{
		\Ebb[\| \Zv^m-\Zhv^m\|_2^2]&= \sum_{i=1}^{\infty} p_i \Ebb [\| \Zv^{(i)}-\Zhv^{(i)}\|_2^2]\\
		&=\sum_{i=1,\, e_i\neq 0}^{\infty} p_i \Ebb[\|\Nv_i\|_2^2]{\changed=\sum_{i=1}^{\infty} p_i\frac{e_i D}{d(\Zv^m)}} =D,\label{eqn: achiev_gauss_d}
	}
	{\changed
	\changedo{where the last identity holds because \ea{d(\Zv^m)= \sum_{i=1, e_i\neq 0}^{\infty} p_i e_i=\sum_{i=1}^{\infty} p_i e_i.} }
	} 
	
	Hence, we shall have that
	\ea{
		R(\Zv^m, D)\leq \Isf(\Zv^m; \Zhv^m). \label{eqn: achieve_less_mut}
	}
	Using the same steps as in \eqref{eqn: decomp_mutu_rate_dist}, we have that
	\ea{
		\Isf(\Zv^m; \Zhv^m)= \Isf(V_m; \Zhv^m)+\sum_{i=1}^{\infty} p_i \Isf(\Zv^{(i)};\Zhv^{(i)}).\label{eqn: decomp_mutu_rate_dist_2}
	}
	For the first term, using \eqref{eqn: achiev_gauss_d} and \cite[Lemma 3]{gyorgy1999rate}, we can write  that
	\ea{
		\lim_{D\to 0} \Isf(V_m; \Zhv^m)= \Hsf(V_m). \label{eqn: approx_est_func}
	}
	For the second term in \eqref{eqn: decomp_mutu_rate_dist_2}, using the fact that an invertible function preserves the value of mutual information, we have that
	\ea{
		\Isf(\Zv^{(i)}; \Zhv^{(i)})&= \Isf(\Cv_i; \Cv_i+\Nv_i) \nonumber \\
		&= \hsf(\Cv_i+\Nv_i)- \hsf(\Cv_i+\Nv_i|\Cv_i) \nonumber \\
		&=\hsf(\Cv_i+\Nv_i)-\hsf(\Nv_i|\Cv_i) \nonumber  \\
		&\overset{(a)}{=}\hsf(\Cv_i+\Nv_i)-\hsf(\Nv_i),\nonumber\\
		& {\changed = \hsf(\Cv_i+\Nv_i)-\frac{e_i}{2}\log \frac{2\pi e D}{d(\Zv^m)}}\label{eqn: decom_mut_single_lateral}
	}
	where $(a)$ is true because $\Cv_i$ is independent of $\Nv_i$. 
	%
	%
	Now, using the assumption that $\hsf(\Cv_i)>-\infty$ and the fact that $\Ebb[\|\Cv_i\|_{\alpha}^{\beta}]<\infty$ (otherwise, $\Ebb[\|\Zv^m\|_{\alpha}^{\beta}]=\infty$), we conclude from Lemma \ref{lem: frob_diff_ent} that
	\ea{
		\lim_{D\to 0} \hsf(\Cv_i+\Nv_i) = \hsf(\Cv_i). \label{eqn: continuity_gauss_entropy}
	}
	Finally, by combining \eqref{eqn: achiev_gauss_d}-\eqref{eqn: continuity_gauss_entropy}, we prove that
	\ea{
		\limsup_{D\to 0} \bigg(&R_2(\Zv^m, D)+\f {d(\Zv^m)}{2}\log \f {2\pi e D}{d(\Zv^m)}\bigg) \nonumber\\&\leq \Hsf(V_m)+\sum_{i=1}^{\infty} p_i \hsf(\Cv_i).
				\label{eq: limsup}
	}		

\end{noindlist}
Since the lower and upper-bounds in \eqref{eq: liminf} and \eqref{eq: limsup}, respectively, coincide, the proof is complete.
}

\section{Proof of Theorem \ref{thm: disc_cont_fat}}\label{app: disc_cont_fat}
{\changed We obtain the RID and DRB of a linear transformation of orthogonally singular RVs in the following sections, respectively.
\subsection{RID of linear transformations of orthogonally singular RVs}
}
Using Lemma \ref{lmm: Ax_has_subsets} (which shall be proved later), we know that $\Yv^m$ has an affinely singular measure as in \eqref{eqn: singular_subsets} with $V_m=\Upsilon_m$, and $\Zv^{(i)}=\Ytv^{(i)}$ for $i\in \Nbb$. In particular, for each 
$$\nuv_0 \in\bigcup_{(\nuv, \xv_{\dm})\in \Tcal_i^{(m)}}\nuv,$$
 the support of $\Yv^m$ contains an affine subspace with dimension $e_0= 	\rank (A_m^{[\nuv_0]})$; the set $\Tcal_i^{(m)}$ was defined in Section \ref{sec:DRB}.

Let  $\Upsilon_m$ be the RV defined in \eqref{eqn: upsilon_m}. We first show that $\Hsf(\Upsilon_m)<\infty$. Using the definition of $\Upsilon_m^{(i)}$ defined in \eqref{eqn: v_prime_t} we have that
\ean{
	-\Pr&(\Upsilon_m=i)\log \Pr(\Upsilon_m=i)\nonumber\\&\leq -\Pr(\Upsilon_m=i)\log \Pr(\Upsilon_m=i)+ \Pr(\Upsilon_m=i)\Hsf(\Upsilon_m^{(i)})\\
	&=-\sum_{(\sv,\xv_{\dm}^{\sov})\in \Tcal_i^{(m)}} {\Pr}_{\nuv,\Xv_{\dm}^{\nuov}}(\sv,\xv_{\dm}^{\sov})\log{\Pr}_{\nuv,\Xv_{\dm}^{\nuov}}(\sv,\xv_{\dm}^{\sov}).
}
As a result, we can write that
\ean{
	\Hsf(V)&\leq \Hsf(\nuv)+\sum_{\sv \in \{0,1\}^d } \Hsf(\Xv_{\dm}^{\overline{\sv}}|\nuv=\sv)\Pr(\nuv=\sv)\\
	&\leq \Hsf(\nuv)+\sum_{\sv \in \{0,1\}^d} \Hsf(\Xv_{\dm}|\nuv=\sv)\Pr(\nuv=\sv)\\
	&=\Hsf(\nuv, \Xv_{\dm}).
}
Since $\Hsf(\nuv, \Xv_{\dm}) < \infty$ by assumption, we conclude that $\Hsf(\Upsilon_m)<\infty$. We now recall Theorem \ref{thm: singular_subset} to conclude the claim that
\ea{
	d(\Yv^m) 	= 	\Ebb_{\nu}[\rank( A_m^{[\nuv]})].
}
{\changed 
\subsection{DRB of linear transformations of orthogonally singular RVs}
{\changed Here, we first prove the following Lemma that states the conditions by means of which $\hsf(A_m \Xv^n)$ is lower bounded.
\begin{lem}\label{lem: h_minus_infinity}
	Let $\Xv^n$ for $n\in \Nbb$ be an RV with $\Ebb[\|\Xv^n\|_{\alpha}^{\beta}]<\infty$. If $A_m$ is a full-rank $k\times n$ matrix, and $\Xv^n$ is an absolutely continuous RV with $\hsf(\Xv^n)>-\infty$, then $\hsf(A_m\Xv^n)>-\infty$.
\end{lem}
\begin{IEEEproof}
	We use the singular value decomposition $A_m$ as
 \ea{
	A_m=U\, D\, V_k^{\dagger},
}
where $U$ is a unitary $k\times k$ matrix, $D$ is $k\times k$ diagonal matrix and $V_k$ is a $n\times k$ matrix (incomplete unitary matrix). Indeed, 
$V_k$ is formed by the first columns of thea unitary matrix $V$ 
\ea{
	V = [V_k, V_{k+1, n}].
}
Using the properties of the differential entropy we know that
\ea{
	\hsf(A_m\Xv^n)	= \hsf(V_k^{\dagger}\Xv^n)+\sum_{i=1}^{k}\log \sigma_i,
}
where $\sigma_i$s are the singular values, and due to the full-rank property of $A_m$, they are strictly positive.
Hence, $\hsf(A_m\Xv^n)	>-\infty$ is equivalent to
\ea{
	\hsf(V_k^{\dagger} \Xv^n) >-\infty.\label{eqn: orthonormal_minus_infinity}
} 
By employing the rotation-invariance property of the differential entropy, we know that
\ea{
	\hsf(\Xv^n) = \hsf(V^{\dagger}\Xv^n)  &= \hsf(V_{k+1,n}^{\dagger}\Xv^n, V_k^{\dagger}\Xv^n)  \\
	&\leq \hsf(V_{k+1,n}^{\dagger}\Xv^n)+ \hsf( V_k^{\dagger}\Xv^n). \label{eqn: bound_unitary}
}
Moreover, we have that
\ea{
	\Ebb\big[\|V_{k+1,n}^{\dagger}\Xv^n\|^{\beta}_{\alpha}\big]&\leq \|V_{k+1,n}\|^{\beta}_{\alpha}\Ebb\big[\|\Xv^n\|_{\alpha}^{\beta}\big]\\&\overset{(a)}{\leq} n^ {\f{\beta(1+\alpha)}{\alpha}} \Ebb\big[\|\Xv^n\|_{\alpha}^{\beta}\big],
}
where $\|V_{k+1,n}\|_{\alpha}$ is the operator norm of matrix $V_{k+1,n}$ induced by vector $\alpha$-norm, and $(a)$ is due to \eqref{eqn: norm_alpha_max}. Using \cite[Corollary 1]{Zamir1994}, we conclude that
\ea{\label{eq:Upper-Bound-V}
	\hsf(V_{k+1,n}^{\dagger}\Xv^n)\leq \log M,
}
where
\ea{
	M = V_{\alpha, n-k}\cdot \Big(\tfrac{\alpha\cdot e\cdot n^{\f{\beta(1+\alpha)}{\alpha}}\Ebb\big[\|\Xv^n\|_{\alpha}^{\beta}\big]}{n-k}\Big)^{\tfrac{n-k}{\beta}}\cdot \Gamma\Big(1+\tfrac{n-k}{\beta}\Big),
}
$V_{\alpha, n-k}$ is the volume of the $\alpha$-norm unit-ball in  $(n-k)$ dimensions, 
and $\Gamma(\cdot)$ is the Gamma function.

Since $\hsf(\Xv^n)>-\infty$, \eqref{eq:Upper-Bound-V} induces a lower-bound on $\hsf(V_k^{\dagger} \Xv^n)$ via \eqref{eqn: bound_unitary}, which completes the proof.


\end{IEEEproof}
}

	Now, we divide the proof of the second part of Theorem \ref{thm: disc_cont_fat} in three steps: (i) we first show that $\hsf(\Ytv_i)>-\infty$, (ii) then, we prove that $\Ebb[\|\Yv^m\|_{\alpha}^{\beta}]<\infty$, and (iii) finally, we utilize Theorem \ref{thm: singular_subset} to complete the proof.
\begin{noindlist}
	\item {\bf Step (i)}:
	Using \eqref{eqn: W_s_x} and \eqref{eqn: y_tilde_i} one can see that
	\ea{ \label{eq:xv_d_inv}
		\hsf\big(\Ytv_i|\Upsilon_m^{(i)}=(\sv, \xv_{\dm}^{\sv})\big)	= \hsf(B_{\sv}\Xv_{\cm}^{\sv}).
	}
	Here, $B_{\sv} = \tilde{D}_{\sv}\underset{\hspace{-1mm}{\leftharpoonup}}{U_{\sv}}^{\dagger}$, and $\underset{\hspace{-1mm}{\leftharpoonup}}{U_{\sv}}^{\dagger} $ and $\tilde{D}_{\sv}$ were previously defined in Section \ref{sec:DRB}. 
	It is not difficult to verify that $B_{\sv}$ is a full row rank matrix. Besides, the RHS in \eqref{eq:xv_d_inv} is invariant to $\xv_{\dm}^{\sv}$. 
	As a result, we have
	\ea{
		\hsf(&\Ytv_i)\geq \hsf(\Ytv_i|\Upsilon_m^{(i)})\\ &= \sum_{(\sv, \xv_{\dm}^{\sv})\in \Tcal_i^{(m)}} \Pr\big(\Upsilon_m^{(i)}=(\sv, \xv_{\dm}^{\sv})\big)\hsf\big(\Ytv_i|\Upsilon_m^{(i)}=(\sv, \xv_{\dm}^{\sv})\big)	\\
		&= \sum_{\sv\in \{0, 1\}^n}\Pr(\boldsymbol{\nu}=\sv) \hsf(B_{\sv} \Xv_{\cm}^{\sv}).\label{eqn: ytilde_lower}
	}
	Next, we form $B_{\sv, {\rm aug}}$ with $n$ columns by including the columns of $B_{\sv}$ in places where $\sv$ is  equal to $1$; the rest of the columns are zero (corresponding to zero locations in $\sv$). In this way, we are able to write
	\ea{
		B_{\sv}\Xv_{\cm}^{\sv} = B_{\sv, {\rm aug}} \Xv_{\cm}.
	}
Since $B_{\sv}$ is of full row rank, $B_{\sv, {\rm aug}}$ has also full row rank (some zeros with fixed pattern are inserted within the rows). 
	Hence, by using Lemma \ref{lem: h_minus_infinity}, $\hsf(\Xv_{\cm})>-\infty$, and $\Ebb[\|\Xv_{\cm}\|_{\alpha}^{\beta}]<\infty$ we have that
	\ea{
		\hsf(B_{\sv}\Xv_{\cm}^{\sv}) = \hsf(B_{\sv, {\rm aug}}\Xv_{\cm})>-\infty.
	}
	In addition, $\sv$ is a discrete RV with at most $2^{n}$ possible values, hence
	\ea{
		\hsf(\Ytv_i)\geq  \min_{\sv} \hsf(B_{\sv}\Xv_{\cm}^{\sv}) >-\infty.
	}
	
	\item {\bf Step (ii)}: 
We can bound $\|\Xv\|_{\alpha}^{\beta}$ as
%
%
%
	\ea{
	\|\Xv^n\|_{\alpha}^{\beta} &= \big(\sum_{i=1}^n |X_i|^{\alpha}\big)^{\beta/\alpha} \nonumber\\
	&\overset{(a)}{=} \big(\sum_{i=1}^n \nu_i|X_{\cm, i}|^{\alpha}+(1-\nu_i)|X_{\dm, i}|^{\alpha}\big)^{\beta/\alpha} \nonumber\\
	&\leq (\|\Xv_{\cm}\|_{\alpha}^{\alpha}+\|\Xv_{\dm}\|_{\alpha}^{\alpha})^{\beta/\alpha}. \label{eqn: bound_xn}
}
where $(a)$ is because of the definition of $\Xv^n$ in \eqref{eqn: indep_disc_cont}. 

Further, general equivalence of norms in finite-dimensional spaces (e.g., see \cite[pp. 517]{foucart2013invitation}) deduces that
\ea{
	\|\vv\|_1\leq C_q\|\vv\|_q, \label{eqn: norm_ineq}
}
for a scalar $C_q\in \Rbb^{+}$.
Hence, by setting $q=\frac{\beta}{\alpha}$ and $\vv_{2\times 1} = (\|\Xv_{\cm}\|_{\alpha}^{\alpha} \,,\,\|\Xv_{\dm}\|_{\alpha}^{\alpha})^{\intercal}$, \eqref{eqn: bound_xn} and \eqref{eqn: norm_ineq} imply that
\ea{
\|\Xv^n\|_{\alpha}^{\beta} &\leq C_{\beta/\alpha} \big(\|\Xv_{\cm}\|_{\alpha}^{\beta}+\|\Xv_{\dm}\|_{\alpha}^{\beta}\big) \nonumber\\
\Rightarrow ~~ 
\Ebb \big[ \|\Xv^n\|_{\alpha}^{\beta} \big] &\leq C_{\beta/\alpha} \big( \Ebb \big[\|\Xv_{\cm}\|_{\alpha}^{\beta} \big]+\Ebb \big[ \|\Xv_{\dm}\|_{\alpha}^{\beta}\big]\big).
}

Hence, since 	$\Ebb[\|\Xv_{\dm}\|_{\alpha}^{\beta}], \Ebb[\|\Xv_{\cm}\|_{\alpha}^{\beta}]<\infty$, we have $\Ebb \big[ \|\Xv^n\|_{\alpha}^{\beta} \big]<\infty$.
%
Thus, with $\Yv^m = A_m\Xv^n$, we know that 
	$\Ebb[\|\Yv^m\|_{\alpha}^{\beta}]<\|A_m\|_{\alpha}^{\beta}\Ebb[\|\Xv^n\|_{\alpha}^{\beta}]<\infty$, where $\|A_m\|_{\alpha}$ is an operator norm of matrix $A_m$ induced by vector $\alpha$-norm. This inequality is followed by the fact that 
	\ea{
		\|A_m\|_{\alpha}\leq \|A_m\|_{\alpha, \infty}\leq m^{1/\alpha} n \max_{i, j} |a_{ij}|,\label{eqn: norm_alpha_max}
	}
	 for $A_m=[a_{ij}]$, where $\|A_m\|_{\alpha, \infty}$ denotes $(\alpha, \infty)$-subordinate norm defined as
	 \ea{
	 	\|A_m\|_{\alpha, \infty} = \max_{\xv^n\neq\boldsymbol{0}} \f{\|A_m\xv^n\|_{\alpha}}{\|\xv\|_{\infty}}.
	 }

	\item {\bf Step (iii)}: 
	The claim easily follows from Theorem \ref{thm: singular_subset} if we substitute $\Yv^m\to \Zv^m$, $\Ytv_i\to\Cv_i$, and $V\to V_m$, where $\Ytv_i$ and $V$ are defined in \eqref{eqn: y_tilde_i} and \eqref{eqn: upsilon_m}, respectively. 
\end{noindlist}
}

\section{Proof of Lemma \ref{thm: lip_opt}}\label{proof: lip_opt}
Since $f$ is a Lipschitz function, $g_{\xv_{\overline{\nuv}}}(\cdot)$ defined via
\ea{
	g_{\xv_{\overline{\nuv}}}(\Xv_{\nuv})= f(\Xv_{\nuv}, \xv_{\overline{\nuv}})
}
is also a Lipschitz function. 
Thus, by using \cite[Theorem 2]{WuThesis}, we achieve\footnote{The direction of the inequality is misprinted in the referenced theorem.}
\ea{
	\overline{d}\big( f(\Xv_{\nuv}, \xv_{\overline{\nuv}})\big)\leq \sum_{i=1}^n\nuv_i .\label{eqn: lip_b_1}
}
Alternatively, \cite[Eqn. 80]{renyi1959dimension} shows that
\ea{
	\overline{d}\big( f(\Xv_{\nuv}, \xv_{\overline{\nuv}})\big)\leq m. \label{eqn: lip_b_2}
}
Now, by combining \cite[Theorem 5]{WuThesis}, \eqref{eqn: lip_b_1}, and \eqref{eqn: lip_b_2} we conclude that 
\ea{
	\overline{d}\big(f(\Xv^n)\big)&\leq \Ebb_{\nuv} \Big[\min\big\{\sum_{i=1}\nu_i \,,\, m \big\}\Big] \nonumber \\
	&= d (A_m\Xv^n),
	\label{eq: d (A Xv)}
}
where the last equality follows from Theorem \ref{thm: disc_cont_fat} and equation \eqref{eqn: rank_vand}.

\section{Proof of Proposition \ref{prop: random_matrix}}\label{app: proof_random_matrix}
{\changed 
		To prove this proposition, first we show that for a binary vector $\sv$, we have $\rank(A^{[\sv]})= \min \big\{m, \sum_{i=1}^n s_i\big\}$ with probability $1$. The reason is twofold. First, it is proved that a polynomial is either trivial ( i.e., $p(x)=0$ everywhere ) or the set of its roots is zero Lebesgue measure. As a result, the determinant of a square matrix with i.i.d continuous random variables is zero with probability $0$, because joint probability measure of i.i.d continuous random variables is absolutely continuous w.r.t the Lebesgue measure, and the determinant is a polynomial function of the matrix elements. In other words, every $k$ number of $m$-dimensional vectors with i.i.d continuous random variables are linearly independent with probability $1$, if we have $k\leq m$. As a result, in the case that $\sum_{i=1}^n s_i\leq m$, we have $\rank(A^{[\sv]})=\sum_{i=1}^n s_i$ with probability $1$. On the other hand, if $\sum_{i=1}^n s_i>m$, the above discussion shows that every sub-matrix of $A^{[\sv]}$ with $m$ number of columns is full-rank with probability $1$. Finally, since $\rank(A^{[\sv]})$ is bounded by its number of rows $m$, one sees that  $\rank(A^{[\sv]})=m$ with probability $1$. This proves the claim  $\rank (A^{[\sv]})=\min \{m, \sum_{i=1}^n s_i\}$. 
		
		Next, we show that $\frac{1}{m} d(A \Xv^n)\to \min\{1, \frac{p}{r}\}$. Firstly, Theorem \ref{thm: disc_cont_fat} shows that
		\ea{
		\frac{1}{m} d(A \Xv^n)= \Ebb_{\nuv}\big[\tfrac{1}{m}\rank (A^{[\nuv]})\big] = \Ebb_{\nuv} \Big[\min \big\{1, \frac{\sum_{i=1}^n \nu_i}{m} \big\}\Big],
		}
		where the last equality is followed by the above discussion. By conditioning on whether $\sum_{i=1}^n \nu_i \leq m$ or not, we have
		\ea{
		\frac{1}{m} d(A \Xv^n)=  &\Pr(\sum_{i=1}^n \nu_i> m) \nonumber\\&+  \Ebb_{\nuv} \Big[  \frac{\sum_{i=1}^n \nu_i}{m} \big| \sum_{i=1}^n \nu_i \leq m\Big] \Pr\Big(\sum_{i=1}^n \nu_i \leq m\Big),\label{eqn: d_conditional}
		}
		
		In the case that $p\geq r$, the proportion $\frac{\sum_{i=1}^n \nu_i}{m}$ concentrates around $\frac{p}{r}\geq 1$, which concludes that $\Pr(\sum_{i=1}^n \nu_i\leq  m)\to 0$. Furthermore, $\Ebb_{\nuv} \Big[  \frac{\sum_{i=1}^n \nu_i}{m} \big| \sum_{i=1}^n \nu_i \leq m\Big]$ is bounded by $\frac{1}{r}$. As a result, asymptotically we have $\frac{1}{m} d(A\Xv^n)\to 1$.
		
		In a case that $p<r$, we rewrite \ref{eqn: d_conditional} as
		\ea{
		\frac{1}{m} &d(A\Xv^n)\nonumber\\&=\Pr(\sum_{i=1}^n \nu_i > m) \cdot \underbrace{\bigg(1-\Ebb_{\nuv} \Big[  \frac{\sum_{i=1}^n \nu_i}{m} \big| \sum_{i=1}^n \nu_i > m\Big]\bigg)}_{A}\nonumber\\ &+\Ebb_{\nuv}\big[\tfrac{\sum_{i=1}^n \nu_i}{m}\big].
		}
		Here, with the same reasoning as in the other case, we have $\Pr\big(\sum_{i=1}^n \nu_i >m\big)\to 0$. Further, the term $A$ is upper- and lower-bounded by $1$ and $1-\frac{1}{r}$, respectively. Moreover, the average of Bernoulli random variables $\Ebb_{\nuv}\big[\tfrac{\sum_{i=1}^n \nu_i}{m}\big]$ converges to $\frac{p}{r}$. This proves that $\frac{1}{m} d(A\Xv^n)\to \frac{p}{r}$.
		
		These two cases show that $\frac{1}{m} d(A\Xv^n)\to \min\{1, \tfrac{p}{r}\}$.
		}
		\section{Proof of Lemma \ref{lem: abs_cont_affine}} \label{app: prf_abs_cont_aff_sing}
	We prove two parts  of this lemma (direct part and converse) as follows. 
	
	First, we assume that $\mu_{\Yv}(\cdot)$ is an affinely singular probability measure on set  $\Scal_{\Yv}\subseteq \Scal_{\Xv}$ of affine subsets, and each of its measures $\mu_{\Yv}^{\Acal}$ on set $\Acal\in \Scal_{\Yv}$ is absolutely continuous with respect to $\mu_{\Xv}^{\Acal}$. Then, for any set $\Ccal$, if we have
	\ea{
		\mu_{\Xv}(\Ccal)=0=\sum_{i} p_i \mu_{\Xv}^{\Acal_i}(\Ccal\cap \Acal_{i} ),
	}
	one concludes that 
	\ea{\mu_{\Xv}^{\Acal_i}(\Ccal \cap \Acal_i )=0,}
	for every $\Acal_i\in \Scal_{\Xv}$. Next, because of absolute continuity of $\mu_{\Yv}^{\Acal_i}$ with respect to $\mu_{\Xv}^{\Acal_i}$, we have
	\ea{
		\mu_{\Yv}^{\Acal_i}(\Ccal \cap \Acal_{i} )=0,
	}
	that results in
	\ea{
		\mu_{\Yv}(\Ccal)=\sum_{i} q_i \mu_{\Yv}^{\Acal_i} (\Ccal\cap \Acal_i ) =0.
	}
	Note that since $\Scal_{\Yv}\subseteq\Scal_{\Xv}$, then $q_i$s could get zero values. 
	The above identity shows that $\mu_{\Yv}(\cdot)$ is absolutely continuous with respect to $\mu_{\Xv}(\cdot)$.
	
	Secondly, we assume that $\mu_{\Yv}(\cdot)$ is absolutely continuous with respect to $\mu_{\Xv}(\cdot)$. We order $\Acal_i$s in a way that $e_i$ is increasing with $i$. Then, we define $\Bcal_i$ as
	$$\Bcal_i := \Acal_i \setminus \cup_{j=1}^{i-1} \Acal_j.$$
	
	One can see that based on such definition, we have $\Bcal_i\cap \Bcal_j=\emptyset$, where $i\neq j$, and $\cup_i \Bcal_i = \cup_{i} \Acal_i $. 
	
	Next, for every set $\Ccal$, we have
	\ea{
		\mu_{\Yv} (\Ccal) &= \mu_{\Yv} (\Ccal \cap \cup_{i}\Bcal_i )+\mu_{\Yv} (\Ccal\setminus \cup_{i}\Bcal_i)\\
		&\overset{(a)}{=} \mu_{\Yv}(\Ccal\cap \cup_{i}\Bcal_i)\\
		& \overset{(b)}{=} \sum_{i} \mu_{\Yv}(\Ccal\cap_i \Bcal_i),
	}
	where $(a)$ holds because $\mu_{\Yv}(\cdot)$ is absolutely continuous with respect to $\mu_{\Xv}(\cdot)$, and $\mu_{\Xv}(\Ccal\setminus \cup_{i} \Bcal_i)=0$, and $(b)$ is followed by mutual exclusion of $\Bcal_i$s. 
	
	Now, we define $q_i$ and $\mu_{\Yv}^{\Acal_i}(\cdot)$ as 
	\ea{
		q_i := \mu_{\Yv}(\Bcal_i),
	}
	and
	\ea{
		\mu_{\Yv}^{\Acal_i}(\cdot) = \frac{\mu_{\Yv}(\cdot \cap \Bcal_i)}{\mu_{\Yv}(\Bcal_i)},
	}
	if $q_i\neq 0$. One can check that $\mu_{\Yv}^{\Acal_i}(\cdot)$ is a probability measure. Further, since $\mu_{\Yv}(\cdot \cap \Bcal_i)$ is absolutely continuous with respect to $\mu_{\Xv}(\cdot \Bcal_i)$, and $\mu_{\Xv}(\cdot\cap\Bcal_i)$ is absolutely continuous with respect to the Lebesgue measure on $\Bcal_i$, and because of transitivity of absolute continuity, one can see that $\mu_{\Yv}^{\Acal_i}(\cdot)$ is absolutely continuous with respect to the Lebesgue measure on $\Bcal_i$. Hence, one can write $\mu_{\Yv}(\Ccal)$ as
	\ea{
		\mu_{\Yv}(\Ccal) = \sum_{i} q_i \mu_{\Yv}^{\Acal_i}(\Ccal\cap \Bcal_i),\label{eqn: mu_y_sum}
	}
	for a set of absolutely continuous probbability measures $\{\mu_{\Yv}^{\Acal_i}(\cdot)\}$ on the collection of affine subsets $\Scal_{\Yv}\subseteq\Scal_{\Xv}$ (these collections could be not equal since $q_i$ could take zero values). The absolute continuity of each $\mu_{\Yv}^{\Acal_i}(\cdot)$ with respect to $\mu_{\Xv}^{\Acal_i}(\cdot)$ is followed by the absolute continuity of $\mu_{\Yv}(\cdot)$ with respect to $\mu_{\Xv}(\cdot)$, which completes the proof. Finally, since $\Acal_i\cap \cup_{j=1}^{i-1} \Acal_j$ is formed by affine sets with dimension strictly lower than $e_i$, and because of absolute continuity of $\mu_{\Yv}^{\Acal_i}(\cdot)$, we have $\mu_{\Yv}^{\Acal_i}(\Acal_i\cap \cup_{j=1}^{i-1} \Acal_j)=0$, therefore, we can rewrite \eqref{eqn: mu_y_sum} as
	\ea{
		\mu_{\Yv}(\Ccal) = \sum_{i} q_i \mu_{\Yv}^{\Acal_i}(\Ccal\cap \Acal_i).
	}
	
	\section{Proof of Lemma \ref{lem: kl}} \label{app: prf_kl}
	The necessary and sufficient condition of existence of $\Dsf(\mu_{\Xv}\|\mu_{\Yv})$ is followed by Lemma \ref{lem: abs_cont_affine} and by the fact that $\Dsf(\mu_{\Xv}\|\mu_{\Yv})$ is defined if and only if $\mu_{\Xv}$ is absolutely continuous with respect to $\mu_{\Yv}$. 
	
	For calculating $\Dsf(\mu_{\Xv}\|\mu_{\Yv})$, firstly we order the collection $\Scal_{\Xv}$ of the affine sets $\{\Acal_{i}\}_i$ in the way that the dimension $e_i$ of the set $\Acal_i$ is an increasing function of $i$. Next, we define $\Bcal_i$ as
	\ea{
		\Bcal_i = \Acal_i \setminus  \cup_{j=1}^{i-1} \Acal_j.
	}
	Note that $\cup_i \Bcal_i=\cup_i \Acal_i$ and $\Bcal_i$s are mutually exclusive (i.e., $\Bcal_i\cap \Bcal_j=\emptyset$ if $i\neq j$).
	
	Next, we write the formulation of $\Dsf(\mu_{\Xv}\|\mu_{\Yv})$ as
	\ean{
		\Dsf(\mu_{\Xv}&\|\mu_{\Yv}) = \int_{\cup_i \Acal_i} \diff \mu_{\Xv} \log \frac{\diff \mu_{\Xv}}{\diff \mu_{\Yv}} \\
		&\overset{(a)}{=} \int_{\cup_i \Bcal_i} \diff \mu_{\Xv}\log \frac{\diff \mu_{\Xv}}{\diff \mu_{\Yv}}\\
		&\overset{(b)}{=} \sum_{i}\int_{\Bcal_i} \diff \mu_{\Xv} \log \frac{\diff \mu_{\Xv}}{\diff \mu_{\Yv}}\\
		&\overset{(c)}{=} \sum_{i} p_{\Xv}^{\Acal_i} \int_{\Bcal_i} \diff\mu_{\Xv}^{\Acal_i}\log \frac{p_{\Xv}^{\Acal_i}  \diff \mu_{\Xv}^{\Acal_i}}{p_{\Yv}^{\Acal_i}  \diff \mu_{\Yv}^{\Acal_i}}\\
		&= \sum_i p_{\Xv}^{\Acal_i}  \log \frac{p_{\Xv}^{\Acal_i} }{p_{\Yv}^{\Acal_i} } +\sum_i p_{\Xv}^{\Acal_i} \int_{\Bcal_i} \diff \mu_{\Xv}^{\Acal_i}\log \frac{\diff\mu_{\Xv}^{\Acal_i}}{\diff\mu_{\Yv}^{\Acal_i}}\\
		&\overset{(d)}{=}\Dsf(p_{\Xv}\|p_{\Yv})+\sum_i p_{\Xv}^{\Acal_i} \int_{\Acal_i}\diff \mu_{\Xv}^{\Acal_i}\log \frac{\diff\mu_{\Xv}^{\Acal_i}}{\diff\mu_{\Yv}^{\Acal_i}}\\
		&=\Dsf(p_{\Xv}\|p_{\Yv}) +\sum_{\Acal\in \Scal_{\Xv}} p_{\Xv}^{\Acal} \Dsf(\mu_{\Xv}^{\Acal}\|\mu_{\Yv}^{\Acal}),
	}
	which completes the proof. Note that $(a)$ holds because $\cup_i \Bcal_i = \cup_i \Acal_i$, $(b)$ is followed by mutual exclusion of $\Bcal_i$s, $(c)$ holds because of the form of affinely singular probability measures $\mu_{\Xv}$ and $\mu_{\Yv}$, which for every set $\Ccal\subseteq \cup_i \Acal_i$ is as
	\ea{
		\mu_{\Xv}(\Ccal) = \sum_i p_{\Xv}^{\Acal_i} \mu_{\Xv}^{\Acal_i}(\Ccal\cap \Acal_i),
	}
	and
	\ea{
		\mu_{\Yv}(\Ccal) = \sum_i p_{\Yv}^{\Acal_i} \mu_{\Yv}^{\Acal_i}(\Ccal\cap \Acal_i).
	}
	Finally, $(d)$ is correct because $\Acal^{-i}$ is mutually singular with respect to $ \cup_{j=1}^{i-1} \Acal_j$, and therefore $\mu_{\Xv}^{\Acal_i}(\Ccal\cap \Bcal_i)=\mu_{\Xv}^{\Acal_i}\big(\Ccal\cap(\Bcal_i\cup \cup_{j=1}^{i-1} \Acal_j\big)= \mu_{\Xv}(\Ccal\cap\Acal_i)$.

\section{Proof of Theorem \ref{thm: concent_Rs}}\label{app: prf_concent_Rs}
Recalling the definition of BID in \eqref{eqn: BID_def} and the result in Theorem \ref{thm: singular_subset},
	\eqref{eqn: BID_di_m} is immediate. 
	To prove \eqref{eqn: Rs_equal}, we continue with the following three steps: (i) we first prove that {\changed $R(\ep)\leq d_B\big(\{\Zv_t\}\big)$, (ii) then, we show that $R^*(\ep)\geq d_B\big(\{\Zv_t\}\big)$}, and (iii) finally we use \cite[Eqn. 75]{wu2010renyi} which  for a general source states that
	\ea{
		R^*(\ep)\leq R_B(\ep)\leq R(\ep). \label{eqn: R_relation}
	}	
	The following definition of rectifiability in \cite[Section 3.2.14]{federer1969geometric} is also used in the development of the proof.
	\begin{defi}\label{def: rectifiability}
		The set $E$ is $m$-rectifiable if there exists a Lipschitz function from a bounded subset of $\Rbb^m$ to $E$.
	\end{defi}
	
	\begin{noindlist}
		\item {\bf Step (i)}: In this step we prove that 
		\ea{
		R(\ep)\leq d_B\big(\{\Zv_t\}\big). \label{eqn: Up_bound_R}
	}
	 In order to do this, we utilize \cite[Lemma 12]{wu2010renyi} that provides a sufficient condition to upper-bound $R(\ep)\leq r$. More specifically, the condition is that for large $m$s, we have an {\changed $\lfloor mr\rfloor $-rectifiable }set $\Rcal_m\in \Rbb^m$, in which $\Pr(\Zv^m\in \Rcal_m)\geq 1-\ep$. In fact, for an arbitrary $\delta\in[0, 1]$ we provide a set $\Rcal_m$ with $r=\Big(d_B\big(\{\Zv_t\}\big)+\delta\Big)$ . Thereby, using the arbitrariness on $\ep$ and $\delta$, and the above discussion we complete the proof.
		
		To begin with, for independent vectors $\{\uv_{i,0}, \uv_{i,1}, \ldots, \uv_{i, e_i}\}$ in $\Rbb^n$, let 
		\ea{
		\Scal_i=\spanM\big(\{\uv_{i, j}\}_{j=1}^{e_i}\big)+\uv_{i, 0} 
		}
		be an affine subset. Obviously, any affine subset can be expressed this way.
Further, any bounded subset of $\Scal_i$ like $\Acal$ is also $e_i$-rectifiable. To check this, note that for each 
		 $\xv\in \Acal$ we can write that
		\ea{
			\xv = \sum_{j=1}^{e_i}a_j\uv_j+\uv_0.
		}
		Hence, $\xv$ is a linear, and thus, Lipschitz function of $\av^{e_i}=(a_1, \ldots, a_{e_i})^{\intercal}$.  
		Moreover, due to the independence of  $\{\uv_i\}$s and the boundedness of $\Acal$, we conclude that $\av^{e_i}$s are also bounded. 
Thus, $\Acal$ is $e_i$-rectifiable (Definition \ref{def: rectifiability}).
		Recalling \cite[Lemma 11]{wu2010renyi}, we conclude that every bounded subset of $\bigcup\limits_{i}\Scal_i$ is $(\max_{k} \{e_k\})$-rectifiable. 
		Similarly, if {\changed 
					\ea{
				\Ical_m = \bigg\{i:\, e_i\leq \lfloor mr\rfloor\bigg\},
			}}
			and 
		\ea{
		\Rcal_m = \Big(\bigcup\limits_{i\in \Ical_m} \Kcal_i \Big) \bigcap \, [-l, l]^{m}	,
		}	
		for arbitrary $l\in \Rbb^{+}$, then, $\Rcal_m$ is $\lfloor mr\rfloor $-rectifiable (See the definition of $\Kcal_i$ in Section \ref{sec:DRB} ).
		Furthermore, using the assumption in \eqref{eqn:Concentration}, for every $\delta, \ep\in \Rbb^{+}$, we know that   there exists a large enough $m$ such that {\changed $e_i< mr$ (or equivalently $e_i\leq \lfloor mr\rfloor$)} with probability at least $1-\ep/2$; formally,  $\Pr(V_m\in \Ical_m)\geq 1-\tfrac{\ep}{2}$, for sufficiently large $m$.
		Besides, one can choose large enough $e$ so as to
		\ea{
			\Pr\big(\Zv^m\in [-l, l]^m| V_m\in\Ical_m\big)\geq 1-\tfrac{\ep}{2}.
		}
		Thus, we have
		\ea{
			\Pr(\Zv^m\in \Rcal_m)=\Pr\big(\Zv^m\in [-l, l]^m, V_m\in\Ical_m\big)\geq 1-\ep,
			\label{eq:rect}
		}
	which completes the proof.

		\item {\bf Step (ii) -- proof by contradiction}: 
		Assume that $R^*(\ep)< d_B\big(\{\Zv_t\}\big)$ and define $\delta = d_B\big(\{\Zv_t\}\big) - R^*(\ep)$. We can check that 
		 $R=d_B\big(\{\Zv_t\}\big)-\delta/2$ satisfies the conditions in  Lemma \ref{lmm: min_R}. 
		In other words, there should exist $M_1\in \Nbb$, $m\geq M_1$, $\Scal^m\in \Rbb^m$, and a  $\lceil (1-R)m\rceil$-dimensional subspace $\Hcal^m\subset \Rbb^m$ such that 
		\ea{
		\Pr(\Zv^m\in \Scal^m)\geq 1-\ep\label{eqn: scal_m_ep}}
		 and $(\Scal^m-\Scal^m)\cap \Hcal^m =\{0\}$. 
Below, we find a non-zero vector in $(\Scal^m-\Scal^m)\cap \Hcal^m$, which contradicts the assumption and proves the claim. 
%
		Let $k_m = \lfloor mR\rfloor$ and $k'_m=\Big\lfloor m\big(d_B\big(\{\Zv_t\}\big)-\delta/4\big)\Big\rfloor$. 
		Since $R< d_B\big(\{\Zv_t\}\big)-\delta/4$, for large enough $m$ we have
		\ea{
			k'_m>k_m.
		}

Next, let us define $\Fcal_m^{k_m'}$ as the set of all affinely singular subsets (in $\Rbb^m$) for the measure of $\Zv^m$  with dimension $k'_m$ or higher. 
Formally, 
		\ea{
			\Fcal_m^{k_m'}:= \{\Fcal\in \Kcal\, ,\,  \dim(\Fcal)\geq k'_m\},
		}
		where $\Kcal$ is defined in Section \ref{sec:DRB}. We recall that due to the definition of affinely singular subsets, $\Zv^m$ has an absolutely continuous distribution on each $\Fcal\in \Fcal_m^{k_m'}$.
		%

For each $\Fcal\in\Fcal_m^{k'_m}$, we know that $\Fcal-\Fcal$ is a subspace of $\Rbb^m$ (since $\Fcal$ is an affine subset of $\Rbb^m$). Furthermore,
\ea{
& \dim(\Hcal^m)= m-k_m>m-k'_m\geq m- \dim(\Fcal-\Fcal) \nonumber\\
\Rightarrow ~~ &  \dim(\Hcal^m) + \dim(\Fcal-\Fcal) > m. \label{eqn: au_Hm_F-F}
}
This shows that  $\Hcal^m \cap (\Fcal-\Fcal)$ is non-trivial; i.e., there exists $\uv \in \Hcal^m \cap (\Fcal-\Fcal)$ with $\|\uv\|_2=1$.
%
As we have{\changed 
\ea{
	\Pr_{V_m}(e_{V_m}\geq k'_m) &\geq \Pr_{V_m}\Big(e_{V_m}\geq m\big(d_B\big(\{\Zv_t\}\big)-\delta/4\big) \Big) \nonumber\\
	&= \Pr_{V_m}\Big(\tfrac{e_{V_m}}{m} - d_B\big(\{\Zv_t\}\big) \geq -\delta/4  \Big) \nonumber\\
	&\geq \Pr_{V_m}\Big(\big|\tfrac{e_{V_m}}{m} - d_B\big(\{\Zv_t\}\big) \big| \leq \delta/4  \Big),
}}
due to the assumption of the theorem, it is possible to set $m$ large enough that the latter probability exceeds $\frac{1}{2}+\frac{\epsilon}{2}$ (for $\epsilon <\frac{1}{2}$):
\ea{
	m\gg 1:~~ \Pr_{V_m}(e_{V_m}\geq k'_m) &\geq \tfrac{1+\epsilon}{2}.
\label{eqn: dv_kmprime}
}
%
		Next, we have
		{\changed
		\ea{
			\Pr\Big(\Zv^m \in  &\big(\bigcup\limits_{\Fcal\in\Kcal}\Fcal\setminus \bigcup\limits_{\Fcal'\in\Kcal\setminus \Fcal_{{\Fcal'}^{k'_m}_{m}}} \Fcal'\big)\cap \Scal^m\Big) \nonumber\\&\overset{(a)}{\geq} -\Pr\big(\Zv^m \notin \bigcup\limits_{\Fcal\in\Kcal}\Fcal\setminus \bigcup\limits_{\Fcal\in\Kcal\setminus {{\Fcal}^{k'_m}_{m}}} \Fcal'\big)  +\Pr(\Zv^m \in \Scal^m)\\&\overset{(b)}{=} -\Pr (\Zv^m \in \bigcup\limits_{\Fcal'\in\Kcal\setminus {{\Fcal}^{k'_m}_{m}}} \Fcal') +\Pr(\Zv^m \in \Scal^m) \\& = -\underbrace{\Pr_{V_m}(e_{V_m}< k'_m)}_{\leq \frac{1-\epsilon}{2}}+\underbrace{\Pr(\Zv^m\in\Scal^m)}_{\geq 1-\epsilon}\\
			&\geq \frac{1-\ep}{2}>0, \label{eqn: low_bound_cap_Fn}
		}
		}
		where $(a)$ is a result of  the fact that $\Pr(A\setminus B)\geq -\Pr(B)+\Pr(A)${\changed, and $(b)$ holds because $\Pr(\Zv^m \notin \bigcup\limits_{\Fcal\in\Kcal}\Fcal)=0$}. 
		The positive probability in \eqref{eqn: low_bound_cap_Fn} (and the fact that $\Fcal_{m}^{k'_m}$ is countable) shows that at least for one $\Fcal\in\Fcal_{m}^{k'_m}$,{\changed if we define $\Lcal$ as
		\ea{
		\Lcal = \big(\Fcal \setminus\bigcup\limits_{\Fcal'\in\Kcal\setminus {{\Fcal}^{k'_m}_{m}}} \Fcal'\big)
		}

		the probability  $\Pr\Big(\Zv^m \in \Lcal\cap \Scal^m\Big)$ (and therefore $\Pr(\Zv^m \in \Lcal)$) is nonzero:
%
		\ea{
		\Pr(\Zv^m\in \Lcal\cap \Scal^m \,|\,\Zv^m\in \Lcal)=\frac{\Pr(\Zv^m\in \Lcal\cap \Scal^m)}{\Pr(\Zv^m\in \Lcal)}>0. \label{eqn: pos_meas}
		}
Since $\Zv^m$ has an absolutely continuous measure on  $\Lcal$, the set $\Ecal=\Lcal\cap \Scal^m$ should have a non-zero Lebesgue measure in $\Rbb^{\dim(\Fcal)}$. Using this fact, and since every $\Fcal' \in \Kcal \setminus \Fcal_{{\Fcal'}^{k'_m}_{m}}$ has zero Lebesgue measure in $\Rbb^{\dim(\Fcal)}$, we can show that $\Gcal = \Fcal \cap \Scal^m$ has non-zero Lebesgue measure in $\Rbb^{\dim (\Fcal)}$. }
%
%
		%
		Now, utilizing Steinhaus Theorem \cite[Theorem VIII]{Steinhaus1920}, we conclude the existence of $a\in \Rbb^{+}$ such that
		\ea{
			{\rm Ball}^{\dim (\Fcal)}_a ~\subset~ \Gcal-\Gcal ~\subset~ (\Fcal - \Fcal) \cap (\Scal^m - \Scal^m).
		}
		Hence, for all $\overline{\uv} \in \Fcal - \Fcal$ with $\| \overline{\uv}\|_2\leq a$, we know that $\overline{\uv} \in \Scal^m - \Scal^m$. Now, if we set $\overline{\uv} = a \uv$ where $\uv \in \Hcal^m \cap (\Fcal-\Fcal)$ and $\|\uv\|_2=1$, we obtain that
				\ea{
\overline{\uv} \in \Hcal^m \cap (\Fcal-\Fcal) \cap (\Scal^m - \Scal^m).
		}
		This reveals that $\Hcal^m$ and $\Scal^m - \Scal^m$ have a non-zero intersection, which completes the proof.

		\item {\bf Step (iii)}:  \eqref{eqn: Rs_equal} follows directly from \eqref{eqn: Up_bound_R}, the fact that $R_B^*(\ep) \geq d_B\big(\{\Zv_t\}\big)$, and \eqref{eqn: R_relation}.
	\end{noindlist}

\section{Proof of Lemma \ref{lmm: concent_MA}}\label{app: prf_concent_MA}
Recalling Lemma \ref{lmm: Ax_has_subsets}, we know that an $i$-dimensional singularity occurs when $\nuv=\sv\in \{0, 1\}^n$ such that $\rank \lb A_m^{[\sv]} \rb = i$. 
	%
		Besides, the columns of $A_m$ indexed in the range $[l_1+l_2+1:n]$ are linearly independent. 
		Hence, if $\rank( A_m^{\sv})=i$, then $\|\sv\|_1\leq l_2+l_1+i$ (at most $i$ columns in the range $[l_1+l_2+1:n]$ can be selected). 
	As a consequence, we have
	\ea{
		\Pr(e_{V_m}\leq k) &\leq \sum_{i=0}^{k+l_1+l_2} {{n}\choose{i}} \alpha^i (1-\alpha)^{n-i}\nonumber  \\
		&\overset{(a)}{\leq} {\rm exp}\Big(-n\Dsf\big(\tfrac{k+l_2+l_1}{n}\|\alpha\big)\Big)	,
	}
	where $(a)$ is obtained from \cite[Theorem 1]{arratia1989} by assuming $\f{k+l_2+l_1}{n}<\alpha$.
	
	In contrast, if we select $k$ columns all in the range $[l_1+l_2+1:n]$, then $\rank A_m^{[\sv]} = k$. Thus,
\ea{
\Pr_{V_m}(e_{V_m}\geq k) & \geq \sum_{i=k}^{n-l_1-l_2}{{n-l_2-l_1}\choose i} \alpha^{i} (1-\alpha)^{n-l_2-l_1} \nonumber  \\
		 	& \overset{(b)}{\geq} {\rm exp}\Big(-(n-l_1-l_2)\Dsf\big(\tfrac{k}{n-l_2-l_1}\|\alpha\big)\Big),
}
	where $(b)$ is also obtained from \cite[Theorem 1]{arratia1989} using $\f{k}{n-l_2-l_1}> \alpha$. Substituting $n=m+l_1+l_2$ in the two previous  inequalities, the desired claim in achievable.

\section{Proof of Lemma \ref{lem: rho}} \label{app: prf_lem_rho}
	 Using \eqref{eqn: sing_drb} and \eqref{eqn: b_decomp}, we have
	\ea{
		n b(X_1)= \Hsf(\nuv, \Xv_{\dm}^{\nuov})+\hsf(\Xv_{\cm}^{\nuv}|\nuv).	
	}
	Theorem \ref{thm: disc_cont_fat} shows that $b(\Yv^m) = \Hsf(\Upsilon_m) + \hsf(\Ytv_{\Upsilon_m}|\Upsilon_m)$. Therefore,
	\ea{
	 b(\Yv^m) \hspace{-1mm}&-\hspace{-1mm} n b(X_1)\\& = \underbrace{\Hsf(\Upsilon_m) \hspace{-1mm}-\hspace{-1mm} \Hsf(\nuv, \Xv_{\dm}^{\nuov})}_{\De_1} + \hsf(\Ytv_{\Upsilon_m}|\Upsilon_m) - \hsf(\Xv_{\cm}^{\nuv}|\nuv)\label{eqn: De1}\\
	& = \De_1 + \underbrace{\hsf(\Ytv_{\Upsilon_m}|\Upsilon_m)-\hsf(\Wv_{\nuv, \Xv_{\dm}^{\nuov}}|\nuv, \Xv_{\dm}^{\nuov}) }_{\De_2} \nonumber\\
	& \phantom{ = \De_1} + \hsf(\Wv_{\nuv, \Xv_{\dm}^{\nuov}}|\nuv, \Xv_{\dm}^{\nuov}) - \hsf(\Xv_{\cm}^{\nuv}|\nuv).
	}
	Thus, we have that
%
	\ea{
		&\frac{\big|b(\Yv^m)-n b(X_1)-m\psi_{m, \nuv}(A_m)\big|}{m}\leq \frac{|\De_1|+|\De_2|+|\De_3|}{m},
	}
	where
	$$\De_3=  \hsf(\Wv_{\nuv, \Xv_{\dm}^{\nuov}}|\nuv, \Xv_{\dm}^{\nuov})-\hsf(\Xv_{\cm}^{\nuv}|\nuv)-m\psi_{m, \nuv}(A_m).$$
	
	Let $\Upsilon_{m}$ and $\Tcal_i^{(m)}$ for $i\in \Nbb$ denote the RV $V_m$ and the sets $\Tcal_i^{(m)}$ defined  in Section \ref{sec:DRB} for the mixing matrix $A_m$. 
	Below, we prove the claim of the lemma for $\alpha_i\neq 0$ in six steps:
	 (i) finding an upper-bound for the probability $\Pr_{V_m}(|\Tcal_{V_m}^{(m)}|\geq 2)$ (this is the probability that a singular subset appears for more than one choice of $\nuv$), (ii) upper-bounding $|\De_1|$, (iii) upper-bounding $|\De_2|$, (iv) bounding the difference between the differential entropy of $T\Xv_{\cm}^{\sv}$ and the differential entropy of $\Xv_{\cm}^{\sv}$ as
	\ea{\label{eq:Delta4}
		\De_4 = \hsf(T\Xv_{\cm}^{\sv})-\hsf(\Xv_{\cm}^{\sv}),	
	}
	in which $T$ is formed by $k<l$ rows of an $l\times l$ unitary matrix, where $l=\sum\limits_{i\in \Nbb} s_i$, (v)  upper-bounding $|\De_3|$, and (vi) combining the previous bounds to conclude the claim for $\alpha_i\neq 0$. Next, we explain how the claim for $\alpha_i= 0$ is obtained by a similar approach.

	\begin{noindlist}
		\item {\bf Step (i)}: Let $A_m=\big[A_m^{[1]}, \ldots, A_{m}^{[n]}\big]$, where $A_m^{[i]}$s stand for the columns. For each $\sv\in \{0, 1\}^n$, we define $\Acal_m^{[\sv]} = \bigcup\limits_{i: s_i=1} \{A_m^{[i]}\}$, and for simplicity of notation, we define $\Acal_m \overset{\De}{=} \Acal_m^{(1, 1, \ldots, 1)}$.
To bound $\Pr_{\Upsilon_m}(|\Tcal^{(m)}_{\Upsilon_m}|\geq 2)$, 
where the randomness in $\Upsilon_m = \digamma((\nu, \Xv_{\dm}^{\nuov}))$ is due to the randomness of $\nu$ and $\Xv_{\dm}^{\nuov}$, 
 note that $|\Tcal^{(m)}_{i}|\geq 2$ implies that for every pair $(\sv,\xv_{\dm}^{\sov})$ in $\Tcal^{(m)}_i$, there exists a distinct pair $(\sv', \xv_{\dm}^{\sov'})$, such that $\spanM(\Acal_m^{[\sv]}) = \spanM(\Acal_m^{[\sv]'}) $. 
There are now two possibilities:
	\begin{enumerate}[(i)]
		\item The set of non-zero locations in $\sv'$ is a subset of the non-zero locations in $\sv$. Hence,
		$A_{m}^{[\sv]}$ is column-rank-deficient (i.e., $I_{\text{fcr}}(A_m^{[\sv]})=0$).
		
		\item There is an $i$ for which we have $s'_i=1$ and $s_i=0$. Hence, 
		$(A_{m}^{[\sv]},A_m^{[i]})$ is column-rank-deficient (i.e., $I_{\text{fcr}}\big((A_{m}^{[\sv]},A_m^{[i]})\big)=0$).
	\end{enumerate}

We can now upper-bound $\Pr_{\Upsilon_m}(|\Tcal^{(m)}_{\Upsilon_m}|\geq 2)$ with
	\ea{
& \Pr_{\nuv}\Big(I_{\text fcr}(A_m^{[\nuv]})=0~~\text {or}\nonumber\\&\phantom{\leq \Pr_{\nuv}\Big(}~~\exists i\in [n]\,,\, \nu_i=0:\, I_{\text{fcr}} \big((A_m^{[\nuv]}\,,\, A_m^{[i]})\big) = 0 \Big) \label{eqn:two_cases} 
}
To evaluate the above upper-bound, let us consider the example $A_m=$\usebox{\smlmat}. As $A_m$ has $4$ columns, $\nuv$ has $2^{4}=16$ possibilities. In Figure \ref{fig: p_TVm}, each vertex represents a submatrix of $A_m$ corresponding to the written $4$-tuple (which columns are present and which columns are dropped). 

\begin{figure}
	\centering
	\scalebox{0.6}{
	\begin{tikzpicture} [scale=1]
	\tikzstyle{vertex2}=[circle,minimum size=20pt,inner sep=0pt, fill=red!24, draw]
	\tikzstyle{vertex1}=[circle,minimum size=20pt,inner sep=0pt, fill=green!24]
	\tikzstyle{vertex0}=[circle,minimum size=20pt,inner sep=0pt, fill=green!24, draw]
	\tikzstyle{edge1} = [draw,line width=2pt,->,blue!50]
	\tikzstyle{edge2} = [draw,line width=2pt,->,orange!50]	
	\tikzstyle{edge3} = [draw,line width=2pt,->,yellow!50]
	\tikzstyle{edge4} = [draw,line width=2pt,->,pink!50]
	\node[vertex1] (v4) at (.5*2,1*2) {$0100$};
	\node[vertex0] (v5) at (-1.5*2,2*2) {$0101$};
	\node[vertex0] (v6) at (.5*2, 2*2) {$0110$};
	\node[vertex2] (v7) at (-1.5*2,3*2) {$0111$};
	\draw[edge1] (v4) to (v5);
	\draw[edge1] (v6) to (v7);
	\draw[edge2] (v4) to (v6);
	\draw[edge2] (v5) to (v7);
	\node[vertex1] (v0) at (0,0) {$0000$};
	\node[vertex1] (v1) at (-1.5*2,1*2) {$0001$};
	\node[vertex1] (v2) at (-.5*2,1*2) {$0010$};
	\node[vertex0] (v3) at (-2.5*2,2*2) {$0011$};
	\draw[edge1] (v0) to (v1);
	\draw[edge1] (v2) to (v3);
	\draw[edge2] (v0) to (v2);
	\draw[edge2] (v1) to (v3);
	\draw[edge3] (v0) to (v4);
	\draw[edge3] (v1) to (v5);
	\draw[edge3] (v2) to (v6);
	\draw[edge3] (v3) to (v7);

	\node[vertex0] (v12) at (2.5*2,2*2) {$1100$};
	\node[vertex0] (v13) at (.5*2,3*2) {$1101$};
	\node[vertex2] (v14) at (1.5*2,3*2) {$1110$};
	\node[vertex2] (v15) at (0, 4*2) {$1111$};
	\draw[edge1] (v12) to (v13);
	\draw[edge1] (v14) to (v15);
	\draw[edge2] (v12) to (v14);
	\draw[edge2] (v13) to (v15);	
	\node[vertex1] (v8) at (1.5*2,1*2) {$1000$};
	\node[vertex1] (v9) at (-.5*2,2*2) {$1001$};
	\node[vertex0] (v10) at (1.5*2,0+2*2) {$1010$};
	\node[vertex0] (v11) at (-.5*2,3*2) {$1011$};
	\draw[edge1] (v8) to (v9);
	\draw[edge1] (v10) to (v11);
	\draw[edge2] (v8) to (v10);
	\draw[edge2] (v9) to (v11);
	\draw[edge3] (v8) to (v12);
	\draw[edge3] (v9) to (v13);
	\draw[edge3] (v10) to (v14);
	\draw[edge3] (v11) to (v15);
	
	\draw[edge4] (v0) to (v8);
	\draw[edge4] (v1) to (v9);
	\draw[edge4] (v2) to (v10);
	\draw[edge4] (v3) to (v11);
	\draw[edge4] (v4) to (v12);
	\draw[edge4] (v5) to (v13);
	\draw[edge4] (v6) to (v14);
	\draw[edge4] (v7) to (v15);

	\end{tikzpicture}	
}
\caption{Each circle $a_1a_2a_3a_4$, where $a_i\in\{0, 1\}^4$ for $i\in [1:4]$, represents the event $\nuv=(a_1\,,\,a_2\,,\,a_3\,,\,a_4)^{\intercal}$. The red circles point out that $A_m^{[\nuv]}$ is not a full column-rank matrix, while green circles stand for full column-rank matrices $A_m^{[\nuv]}$. Each directional edge which goes out from a node, denotes the augmentation of the associated matrix with a specific column.   }\label{fig: p_TVm}
\end{figure}
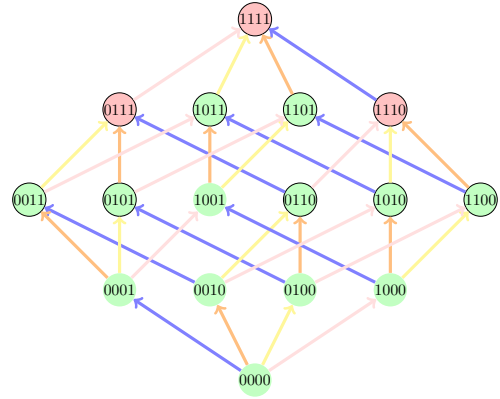

In this graph, vertex $i$ is connected to vertex $j$ (with a directed edge) if $A_m^{[\nuv_j]}$ has exactly one additional column with respect to $A_m^{[\nuv_i]}$ (the color of the edges in Figure \ref{fig: p_TVm} distinguish this additional column). 
Red vertices in this graph are column-rank-deficient matrices. 
The upper-bound in \eqref{eqn:two_cases} actually sums over the probability of red vertices as well as the vertices connected to the red vertices. Note that the probability of a vertex whose $4$-tuple has $i$ ones is $\alpha^{i}(1-\alpha)^{n-i}$ ($n=4$ in this example). Let $L_i$ denote the number of red vertices whose $n$-tuple has $i$ ones. Therefore, the probability of rank-deficiency $\rhom$ is given as
\ea{
\rhom = \sum_{i=0}^{n} L_i \alpha^i (1-\alpha)^{n-i}.
}
We now express the upper-bound in \eqref{eqn:two_cases} in terms of $L_i$s. Note that there are $i$ vertices connected to a red vertex whose $n$-tuple has $i$ ones, and also their $n$-tuples have $i-1$ ones. Hence, the probability of vertices connected to red vertices is upper-bounded by $\sum_{i=1}^{n} i L_i \alpha^{i-1} (1-\alpha)^{n-i+1}$; vertices could be counted more than once in this expression. However, as red vertices  are connected only to red vertices, we know that the latter expression contains the term $\sum_{i=1}^{n} (n-i) L_i \alpha^{i} (1-\alpha)^{n-i}$ just for red vertices. As a result, we have that
\ea{\label{eqn: pv}
\Pr_{V_m}(|\Tcal^{(m)}_{V_m}|\geq 2) 
&\leq \rhom
+ \sum_{i=1}^{n} i L_i \alpha^{i-1} (1-\alpha)^{n-i+1} \nonumber\\
& \phantom{\rhom} - \sum_{i=1}^{n} (n-i) L_i \alpha^{i} (1-\alpha)^{n-i} \nonumber\\
=& \sum_{i=0}^{n} \big(\underbrace{\tfrac{i}{\alpha}}_{\leq \frac{n}{\alpha}} - n +1\big) L_i \alpha^{i} (1-\alpha)^{n-i} \nonumber\\
=& \big(\tfrac{n}{\alpha}-n+1\big) \rhom \leq n C_0'\rhom,
}
where $C_0' = \frac{1}{\alpha}$.

	\item {\bf Step (ii)}: 
	Since $X_{\dm, i}$ is supported on a finite set ($\Xcal_{\dm}$), the cardinality of the sample spaces of $(\nuv, \Xv_{\dm}^{\nuov})$, and  $V_m$ is upper-bounded by 
	 $2^{C'_1n}$ where $C'_1 = \log (|\Xcal_{\dm}|+1)>0$ (note that $V_m$ is a probability distribution on a partition of pairs $\big\{(\sv, \xv_{\dm}^{\sov}):\, \sv\in \{0, 1\}^n,\, \xv_{\dm}^{\sov}\in \Xcal_{\dm}^{\bigotimes \|\sv\|_1}\big\}$).
	
	Let us define $g:\Nbb\to \big(\cup_{i} \Tcal_i^{(m)}\big) \cup \Nbb$ as
	{\changed
	\ea{
	g(i) :=\lcb \begin{array}{c c}
		\digamma^{-1}(i), & |\Tcal_i^{(m)}|=1,\\
		i, & \text{otherwise},
	\end{array}\right.
	} 
	}
	where $\digamma(\cdot)$ was defined in Section \ref{sec:DRB}. This implies that when $|\Tcal_{\Upsilon_m}^{(m)}|=1$, we have $g(\Upsilon_{m})=(\nuv, \nuov)$; when $|\Tcal_{\Upsilon_m}^{(m)}|>1$, $g(\Upsilon_{m})$ is an integer (not a pair). 
%
	 As a result, the total variation distance between the distributions of $g(\Upsilon_m)$ and $(\nuv, \Xv_{\dm}^{\nuov})$ does not exceed $2\Pr_{\Upsilon_m}\big(|\Tcal_{\Upsilon_m}^{(m)}|\geq 2\big)$. 
	 %
	Using the assumption that $\Pr_V\big(|\Tcal_{\Upsilon_m}^{(m)}|\geq 2\big)\leq nC_0'\rhom\leq 1/{2 e}$,  \cite[Lemma 2.7]{csiszar2011information} provides the following bound on the distance of the entropies:
	\ea{
		&|\Hsf(\Upsilon_m) - \Hsf(\nuv , \Xv_{\dm}^{\nuov}) | \nonumber\\
		& = |\De_1| \leq -	2\Pr_{\Upsilon_m}(\scaleto{|\Tcal_{\Upsilon_m}^{(m)}|\geq 2}{8pt}) \Big(\log \big(2\Pr_{\Upsilon_m}(\scaleto{|\Tcal_{\Upsilon_m}^{(m)}|\geq 2}{8pt})\big)-C'_1n\Big). \label{eqn: De1_ub}
		}
	Now, by using \eqref{eqn: pv}, the above inequality results in
	\ea{
			|\De_1| \leq& -2 n C'_0\rhom \times \nonumber\\
			&\Big(\log \big(\rhom\big) + \log(\underbrace{n C'_0}_{\geq 0})  - C'_1 n \Big)\nonumber\\
			\leq& -2 n C'_0\rhom (\log \big(\rhom\big)  -C'_1 n).
			\label{eqn: dee1}
	}

	\item {\bf Step (iii)}: Using the definition of $\Ytv_i$ in \eqref{eqn: ytv_i}, one can see that
	\ea{
		0&\leq \De_2 =\Ebb_{\Upsilon_m}\big[ \Isf(\Ytv_{\Upsilon_m}; \Upsilon_m^{(\Upsilon_m)})\big]	\leq \Ebb_{\Upsilon_m}\big[\Hsf(\Upsilon_m^{(\Upsilon_m)})\big]\nonumber\\
		&\leq \Ebb_{\Upsilon_m}\big[\log\big(\big|\text{sample space of }\Upsilon_m^{(\Upsilon_m)}\big|\big)\big]  \nonumber\\
		&\leq \Ebb_{\Upsilon_m}\big[\log\big(2^{C_1' n}\big) \,\, \big| \,\, |\Tcal_{\Upsilon_m}^{(m)}|\geq 2 \big]\cdot \Pr_{\Upsilon_m}\big(|\Tcal_{\Upsilon_m}^{(m)}|\geq 2\big)  \nonumber\\
		&\leq n C'_1\Pr_{\Upsilon_m}\big(|\Tcal_{\Upsilon_m}^{(m)}|\geq 2\big). \label{eqn: de2}
		}
%
Again we recall \eqref{eqn: pv} to bound the latter probability
	\ea{
	|\De_2|\leq n^2C'_1C'_0\rhom. \label{eqn: dee2}
	}

	\item {\bf Step (iv)}: We start by studying the moments of $T\Xv_{\cm}^{\sv}$, when $T$ is a submatrix of a unitary matrix:
	\changedo{\ea{
		\Ebb\Big[\|T\Xv_{\cm}^{\sv}\|_{\alpha}^{\beta} \Big]&\overset{(a)}{\leq} k^{\beta/\alpha} l^{\beta} \Ebb\big[\|\Xv_{\cm}^{\sv}\|_{\alpha}^{\beta}\big]	\\
		&\overset{(b)}{\leq} k^{\beta/\alpha} l^{\beta+\beta/\alpha-1} M_{X_{\cm, 1}}(\beta)\\
		&\leq n^{C'_2+1} M_{X_{\cm, 1}}(\beta),
	}}
	where $C'_2 =\beta(2/\alpha+1)-2$. The validity of $(a)$ is due to \eqref{eqn: norm_alpha_max} and the fact that the entries of $T$ do not exceed $1$ in absolute value ($T$ is a submatrix of a unitary matrix). $(b)$ is also achieved via \eqref{eqn: norm_ineq}.
	
	Since $T\Xv_{\cm}^{\sv}$ has some bounded moments, $\hsf(T\Xv_{\cm}^{\sv})$ is also finite. In particular, \cite[Corollary 1]{Zamir1994} establishes
	\ea{
		\hsf(T\Xv_{\cm}^{\sv})\leq \log M,	\label{eqn: h_log_M}
	}
	where $M$ is 
	\changedo{\ea{
		M 	= 	V_{\alpha, n} \Big({\alpha e n^{C'_2} M_{X_{\cm, 1}}(\beta)}\Big)^{n/\beta} \Gamma(1+\tfrac{n}{\beta}),
	}}
	in which $V_{\alpha, n}$ indicates the volume of the  $n$-dimensional unit-ball associated with the $\alpha$-norm. 
	Obviously, $V_{\alpha, n}\leq V_{\infty, n}<2^{n}$. Now, for $n\geq \beta e$ we have
	\ea{
		M &\leq 	\big({2^{\beta}\alpha e M_{X_{\cm, 1}}(\beta)n^{C'_2}}\big)^{n/\beta} \Gamma(1+\tfrac{n}{\beta}) \nonumber\\
		&\leq \big({2^{\beta}\alpha e M_{X_{\cm, 1}}(\beta)n^{C'_2}}\big)^{n/\beta} \lceil \tfrac{n}{\beta}\rceil! \nonumber\\
		&\overset{(a)}{\leq}  \big({2^{\beta}\alpha e M_{X_{\cm, 1}}(\beta)n^{C'_2}}\big)^{n/\beta} 
e^{\frac{3}{2}}  \Big(\tfrac{\lceil \frac{n}{\beta}\rceil}{e}		\Big) ^{\lceil \frac{n}{\beta}\rceil + \frac{1}{2}} \nonumber\\
		%
& \leq \big({2^{\beta}\alpha e M_{X_{\cm, 1}}(\beta)n^{C'_2}}\big)^{n/\beta} 
e^{\frac{3}{2}}  \Big(\frac{2 n}{\beta e}		\Big) ^{\frac{n}{\beta} + \frac{3}{2}} \nonumber\\
		&\leq\big(\tfrac{2}{\beta}\big)^{\frac{3}{2}} (C'_3)^n n^{C'_4n},\label{eqn: M_upper}
	}
for $C'_3= \sqrt[\beta]{\tfrac{2^{\beta+1} \alpha M_{X_{\cm, 1}}(\beta)}{\beta}}$ and $C'_4= \tfrac{\alpha+2}{\alpha}\geq \frac{C'_2}{\beta} + \frac{1}{\beta}+\frac{3}{2 n}$ (note that $\frac{n}{\beta}\geq \frac{3}{2}\geq 1$). $(a)$ is followed by the Stirling's approximation.
%
If $n\geq \frac{3}{2}\log \big(\frac{2}{\beta}\big)$, we can use \eqref{eqn: M_upper} in \eqref{eqn: h_log_M} as
\ea{
	\hsf(T\Xv_{\cm}^{\sv})&\leq \tfrac{3}{2}\log \big(\tfrac{2}{\beta}\big)+ n\log C'_3+C'_4n\log n\leq C'_5n\log n,	 \label{eqn: tx_upper}
}
for $C'_5 =1+|\log C'_3|+C'_4$.

Returning to \eqref{eq:Delta4}, we know
\ea{
	\De_4&\leq C'_5\log n + |\hsf(\Xv_{\cm}^{\sv})| \leq 
	C'_5\log n + 	 |\hsf(X_{\cm, 1})|n \nonumber\\
	&\leq C_6' n\log n,\label{eqn: de4_1}
}
where $C_6'=|\hsf(X_{\cm, 1})|+C'_5$ ($\Xv_{\cm}^{\sv}$ consists of a number of $X_{\cm, i}$s, and $X_{\cm, i}$s are i.i.d.).
%

Since $T$ is a submatrix of a unitary matrix, we can form $\Tt$ such that $(T; \Tt)$ is this unitary matrix. This implies that $\hsf\big((T; \Tt)\Xv_{\cm}^{\sv}\big) = \hsf(\Xv_{\cm}^{\sv})$. Besides,
\ea{
		\hsf\big((T; \Tt)\Xv_{\cm}^{\sv}\big) \leq \hsf(T\Xv_{\cm}^{\sv}) + \hsf(\Tt \Xv_{\cm}^{\sv}).
}
Hence,
\ea{
 -\De_4 &= \hsf(\Xv_{\cm}^{\sv}) - \hsf(T\Xv_{\cm}^{\sv}) = \hsf\big((T; \Tt)\Xv_{\cm}^{\sv}\big)  - \hsf(T\Xv_{\cm}^{\sv}) \nonumber\\
 &\leq \hsf(\Tt \Xv_{\cm}^{\sv}) \leq C'_5n\log n. \label{eqn: de4_2}
}
The latter inequality uses the same technique	 as in \eqref{eqn: tx_upper} for $\Tt$.
Combining \eqref{eqn: de4_1} and \eqref{eqn: de4_2}, we obtain
\ea{
	|\De_4| \leq 	C_6'n\log n. \label{eqn: de4}
}

\item {\bf Step (v)}:
Recalling $\psi_{m, \nuv}(A_m) = \tfrac{1}{m}\Ebb_{\nuv}[\log \det^{+} A_m^{[\nuv]}]$ and the definition of $\Wv_{\sv, \xv_{\dm}^{\sov}}$ in \eqref{eqn: W_s_x}, we have that
\ea{
\hsf(\Wv_{\nuv, \Xv_{\dm}^{\nuov}}|\nuv, \Xv_{\dm}^{\nuov}) = m \psi_{m, \nuv}(A_m)+\Ebb_{\nuv}\big[\hsf( \underset{\hspace{-1mm}{\leftharpoonup}}{\Ut_{\nuv}}^{\dagger} X_c^{\nuv})\big]	,
}
where $\underset{\hspace{-1mm}{\leftharpoonup}}{\Ut_{\nuv}}$ is the matrix formed by the first $\rank (A_m^{[\nuv]})$ left singular vectors of $A_m^{[\nuv]}$. As a result 
\ea{
|\De_3| &=\Big|\Ebb_{\nuv}\big[\hsf( \underset{\hspace{-1mm}{\leftharpoonup}}{\Ut_{\nuv}}^{\dagger} \Xv_{\cm}^{\nuv}) \big]-\hsf(\Xv_{\cm}^{\nuv} \big| \nuv)	\Big| \nonumber\\
&=\Big|\Ebb_{\nuv}\big[\hsf( \underset{\hspace{-1mm}{\leftharpoonup}}{\Ut_{\nuv}}^{\dagger} \Xv_{\cm}^{\nuv}) -\hsf(\Xv_{\cm}^{\nuv}) \big]	\Big| \nonumber\\
&\leq \Ebb_{\nuv}\Big[\big|\hsf( \underset{\hspace{-1mm}{\leftharpoonup}}{\Ut_{\nuv}}^{\dagger} \Xv_{\cm}^{\nuv})-\hsf(\Xv_{\cm}^{\nuv})\big|\Big] \nonumber\\
&{\leq} C_6' n\log (n)\rhom, \label{eqn: dee3}
}
where we used \eqref{eqn: de4} for $T=\underset{\hspace{-1mm}{\leftharpoonup}}{\Ut_{\nuv}}^{\dagger}$ in the last inequality. Note that $\rhom$ is the probability that $A_m^{[\nuv]}$ is not of full column-rank. When $A_m^{[\nuv]}$ has full  column-rank, $\underset{\hspace{-1mm}{\leftharpoonup}}{\Ut_{\nuv}}$ is a square unitary matrix and $\hsf( \underset{\hspace{-1mm}{\leftharpoonup}}{\Ut_{\nuv}}^{\dagger} \Xv_{\cm}^{\nuv}) = \hsf(\Xv_{\cm}^{\nuv})$.


\item {\bf Step (vi)}: 
For $\alpha_1\neq 0$, it is now easy to complete the proof using \eqref{eqn: dee1}, \eqref{eqn: dee2}, \eqref{eqn: dee3}, and by setting $C_1= C_6'+3C'_1C'_0$ and $C_2=-2C'_0$.
	\end{noindlist}

In the case of $\alpha_1=0$,  $X_1$ and $\Yv^m$ (i.e., a linear combination of $X_i$s) are purely discrete.
%
	Hence, by \eqref{eqn: sing_drb} and \eqref{eqn: b_decomp}, we have
	\ea{
		nb(X_1)=\Hsf(\nuv, \Xv_{\dm}^{\nuov}).	
	}
	Furthermore, Theorem \ref{thm: disc_cont_fat} implies that
	\ea{
		b(\Yv^m) 	= 	\Hsf(\Upsilon_m).	
	}
	As a result, we have
	\ea{
		\frac{\big|b(\Yv^m)-n b(X_1)\big|}{m}\leq \frac{|\De_1|}{m},
	}
	for $\De_1= \Hsf(\Upsilon_m)- \Hsf(\nuv, \Xv_{\dm}^{\nuov})$ as also defined in \eqref{eqn: De1}.

Similar to Step (i), we first bound $\Pr_{\Upsilon_m}\big(|\Tcal^{(m)}_{\Upsilon_m}\geq 2\big)$, and then, follow the same approach as in  Step (ii) to bound $|\De_1|$. 
%
%
Since $\nuv = 0_{1\times n}$, $|\Tcal_{i}^{(m)}|\geq 2$ implies that for every pair $(0_{1\times n}, \xv_{\dm}) \in \Tcal_i^{(m)}$, there exists $\xv_{\dm}'\neq \xv_{\dm}$, such that $(0_{1\times n}, \xv_{\dm}') \in \Tcal_i^{(m)}$ and $A_m \xv_{\dm} = A_m \xv_{\dm}'$. This only happens when $\spanM (\Acal_m^{\tv})= \spanM (\Acal_m^{\tv'})$, where $\tv$ and $\tv'$ are $n$-dimensional indicators to show the non-zero status of the elements in $\xv_{\dm}$ and $\xv_{\dm}'$, respectively. One can verify that $\tv$ and $\tv'$ are distributed as $\xiv$ which is defined in  Lemma \ref{lem: rho}. 
Consequently, similar to Step (i), we can prove that
	\ea{
	\Pr_{V_m}(|\Tcal_{V_m}^{(m)}|\geq 2)\leq nC'_0 \rhotm,
}
where $\rhotm$ is defined in \eqref{eqn: rhotm}.
Finally, by setting $\Ct_1=-2C_0'$ and $\Ct_2=2C_0'C_1'$ in \eqref{eqn: De1_ub}, and using the latter inequality, 
we can achieve \eqref{eqn: disc_b} that completes the proof.

\section{Proof of Proposition \ref{Prop: Psi}} \label{app: prf_prop_psi}
We prove this theorem in five steps: 
\begin{enumerate}[(i)]
\item \label{step:length}
for an $n$-dimensional random vector $\nuv$ with i.i.d. Bernoulli elements, we examine the distribution of the length of maximal sub-blocks with all $1$ elements, 

\item \label{step:pseudo-determinant}
we study the pseudo-determinant of four classes of square matrices; we provide lower- and upper-bounds on some of them, while for the rest, we derive closed-form expressions,

\item
for a given $\sv$, 
we express $\det^{+}(A_m^{[\sv]})$ 
 via the block-lengths in Step (\ref{step:length}) and the pseudo-determinants introduced in Step (\ref{step:pseudo-determinant}),

\item
we prove the existence of the expected value of the logarithm of the pseudo-determinants introduced in Step (\ref{step:pseudo-determinant}) by bounding them, and

\item
we find \changedo{$\Psi_{\alpha}(c_1, c_2)$} in terms of an expected value of a function of an $\alpha$-geometric RV.

\end{enumerate} 
\changedo{To simplify the results, we set $c_1=a$ and $c_2=b$.}

\begin{noindlist}
	\item {\bf Step (i)}: 
	\begin{defi}
		For an $n$-dimensional binary vector $\sv$ and integers $1\leq i\leq j\leq n$, let $E_i^j(\sv)$ be the event that $s_i=\dots=s_j = 1$ while $s_{i-1}=s_{j+1}=0$; by convention, we assume $s_0=s_{n+1}=0$. 
		Besides, we denote the the indicator of $E_{i}^{j}$  by  $\varphi_i^j(\sv)$:
		\ea{
			\varphi_{i}^j(\sv)	= (1-{s_{i-1}})(1-s_{j+1})\prod\limits_{k=i}^{k=j} s_k.
		}
To account for the length of the $1$-blocks, we define
			\ea{
			\varphi_i(\sv) := \left\{\begin{array}{ll}
			\sum_{k=i}^{n-1} (k-i+1)\varphi_{i}^k(\sv), & 1\leq i\leq n-1,\\
			\sum_{k=2}^{n} (n-k+1)\varphi_{k}^n(\sv), & i= n \phantom{\Big|}, \\
			1+ (n-1) \varphi_{1}^n(\sv), & i= n+1.
			\end{array}
			\right. \label{eqn: varphi_i}
		}
		Indeed, for $1\leq i< n$, $\varphi_i(\sv)$ indicates the length of the $1$-block starting at the $i$th location (except when the vector consists solely of $1$s),  $\varphi_n(\sv)$ shows the length of the $1$-block ending at the last location (except when the vector consists solely of $1$s), and $\varphi_{n+1}(\sv)$ takes the value $n$ if the vector consists solely of $1$s, and $1$ otherwise. Note that if the $i$th location ($i<n$) is not the starting place of a $1$-block, then, $\varphi_i(\sv)=0$. 
		
	\end{defi}

	\begin{rem} \label{rem: varphi}
		Let $\nuv$ be an $n$-dimensional vector with i.i.d. elements distributed as ${\rm Bern}(\alpha)$. The distribution of $\varphi_i(\nuv)$ can be obtained as
		\ea{\label{eq:PrPhi}
			\Pr\big(\varphi_i(\nuv) = x\big){=} \left\{\begin{array}{cl}
				\alpha^{x}(1-\alpha), & i=1, n, ~ 1\leq x \leq n-1, \\
				\alpha^{n}, & i=n+1, ~ x=n, \phantom{\Big|}\\
				1-\alpha^{n}, & i=n+1, ~ x=1, \phantom{\Big|}\\
				 \alpha^{x}(1-\alpha)^2, & 1< i < n , ~ 1\leq x\leq n-i, \\
			\end{array}\right.
		}
		which includes all possible non-zero values of $x$.
%
	\end{rem}

	\item {\bf Step (ii)}: To simplify the below arguments we define $b_k(x_1, x_2) := \det^{+} B_k(x_1, x_2)$ for the $k\times k$ tridiagonal matrix $B_k(x_1, x_2)$ given as
	\ea{
		B_k(x_1, x_2):= \left( \begin{array}{c c c c c}
			x_1& ab & \ldots & \ldots & 0\\
			ab &a^2+b^2 & \ddots & \vdots & \vdots\\
			\vdots& \ddots & \ddots & \ddots & \vdots\\
			\vdots & \ldots & \ddots & a^2+b^2 & ab\\
			0 & \ldots & \ldots & ab & x_2
		\end{array}\right),
	}
	where $a,b$ are non-zero real numbers. 
	We need to evaluate and bound
\ea{
E_k &:=b_k(a^2+b^2 \,,\, a^2+b^2), \nonumber\\
F_k &:=b_k(a^2 \,,\, a^2+b^2), \nonumber\\
G_k &:=b_k(a^2+b^2 \,,\, b^2), \nonumber\\
H_k &:=b_k(a^2 \,,\, b^2). \label{eq:EFGH}
}	
First note that if 
\ea{
A_{k} =\left( \begin{array}{c c c c c c c}
			a 			& b 			& 0 			& 0 			& \dots 	& 0			& 0		\\
			0 			& a 			& b 			& 0 			& \dots 	& 	0 			& 0		\\
			\vdots 	& \vdots 	& \vdots 	& \vdots	& \ddots 	& \vdots	& 0		\\
			0 			& 0 			& 0 			& 0 			& \dots 	& 	a 			& b 		\\
		\end{array}\right)_{k\times (k+1)},
}
then, $A_k^{T} \times A_k = B_{k+1}(a^2 \,,\, b^2)$ and $A_k \times A_k^{T} = B_{k}(a^2+b^2 \,,\, a^2+b^2)$. Thus, $B_{k+1}(a^2 \,,\, b^2)$ and $B_{k}(a^2+b^2 \,,\, a^2+b^2)$ are non-negative-definite matrices. Since  
$\det^{+} \big(A_k^{T} \times A_k) = \det^{+} \big(A_k \times A_k^{\intercal})$, we conclude that 
\ea{
H_{k+1} = E_{k}.  \label{eq:HE}
}
Besides, as $B_{k}(a^2 \,,\, a^2+b^2)$ and $B_{k}(a^2+b^2 \,,\, b^2)$ are principal submatrices of $B_{k+1}(a^2 \,,\, b^2)$, they are also non-negative-definite as well. 
The non-negative-definite property of $B_{k}(a^2+b^2 \,,\, a^2+b^2)$, $B_{k}(a^2 \,,\, a^2+b^2)$ and $B_{k}(a^2+b^2 \,,\, b^2)$ implies that their determinants coincide with their $\det^{+}$ if they are full-rank. Below, we evaluate the determinants and as they are non-zero, they coincide with their $\det^{+}$.

	We start by the recurrence relation of $E_k$ (the $\det$ version) as
	\ea{
		\left\{ 
		\begin{array}{cc}
			E_1 = & a^2 + b^2\\
			E_2 = & a^4 + b^4 - a^2b^2\\
			E_k =& (a^2+b^2) E_{k-1} - a^2 b^2 E_{k-2}
		\end{array}
		\right. .
	}
	It is not difficult to solve the above constant-coefficient linear recursive equation to find
	\ea{
		E_k = \frac{a^{2k+2}-b^{2k+2}}{a^2-b^2}, \label{eqn: EK1}
	}
	if $a^2\neq b^2$, and
	\ea{
		E_k = a^{2k} (1+k), \label{eqn: EK2}
	}
	if $a^2=b^2$. Using \eqref{eq:HE}, $H_k$ is also obtained as
	\ea{
		H_k = \left\{\begin{array}{ll}
		\frac{a^{2k}-b^{2k}}{a^2-b^2} , & a\neq b, \phantom{\Big|}\\
		k \, a^{2k-2}, & a=b. \phantom{\Big|}
		\end{array}
		\right. \label{eqn:HK}
	}

	Using these identities, one can show that
	\ea{
		e_l^k\leq E_k = H_{k+1}\leq e_u^k, \label{eqn: Ek_bound1}
	}
	where $e_l := a^2$ and $e_u:=a^2+b^2$.
	
	If we expand the determinant with respect to the first row of the matrix, we can express $F_k$ and $G_k$ in terms of $E_k$ as
%
	\ea{
		F_k=a^2 E_{k-1} -a^2b^2 E_{k-2} = a^{2k}, \label{eqn: Fk}
	}
	and
	\ea{
		G_{k} = b^2 E_{k-1} - a^2 b^2 E_{k-2} = b^{2k}. \label{eqn: Gk}
	}
	By convention, we set $E_0=F_0=G_0 = 1$.
%
%
%

		\item {\bf Step (iii)}: 
		Let $\varsigma$ denote the Gramian matrix $A_{n}^{\intercal} A_{n} = B_{n+1}(a^2, b^2)$. 
	We know that
	\ea{
		(\det^{+} A_n^{[\sv]})^2 = \det^{+}  \big((A_n^{[\sv]})^{\intercal} A_n^{[\sv]} \big)= \det^{+} \varsigma_{\sv}, \label{eqn: pseudo_gram}
	}
	where $\varsigma_{\sv}$ is the principal submatrix of $\varsigma$ formed by columns and rows for which $s_i=1$.
	
	Because $\varsigma$ is a tridiagonal matrix, $\varsigma_{\sv}$ is a block-diagonal matrix consisting of blocks associated with $1$-blocks in $\sv$. 
	More specifically, by definition of $\varphi_i(\sv)$ in \eqref{eqn: varphi_i}  and $E_k$, $F_k$, $G_k$, $H_k$ in Step (ii), we conclude that
	\ea{
		\det^{+} \varsigma_{\sv} = 
		 F_{\varphi_1(\sv)} G_{\varphi_n(\sv)} H_{\varphi_{n+1}(\sv)} \prod\limits_{i\in [2:n-1]} E_{\varphi_i(\sv)}.
		\label{eqn: gram_s}
	} 

	\item {\bf Step (iv)}: In this step, we show that $\Ebb_Y[\log E_Y]$ exists and is bounded when $Y$ has the following $\alpha$-geometric distribution:
	\ea{\label{eq:Geometric}
	\mathop{\Pr(Y=k)}_{k=0,1,\dots} = \alpha^{k} (1-\alpha).
	} 
	Due to \eqref{eqn: Ek_bound1}, we know that
	\ea{
		k \log e_l \leq \log E_k \leq k \log e_u .
	}
	Therefore,
		\ea{
		\underbrace{\Ebb_{Y}[Y]}_{\frac{\alpha}{1-\alpha}} \log e_l \leq \Ebb_{Y}[\log E_Y] \leq \underbrace{\Ebb_{Y}[Y]}_{\frac{\alpha}{1-\alpha}}  \log e_u .
	}
	Note that $\Ebb_{Y}[\log E_Y]  = \sum_{k=0}^{\infty} \Pr(Y=k) \log E_k$. 
	Based on \eqref{eqn: EK1} and \eqref{eqn: EK2}, we know that $E_k$ is increasing for large values of $k$ if $\max(|a|\,,\, |b|) \geq 1$ and decreasing for large values of $k$ otherwise. Thereby, the sign of $\log E_k$ remains the same for large $k$s. This shows that the terms in $\sum_{k=0}^{\infty} \Pr(Y=k) \log E_k$ after some $k$, are all positive or negative. Besides, the sum is bounded; hence, because of the monotone convergence theorem \cite[pp. 125]{Pugh15} the sum is convergent and bounded.
	
%
%
%
%

\item {\bf Step (v)}:
Our goal is to show that for $\alpha<1$ (the distribution is not absolutely continuous),  $\frac{1}{n}\Ebb_{\nuv}\big[\log \det^{+}(A^{[\nuv]})\big]$ asymptotically approaches $\frac{1-\alpha}{2}\Ebb_{Y}[\log E_Y]$ when $n\to \infty$, where $Y$ is an $\alpha$-geometric RV. To show this, first note that
\ea{
& \Ebb_{\nuv}\big[\log \det^{+}(A^{[\nuv]})\big] =  
 \tfrac{1}{2}\Ebb_{\nuv}[\log F_{\varphi_1(\nuv)}+\log G_{\varphi_n(\nuv)} ] \nonumber\\
&\hspace{2cm} + \tfrac{1}{2}\alpha^{n} \log H_n   + \tfrac{1}{2}\sum_{i=2}^{n-1} \Ebb_{\nuv}[\log E_{\varphi_i(\nuv)}].
}
We evaluate the involved terms separately:
\ea{
\Ebb_{\nuv}[\log F_{\varphi_1(\nuv)} &+ \log G_{\varphi_n(\nuv)}] = \sum_{k=1}^{n-1} 2k \alpha^{k}(1-\alpha) \log (|ab|) \nonumber\\
&= \tfrac{2\alpha \big(1 - n \alpha^{n-1} + (n-1)\alpha^n \big)}{1-\alpha} \log(|ab|).
}
As $\alpha<1$, it is easy to check that $\lim_{n\to \infty} \frac{1}{n}\Ebb_{\nuv}[\log F_{\varphi_1(\nuv)} + \log G_{\varphi_n(\nuv)}] =0$. Similarly, from \eqref{eqn: Ek_bound1}, we have that
\ea{
 \alpha^n (n-1) \log(e_l) \leq \alpha^n \log(H_n) \leq \alpha^n (n-1) \log(e_h) .
}
Thus, $\lim_{n\to \infty} \frac{1}{n} \alpha^n \log(H_n) = 0$.

Finally, we consider
\ea{
&\sum_{i=2}^{n-1} \Ebb_{\nuv}[\log E_{\varphi_i(\sv)} ] = \sum_{i=2}^{n-1} \sum_{k=1}^{n-i} \alpha^{k}(1-\alpha)^2 \log(E_k)     \nonumber\\
&=\sum_{k=1}^{n-2} (n-k)\alpha^{k}(1-\alpha)^2 \log(E_k)   \nonumber\\
&= n (1-\alpha)\sum_{k=1}^{n-2} \alpha^{k}(1-\alpha) \log(E_k) - (1-\alpha)^2 \sum_{k=1}^{n-2} k\alpha^{k} \log(E_k) \nonumber\\
&= n (1-\alpha) \Ebb_{Y}[\log E_Y] - (1-\alpha)^2 \underbrace{\sum_{k=1}^{n-2} k\alpha^{k} \log(E_k)}_{\Delta_1} \nonumber\\
& \hspace{2cm} - n (1-\alpha) \underbrace{\sum_{k=n-1}^{\infty} \alpha^{k}(1-\alpha) \log(E_k)}_{\Delta_2}.
}
To complete the proof, we need to show $\lim_{n\to \infty} \frac{\Delta_1}{n} = \lim_{n\to \infty } \Delta_2 = 0$. 
The following two qualities are helpful for this purpose:
\ea{
\ell_1(n) &:= \sum_{k=n}^{\infty} k\alpha^k \nonumber\\
& = \tfrac{-n \alpha^{n+1} + \alpha^{n+1} + n\alpha^{n}}{(1-\alpha)^2}, \\
\ell_2(n) & := \sum_{k=1}^{n} k^2 \alpha^k \nonumber\\
& = \tfrac{-n^2\alpha^{n+3} +(2n^2+2n-1)\alpha^{n+2} - (n+1)^2 \alpha^{n+1} +\alpha^2+\alpha}{(1-\alpha)^3}.
}
Now, based on \eqref{eqn: Ek_bound1}, we have that
\ea{
\ell_2(n-2) \log(e_l) \leq \Delta_1 \leq \ell_2(n-2) \log(e_u)
}
and
\ea{
\ell_1(n-1) \log(e_l) \leq \Delta_2 \leq \ell_1(n-1) \log(e_u).
}
For both cases, it is now straightforward to conclude that $\lim_{n\to \infty} \frac{\Delta_1}{n} = \lim_{n\to \infty } \Delta_2 = 0$.
%

\end{noindlist}

{\changed

\end{document}